\documentclass[iop,numberedappendix,twocolappendix]{emulateapj}

\usepackage[morefloats=7]{morefloats}
\usepackage[breaklinks,colorlinks,citecolor=blue,bookmarks=false]{hyperref}
\usepackage[all]{hypcap}
\usepackage{float}
\usepackage{color}
\providecommand{\sorthelp}[1]{}

\newcommand{\planck}{{\em Planck}}
\newcommand{\herschel}{{\em Herschel}}

\newcommand{\blastpol}{BLASTPol}

\newcommand{\um}{$\mu$m}                            

\newcommand{\msun}{$M_{\sun}$} 

\newcommand{\prs}{$Z_x$}

\newcommand{\ppol}{$\ppol$}

\def\3he{$^3{\rm He}$}
\def\4he{$^4{\rm He}$}

\newcommand{\mzero}{$I$}
\newcommand{\mone}{$\left< v \right>$}
\newcommand{\mtwo}{$\Delta v$}
\newcommand{\mmzero}{I}
\newcommand{\mmone}{\left< v \right>}
\newcommand{\mmtwo}{\Delta v}
\newcommand{\joz}{$J$\,=\,1\,$\rightarrow$\,0}

\newcommand{\bpos}{$\mathbf{\langle\hat{B}_{\perp}\rangle}$}
\newcommand{\edir}{$\mathbf{\hat{E}}$}
\newcommand{\mbpos}{\mathbf{\langle\hat{B}_{\perp}\rangle}}
\newcommand{\nradex}{$n_{\mathrm{H}_2\,\mathrm{rad}}$}
\newcommand{\ndepth}{$n_{\mathrm{H}_2\,\mathrm{x-sect}}$}
\newcommand{\nind}{$n_{\mathrm{ind}}$}
\newcommand{\mnind}{n_{\mathrm{ind}}}
\newcommand{\mndepth}{n_{\mathrm{H}_2\,\mathrm{x-sect}}}

\newcommand{\nh}{$N_\mathrm{H}$}

\newcommand{\tex}{$T_{\mathrm{ex}}$}

\newcommand{\tkin}{$T_{\mathrm{k}}$}
\def\,{\thinspace}
\def\lsim{\mathrel{\raise .4ex\hbox{\rlap{$<$}\lower 1.2ex\hbox{$\sim$}}}}
\def\gsim{\mathrel{\raise .4ex\hbox{\rlap{$>$}\lower 1.2ex\hbox{$\sim$}}}}

\def\simprop{\mathrel{\raise .4ex\hbox{\rlap{$\propto$}\lower 1.2ex\hbox{$\sim$}}}}
\def\deg{\ifmmode^\circ\else$^\circ$\fi}
\def\pdeg{\ifmmode $\setbox0=\hbox{$^{\circ}$}\rlap{\hskip.11\wd0 .}$^{\circ}
          \else \setbox0=\hbox{$^{\circ}$}\rlap{\hskip.11\wd0 .}$^{\circ}$\fi}
\def\arcs{\ifmmode {^{\scriptstyle\prime\prime}}
          \else $^{\scriptstyle\prime\prime}$\fi}
\def\arcm{\ifmmode {^{\scriptstyle\prime}}
          \else $^{\scriptstyle\prime}$\fi}
\newdimen\sa  \newdimen\sb
\def\parcs{\sa=.07em \sb=.03em
     \ifmmode \hbox{\rlap{.}}^{\scriptstyle\prime\kern -\sb\prime}\hbox{\kern -\sa}
     \else \rlap{.}$^{\scriptstyle\prime\kern -\sb\prime}$\kern -\sa\fi}
\def\parcm{\sa=.08em \sb=.03em
     \ifmmode \hbox{\rlap{.}\kern\sa}^{\scriptstyle\prime}\hbox{\kern-\sb}
     \else \rlap{.}\kern\sa$^{\scriptstyle\prime}$\kern-\sb\fi}

\slugcomment{accepted for publication in the Astrophysical Journal}

\shorttitle{Magnetic Field Orientation vs Cloud Density in Vela C }
\shortauthors{Fissel et al.}

\begin{document}

\title{Relative Alignment Between the Magnetic Field and Molecular Gas Structure in the Vela C Giant Molecular Cloud using Low and High Density Tracers}

\author{Laura M. Fissel\altaffilmark{1},
Peter A. R. Ade\altaffilmark{2},
Francesco E. Angil\`e\altaffilmark{3},
Peter Ashton\altaffilmark{4},
Steven J. Benton\altaffilmark{5},
Che-Yu Chen\altaffilmark{6},
Maria Cunningham\altaffilmark{7},
Mark J. Devlin\altaffilmark{3},
Bradley Dober\altaffilmark{8}, 
Rachel Friesen\altaffilmark{1},
Yasuo Fukui\altaffilmark{9},
Nicholas Galitzki\altaffilmark{10}, 
Natalie N. Gandilo\altaffilmark{11}, 
Alyssa Goodman\altaffilmark{12},
Claire-Elise Green\altaffilmark{7},
Paul Jones\altaffilmark{7}, 
Jeffrey Klein\altaffilmark{3},
Patrick King\altaffilmark{6},
Andrei L. Korotkov\altaffilmark{13},
Zhi-Yun Li\altaffilmark{6},
Vicki Lowe\altaffilmark{7},
Peter G. Martin\altaffilmark{14},
Tristan G. Matthews\altaffilmark{15},
Lorenzo Moncelsi\altaffilmark{16},
Fumitaka Nakamura\altaffilmark{17},
Calvin B. Netterfield\altaffilmark{18,19},
Amanda Newmark\altaffilmark{20},
Giles Novak\altaffilmark{15},
Enzo Pascale\altaffilmark{21},
Fr{\'e}d{\'e}rick Poidevin\altaffilmark{22,23},
Fabio P. Santos\altaffilmark{24},
Giorgio Savini\altaffilmark{25},
Douglas Scott\altaffilmark{26},
Jamil A. Shariff\altaffilmark{14},
Juan D. Soler\altaffilmark{24},
Nicholas E. Thomas\altaffilmark{27},
Carole E. Tucker\altaffilmark{2},
Gregory S. Tucker\altaffilmark{13},
Derek Ward-Thompson\altaffilmark{28},
Catherine Zucker\altaffilmark{12}}

\altaffiltext{1}{National Radio Astronomy Observatory, 520 Edgemont Rd, Charlottesville, VA, 22903, U.S.A.}
\altaffiltext{2}{Cardiff University, School of Physics \& Astronomy, Queens Buildings, The Parade, Cardiff, CF24 3AA, U.K.} 
\altaffiltext{3}{Department of Physics \& Astronomy, University of Pennsylvania, 209 South 33rd Street, Philadelphia, PA, 19104, U.S.A.} 
\altaffiltext{4}{Department of Physics, University of California, Berkeley, 366 LeConte Hall, Berkeley, CA 94720, U.S.A.}
\altaffiltext{5}{Department of Physics, Princeton University, Jadwin Hall, Princeton, NJ 08544, U.S.A.}
\altaffiltext{6}{Department of Astronomy, University of Virginia, 530 McCormick Rd, Charlottesville, VA 22904, U.S.A.}
\altaffiltext{7}{School of Physics, University of New South Wales, Sydney NSW 2052, Australia}
\altaffiltext{8}{National Institute for Standards and Technology,  325 Broadway, Boulder, CO 80305, U.S.A.}
\altaffiltext{9}{Department of Physics and Astrophysics, Nagoya University, Nagoya 464-8602, Japan}
\altaffiltext{10}{Center for Astrophysics and Space Sciences, University of California, San Diego, 9500 Gilman Drive \#0424, La Jolla, CA, 92093, U.S.A.}
\altaffiltext{11}{Arizona Radio Observatory, University of Arizona, 933 North Cherry Avenue, Rm.~N204 , Tucson, AZ 85721-0065, U.S.A.}
\altaffiltext{12}{Harvard Astronomy, Harvard-Smithsonian Center for Astrophysics, 60 Garden St., Cambridge, MA 02138, USA}
\altaffiltext{13}{Department of Physics, Brown University, 182 Hope Street, Providence, RI, 02912, U.S.A.}
\altaffiltext{14}{CITA, University of Toronto, 60 St.~George St., Toronto, ON M5S 3H8, Canada}
\altaffiltext{15}{Center for Interdisciplinary Exploration and Research in Astrophysics (CIERA) and Department\ of Physics \& Astronomy, Northwestern University, 2145 Sheridan Road, Evanston, IL 60208, U.S.A.}
\altaffiltext{16}{California Institute of Technology, 1200 E. California Blvd., Pasadena, CA, 91125, U.S.A.}
\altaffiltext{17}{National Astronomical Observatory, Mitaka, Tokyo 181-8588, Japan}
\altaffiltext{18}{Department of Physics, University of Toronto, 60 St. George Street Toronto, ON M5S 1A7, Canada}
\altaffiltext{19}{Department of Astronomy \& Astrophysics, University of Toronto, 50 St. George Street Toronto, ON M5S 3H4, Canada}
\altaffiltext{20}{Department of Astrophysical Sciences, Princeton University, 4 Ivy Lane,
Princeton, NJ 08544 U.S.A.}
\altaffiltext{21}{Department of Physics, La Sapienza Universit{\`a} di Roma, Piazzale Aldo Moro 2, 00185 Roma, Italy}
\altaffiltext{22}{Department of Astrophysics Research, Instituto de Astrofísica de Canarias, E-38200 La Laguna, Tenerife, Spain}
\altaffiltext{23}{Universidad de La Laguna, Departamento de Astrofisica, E-38206, La Laguna, Tenerife, Spain}
\altaffiltext{24}{Max-Planck-Institute for Astronomy, Konigstuhl 17, 69117, Heidelberg, Germany}
\altaffiltext{25}{Department of Physics \& Astronomy, University College London, Gower Street, London, WC1E 6BT, U.K.}
\altaffiltext{26}{Department of Physics \& Astronomy, University of British Columbia, 6224 Agricultural Road, Vancouver, BC V6T 1Z1, Canada}
\altaffiltext{27}{NASA/Goddard Space Flight Center, Greenbelt , MD 20771, U.S.A.}
\altaffiltext{28}{Jeremiah Horrocks Institute, University of Central Lancashire, PR1 2HE, U.K.}

\begin{abstract}
We compare the magnetic field orientation for the young giant molecular cloud Vela\,C inferred from 500-$\mu$m~polarization maps made with the BLASTPol balloon-borne polarimeter to the orientation of structures in the integrated line emission maps from Mopra observations.  Averaging over the entire cloud we find that elongated structures in integrated line-intensity, or zeroth-moment maps, for low density tracers such as $^{12}$CO and $^{13}$CO~$J$\,$\rightarrow$\,1\,--\,0 are statistically more likely to align parallel to the magnetic field, while intermediate or high density tracers show (on average) a tendency for alignment perpendicular to the magnetic field.  This observation agrees with previous studies of the change in relative orientation with column density in Vela\,C, and supports a model where the magnetic field is strong enough to have influenced the formation of dense gas structures within Vela\,C. The transition from parallel to no preferred/perpendicular orientation appears to happen between the densities traced by $^{13}$CO and by C$^{18}$O~$J$\,$\rightarrow$\,1\,--\,0. Using RADEX radiative transfer models to estimate the characteristic number density traced by each molecular line we find that  the transition occurs at a molecular hydrogen number density of approximately\,$10^3$\,cm$^{-3}$.  We also see that the Centre-Ridge (the highest column density and most active star-forming region within Vela\,C) appears to have a transition at a lower number density, suggesting that this may depend on the evolutionary state of the cloud.
\end{abstract}


\section{Introduction}\label{sect:intro}

Molecular clouds form out of the diffuse gas in the interstellar medium, which is both turbulent and magnetized.  In the process of cloud formation the magnetic fields may play an important role in determining how quickly dense gravitationally unstable molecular gas forms \citep{mckee_2007}.

Direct measurement of magnetic field strength in molecular clouds is possible only through observations of Zeeman splitting in a few molecular line species.  
However, because Doppler line broadening is typically much larger than the Zeeman splitting width, only a few dozen detections of Zeeman splitting in molecular gas have been made to date \citep{crutcher_2012}, and at present there is no efficient way of creating large maps of the magnetic fields within molecular clouds using Zeeman observations.

An alternative method for studying magnetic fields in molecular clouds is to measure the magnetic field morphology through observations of linearly polarized radiation emitted by dust grains within the clouds.  Dust grains are known to align with their long axes on average perpendicular to the local magnetic field (see \citealt{andersson_2015} for a recent review).  Observations of stars at optical or near-IR wavelengths located behind the cloud show polarization parallel to the direction of the magnetic field projected on to the plane of the sky, \bpos, due to differential extinction. Thermal dust emission, in contrast, should be linearly polarized, with an orientation perpendicular to \bpos, and can be used to probe the magnetic field in the higher column density cloud material.  Polarized dust emission can therefore be used to construct a detailed ``portrait'' of the cloud magnetic field morphology, weighted by density, dust emissivity, and grain alignment efficiency.

Comparisons of the orientation of molecular cloud structure to the orientation of the magnetic field inferred from polarization are often used to study the role played by magnetic fields in the formation and evolution of dense molecular cloud structures (e.g., \citealt{tassis_2009,li_2013}).  \cite{goldsmith_2008} observed elongated molecular gas ``striations'' in the diffuse envelope of the Taurus molecular cloud that are parallel to the cloud magnetic field traced by polarization.  \cite{heyer_2008} later measured the velocity anisotropy associated with the Taurus $^{12}$CO\,$J$\,=\,1\,$\rightarrow$\,0 observations and concluded that the envelope of Taurus is magnetically subcritical (i.e., magnetically supported against self-gravity).

\cite{soler_2013} introduced the Histograms of Relative Orientation (hereafter HRO) technique, a method that statistically compares the orientation of \bpos~to the local orientation of structures in maps of hydrogen column density (\nh), as characterized by the \nh~gradient field. Applying the HRO method to synthetic observations of 4-pc$^3$~3D MHD RAMSES numerical simulations, \cite{soler_2013} showed that for weakly magnetized gas (where the squared ratio of the sound speed to Alfv\'en speed, $\beta$\,=\,$c^2_{\mathrm{s}}$/$v_{\mathrm{A}}^2$\,=\,100), the magnetic field is preferentially oriented parallel to iso-column density contours for all values of \nh.   In contrast, strong field simulations ($\beta$\,=\,0.1) showed a change in relative orientation between the magnetic field and iso-\nh~contours with increasing \nh~from parallel (for \nh\,$\lsim\,10^{22}$\,cm$^{-2}$) to perpendicular (for \nh\,$\gsim\,10^{22}$\,cm$^{-2}$). Similar results were obtained for strongly magnetized clouds by \cite{chen_2016}. 

Applying the HRO method to actual polarimetry data generally requires a large sample of inferred magnetic field measurements over a wide range in column density.  \cite{planck2016-XXXV} first applied this method to \planck~satellite 353-GHz polarization maps of 10 nearby ($d$\,$<$\,400\,pc) molecular clouds with 10\arcmin~resolution.  They showed that the relative orientation between \bpos~and elongated structures in dust images~changes progressively from preferentially parallel at low \nh~to preferentially perpendicular (or no preferred orientation) at high \nh, with the $\log\left(N_{\mathrm H}\right)$~of the transition ranging from 21.7 (Chamaeleon-Musca) to 24.1 (Corona Australis), though the precise value of the transition depends on the dust opacity assumed.   The change in relative orientation observed by \cite{planck2016-XXXV}~is most consistent with the intermediate or high magnetic field strength simulations from \cite{soler_2013}, suggesting that the global magnetic field strength in most molecular clouds is of sufficient strength to play an important role in the overall cloud dynamics.  However, this study included only one high-mass star-forming region, the Orion Molecular Cloud, which is a highly evolved cloud complex where the magnetic field has likely been altered by feedback from previous generations of massive stars \citep{bally_2008}.

In \cite{soler_2017}~the HRO technique was applied to a more distant and younger giant molecular cloud, namely Vela\,C, using detailed polarization maps at 250, 350, and 500\,$\mu$m from the BLASTPol balloon-borne telescope.  Vela\,C was discovered by \cite{murphy_1991} and has $>$10$^5$\,\msun~of molecular gas with $M\,\approx\,5\,\times\,10^4$\msun~of dense gas as traced by the C$^{18}$O~$J$\,=\,1\,$\rightarrow$\,0 observations of \cite{yamaguchi_1999}.   Far-IR and sub-mm studies of Vela\,C from the BLAST and {\em Herschel}~telescopes indicate a cloud that appears to  be mostly cold ($T_{\mathrm{dust}}\,\simeq\,$10--16\,K) with a few areas of recent and ongoing star formation \citep{netterfield_2009, hill_2011}, most prominently near the compact \ion{H}{2}~region RCW\,36, which harbors three late O-type/early B-type stars as well as a large number of lower mass protostars \citep{ellerbroek_2013}.

We adopt an distance to Vela\,C based on a GAIA-DR2 informed reddening distance, described in Appendix \ref{sect:gaia_dist}, of 933\,$\pm$\,94\,pc.  This distance estimate is somewhat larger than the 700$\,\pm\,$200\,pc Vela\,C distance estimate from \cite{liseau_1992}, used in \cite{fissel_2016} and \cite{soler_2017}.

Comparing the 3\farcm0 FWHM resolution maps of inferred magnetic field morphology to the orientation of structures in the $\nabla N_{\mathrm{H}}$~map made from Herschel-derived dust column density maps at 36\arcsec~(0.16\,pc) FWHM resolution, \cite{soler_2017}~found a preference for iso-$N_{\mathrm{H}}$~ contours to be aligned parallel to  \bpos~for low $N_{\mathrm{H}}$~sightlines and perpendicular for high $N_{\mathrm{H}}$~sightlines.  The result was later confirmed by \cite{jow_2018} using the projected Rayleigh statistic, a more robust statistic for the measurement of preferential alignment between two sets of orientation angles.  These results suggest that in Vela\,C too the magnetic field is strong enough to affect the formation of high density structures within the cloud.  The \nh~value corresponding to  the transition from parallel to perpendicular relative orientation ranged over 22.2\,$<$\,log(\nh)\,$<$\,22.6 for most cloud regions in Vela\,C, though a much lower transition \nh~was found for the most evolved cloud regions near RCW\,36. This $N_{\mathrm{H}}\,\simeq\,10^{22}$~cm$^{-2}$~threshold is similar to the column density above which \cite{crutcher_2010} found that Zeeman observations of magnetic field strength indicate a transition from subcritical (magnetic fields are strong enough to prevent gravitational collapse) to supercritical~(magnetic fields alone cannot prevent gravitational collapse), which suggests that the two transitions could be physically related.

In this paper we further examine the relationship between molecular gas and the magnetic field in Vela\,C by studying the relative orientation of structures in integrated line-intensity maps from Mopra telescope observations of nine different rotational  molecular lines.  Our goal is to determine whether the change in relative orientation with column density observed by \cite{soler_2017} is caused by an underlying change in  relative orientation of cloud structures within different volume density regimes.  

We begin by describing the Mopra, BLASTPol, and \herschel~derived-maps used in our analysis in Section \ref{sect:observations}, then examine in detail both the line-of-sight velocity structure and low-order moment maps for each Mopra molecular line in Section \ref{sect:molecular}.  In Section \ref{sect:methods} we describe the calculation of relative orientation angles, introduce the projected Rayleigh statistic as a tool to quantify the  statistical degree of alignment between the magnetic field and the  structures in zeroth-moment (\mzero) maps, and show that low density tracers tend to have cloud morphology that is preferentially parallel to the cloud-scale magnetic field, while high or intermediate density tracers have a weak preference to align perpendicular to the magnetic field.  We also estimate the characteristic density traced by each molecular line. We then examine the change in relative orientation with density, look for regional variations, and discuss the implications of our findings in Section \ref{sect:discussion}.  A brief summary of our results is given in Section \ref{sect:summary}.

\section{Observations} \label{sect:observations}

\capstartfalse
\begin{deluxetable*}{cccccccccccc}
\tabletypesize{\footnotesize}
\tablecaption{Mopra Molecular Line Data Cube and Moment Map Parameters
\label{tab:mom_params}}
\tablewidth{0pt}
\tablehead{
\colhead{Molecular Line} & \colhead{Rest freq.} & \colhead{Vel. range\footnotemark[1]} &  \colhead{ $v_{\mathrm{LSR}}$~res.\footnotemark[2]} & \colhead{ \mzero~SNR}\footnotemark[3] & \colhead{ \mzero~SNR}\footnotemark[4] & \colhead{ $\sigma_{T_R}$\footnotemark[5]} & 
\colhead{ $\eta_{xb}$\footnotemark[6]} & \colhead{$\theta_{beam}$\footnotemark[7]} & \colhead{$\theta_{sm}$\footnotemark[8]} & \colhead{$\theta_{gr}$\footnotemark[9]} & \colhead{pixel size\footnotemark[10]} \\
 & \colhead{[GHz]} & \colhead{$v_0-v_1$\,[km\,s$^{-1}$]}   & \colhead{[km\,s$^{-1}$]} & \colhead{thresh(0,1)} & \colhead{thresh(2)} & \colhead{[K]} &  \colhead{[K]} &  \colhead{[arcsec]} & \colhead{[arcsec]} & \colhead{[arcsec]}  & \colhead{[arcsec]}
}
\startdata
$^{12}$CO $J$\,=\,1\,$\rightarrow$\,0  & 115.2712 & ~~0\,--\,+12  & 0.18 & 8 & 10 & 0.113 & 0.55 &  33 & 120 &  45 & 12 \\
$^{13}$CO $J$\,=\,1\,$\rightarrow$\,0  & 110.2013 & ~~0\,--\,+12  & 0.18 & 8 & 20 & 0.053 & 0.55 & 33 & 120 &  45 & 12  \\
C$^{18}$O $J$\,=\,1\,$\rightarrow$\,0  & 109.7822 & +2\,--\,+10  & 0.18 & 8 & 10 & 0.053 & 0.55 & 33 & 120 &  45 & 12  \\
N$_2$H$^+$~$J$\,=\,1\,$\rightarrow$\,0 & ~93.1730 & $-$6\,--\,+14 & 0.21 & 6 & 10 & 0.016 & 0.65 & 36 & 120 &  45 & 12  \\
HNC $J$\,=\,1\,$\rightarrow$\,0        & ~90.6636 & +2\,--\,+10  & 0.22 & 8 & 10 & 0.039 & 0.65 & 36 & 120 &  45 & 12  \\
HCO$^+$ $J$\,=\,1\,$\rightarrow$\,0    & ~89.1885 & +2\,--\,+10  & 0.23 & 8 & 10 & 0.018 & 0.65 & 36 & 120 &  45 & 12  \\
HCN $J$\,=\,1\,$\rightarrow$\,0        & ~88.6319 & $-$5\,--\,+15  & 0.23 & 8 & 10 & 0.019 & 0.65 & 36 & 120 &  45 & 12  \\
CS $J$\,=\,1\,$\rightarrow$\,0         & ~48.9910 & +2\,--\,+10  & 0.20 & 8 & 20 & 0.095 & 0.56 & 60 & 120 &  84 & 24  \\
NH$_3$~(1,1)           & ~23.6945 & +2\,--\,+10  & 0.43 & 5 & 10 & 0.059 & 0.65 & 132 & 150 & 150 & 40 
\enddata
\tablecomments{The NH$_3$~(1,1), and N$^2$H$^+$~and HCN $J$\,=\,1\,$\rightarrow$\,0 lines have hyperfine structure.  For the N$^2$H$^+$~and HCN lines we integrate over all the hyperfine components to make the zeroth and first moment maps; however, for the second moment maps we use a narrower velocity integration range of +2\,--\,+8.2 and +2\,--\,+10\,km\,s$^{-1}$, respectively to center on the narrowest possible resolved spectral peak. For the NH$_3$\,(1,1) line we integrate over only the central spectral peak for all moment maps.}
\footnotetext[1]{$v_{\mathrm{LSR}}$ range over which the zeroth-moment (\mzero, Equation \ref{eqn:mom0}), first moment (\mone, Equation \ref{eqn:mom1}), and (for most lines) second moment (\mtwo, Equation \ref{eqn:mom2}) values are calculated (see above note).}  
\footnotetext[2]{Velocity resolution for each molecular line cube.}  
\footnotetext[3]{\mzero~signal-to-noise threshold required for both \mzero~and \mone~maps.}  
\footnotetext[4]{\mzero~signal-to-noise threshold required for \mtwo~maps.}  
\footnotetext[5]{Per channel noise level of $T_{\mathrm{R}}$~after the data cubes were smoothed to $\theta_{sm}$\,FWHM resolution.}  
\footnotetext[6]{Beam efficiency correction factor for extended emission used to convert antenna temperature to radiation temperature ($T_{\mathrm{R}}\,=\,T_{\mathrm{A}}/\eta_{\mathrm{xb}}$).  Measurements of $\eta_{\mathrm{xb}}$~were obtained by \cite{urquhart_2010} (7\,mm and 12\,mm lines), and \cite{ladd_2005} (3\,mm and CO isotopologues). }  
\footnotetext[7]{Telescope beam FWHM without any additional smoothing \citep{ladd_2005,urquhart_2010}.}  
\footnotetext[8]{FWHM resolution of Gaussian smoothed data cubes used to make the moment maps.}
\footnotetext[9]{FWHM of Gaussian derivative kernel used to calculate the gradient angles described in Section \ref{sect:prs}.}.
\footnotetext[10]{ Size of the map pixels for both the original Mopra data cubes and moment maps made from the smoothed Mopra data.}  
\end{deluxetable*}
\capstarttrue

\subsection{\blastpol~Polarization Observations}

For the analysis in this work we utilize the magnetic field orientation inferred from linearly polarized dust emission measured by the \blastpol~balloon-borne polarimeter, during its last Antarctic science flight in December 2012 \citep{galitzki_2014}. BLASTPol observed Vela\,C in three sub-mm bands centered at 250, 350, and 500\,$\mu$m, for a total of 54 hours.  Due to a non-Gaussian telescope beam the maps required additional smoothing. In this paper we focus solely on the 2\farcm5-FWHM-resolution 500-$\mu$m~maps previously presented in \cite{fissel_2016}.\footnote{We note, however, that the inferred magnetic field orientation angles are largely consistent between the three BLAST bands, as discussed in \cite{soler_2017}.}   This resolution corresponds to 0.7\,pc at the distance of Vela\,C. 

We assume that the orientation of \bpos, the magnetic field orientation projected on the plane of the sky, can be calculated from the Stokes parameters as  
\begin{equation}
\mbpos\,=\,\frac{1}{2}\,\arctan\left(U,Q\right)\,+\,\frac{\pi}{2},
\end{equation}
which corresponds to the polarization orientation \edir~derived from the BLASTPol 500\,\um~Stokes $Q$~and $U$~data rotated by $\pi$/2 radians.\footnote{In our coordinate system a polarization orientation angle of 0\deg~implies a Galactic North-South orientation, where the angle value increases with a counter-clockwise rotation towards Galactic East-West.}
Only BLASTPol measurements with an uncertainty in the polarization angle of less than 10\deg~are used in this analysis.

\cite{fissel_2016}\ discussed the several different methods for separating polarized emission due to diffuse ISM dust along the same sightlines as Vela\,C.  This correction is important as the Vela\,C cloud is at a low Galactic latitude ($b$\,=\,0.5--2\deg).  For our analysis, we use the ``Intermediate'' subtraction method from \cite{fissel_2016}.  In Appendix \ref{sect:refregion}, we show that the choice of diffuse emission subtraction method does not change our final results.

\subsection{Mopra Observations}\label{sect:mopra_obs}

To study the density and velocity structure of Vela\,C we 
compare the BLASTPol data to results from a
large-scale molecular line survey of Vela\,C made with the
22-m Mopra Telescope over the period from 2009 to
2013.  The Mopra data presented here are the combination of two surveys: M401 
(PI: Cunningham), which covered
molecular lines at 3, 7, and 12 mm, and M635
(PI: Fissel), which mapped Vela C in the $J$\,=\,1\,$\rightarrow$\,0 lines of $^{12}$CO and 
isotopologues $^{13}$CO and C$^{18}$O. 
For the
M401 observations the cloud was mapped in a series of
square raster maps (5\arcmin, 10\arcmin, and 15\arcmin~respectively, for the 3-, 7- and 12-mm observations), 
while the M635 observations
were taken using the Mopra fast-scanning mode, scanning the telescope in long rectangular
strips of 6\arcmin~height in
both the Galactic longitude and
latitude directions. 

For both surveys the UNSW-MOPS\footnote{The University of New South Wales Digital Filter Bank used for the observations with the Mopra Telescope was provided with support from the Australian Research Council.} digital 
filterbank backend and
the MMIC receiver were used, with multiple zoom bands 
covering 
137.5\,MHz each, with 4096 channels within the 8-GHz bandwidth.  In this paper we present observations of the nine molecular rotational lines for which there is significant extended emission: the $^{12}$CO, $^{13}$CO, C$^{18}$O, N$_2$H$^+$, HNC, HCO$^{+}$, HNC, and CS $J$\,=\,1\,$\rightarrow$\,0 lines, as well as the NH$_3$(1,1) inversion line.  Table \ref{tab:mom_params} summarizes the observed lines including velocity resolution and beam FWHM $\theta_{\mathrm{beam}}$, which ranges from 33\arcsec\,FWHM for the CO $J$\,=\,1\,$\rightarrow$\,0 observations to 132\arcsec\,FWHM~for NH$_3$\,(1,1).
Our Mopra observations were bandpass corrected, using off source spectra
with the {\tt livedata} package, and gridded into FITS cubes using the
{\tt gridzilla} package\footnote{\url{http://www.atnf.csiro.au/computing/software/livedata/index.html}}.
Extra polynomial bandpass fitting was done with the {\tt miriad} package,\footnote{\url{http://www.atnf.csiro.au/computing/software/miriad/}}
and Hanning smoothing was carried out in velocity.

\subsection{Herschel-derived Column Density Maps} \label{sect:col_dens_nh}
We compare the observed molecular line emission to the total hydrogen column density map $N_{\mathrm{H}}$~(in units of hydrogen nucleons per cm$^{-2}$) first presented in Section 4 of \cite{fissel_2016}.\footnote{Note that in this paper $N_{\mathrm{H}}$~and $n_{\mathrm{H}}$~refer respectively to the column density and number density of hydrogen nucleons, while $N_{\mathrm{H_2}}$~and $n_{\mathrm{H_2}}$~refer to the molecular hydrogen column and number density.  Assuming all of the hydrogen is in molecular form at the densities probed in this work the conversion is $n_{\mathrm{H_2}}\,=\,n_{\mathrm{H}}/2$.}  These maps are also used in Section \ref{sect:density_deriv} and Appendix \ref{sec:app_cd} to estimate the abundances of our observed molecules.  The \nh~maps are based on dust spectral fits to four far-IR/sub-mm dust emission maps: \herschel-SPIRE maps at 250, 350, and 500\,$\mu$m; and a \herschel-PACS map at 160\,$\mu$m. Each \herschel\footnote{\herschel~is an ESA space observatory with science instruments provided by European-led Principal Investigator consortia and with important participation from NASA.}~dust map was smoothed to match the \blastpol~500-$\mu$m FWHM resolution of 2\farcm5 before spectral fitting.

\section{The Molecular Structure of Vela C} \label{sect:molecular}

\begin{figure*}
\epsscale{0.57}
\plotone{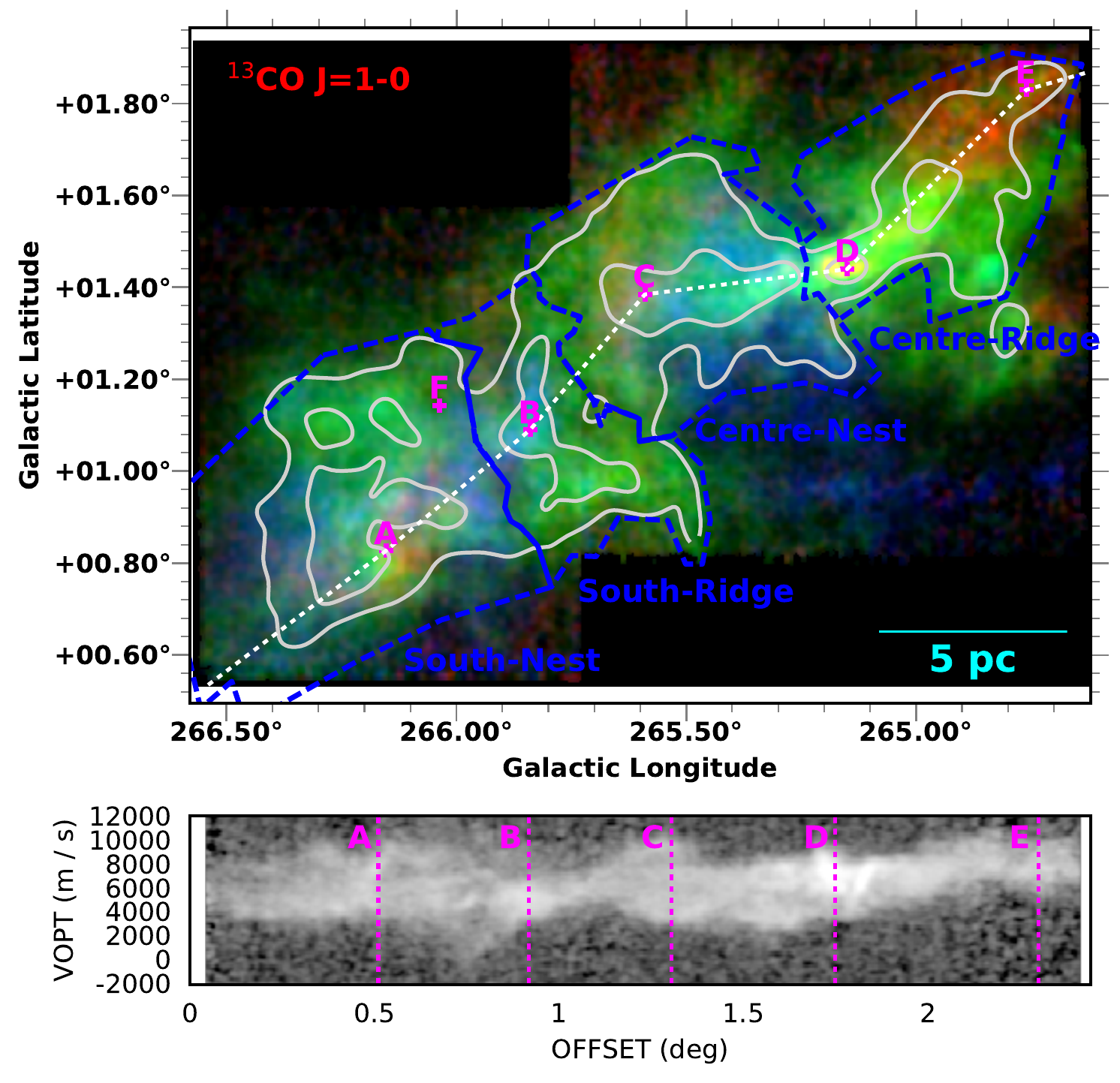}
\plotone{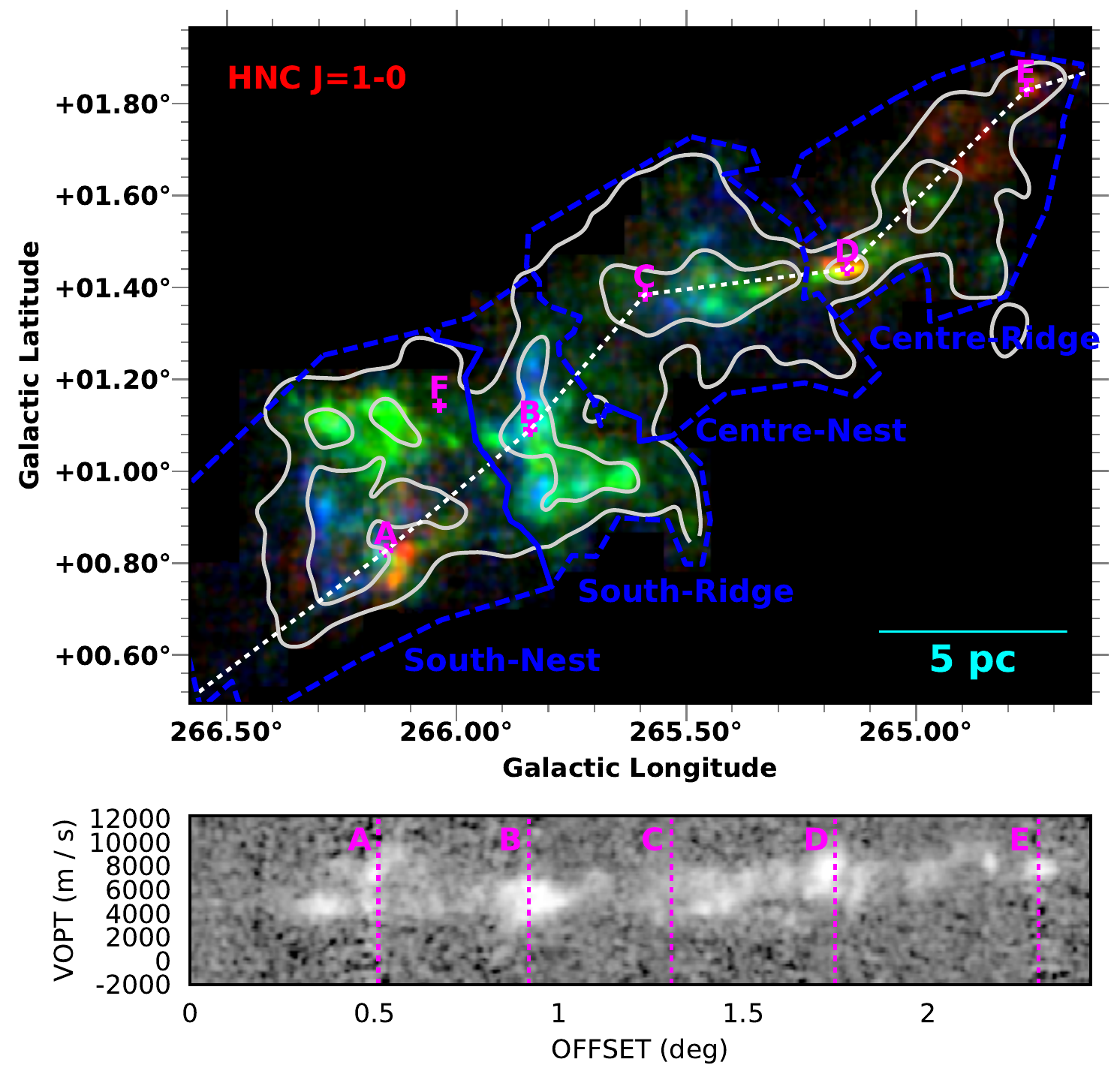}
\caption{Line-of-sight velocity structure of the Vela\,C molecular cloud. {\em Top panels:} RGB images of the $^{13}$CO $J$\,=\,1\,$\rightarrow$\,0 line (left) and HNC $J$\,=\,1\,$\rightarrow$\,0 line (right).  Each color represents emission integrated over a different range in velocities: $-$5.0 to 5.0\,km\,s$^{-1}$~(blue), 5.0 to 7.5\,km\,s$^{-1}$~(green), and  7.5 to 25\,km\,s$^{-1}$ (red).  Contours show the Herschel-derived total hydrogen column density (described in Section \ref{sect:col_dens_nh}) for \nh\,=\,1.2 and 3.6 $\times$\,10$^{22}$\,hydrogen atoms cm$^{-2}$. The labeled positions correspond to the locations where spectra are shown in Figure \ref{fig:spectra}.
These include a sightline towards the ionizing source powering the RCW\,36 HII region (D) and a sightline towards the background cluster G266.0349+01.1450 (F). Dashed blue lines indicate the boundaries of four of the sub-regions of Vela\,C identified in \cite{hill_2011}. {\em Bottom panels:} Position-velocity diagrams sampled along the dotted white line shown in the upper panels. The dotted vertical lines indicate the locations of the  positions labeled in the top panel.\label{fig:vel_struct}
}
\end{figure*}

\begin{figure*}
\epsscale{1.0}
\plotone{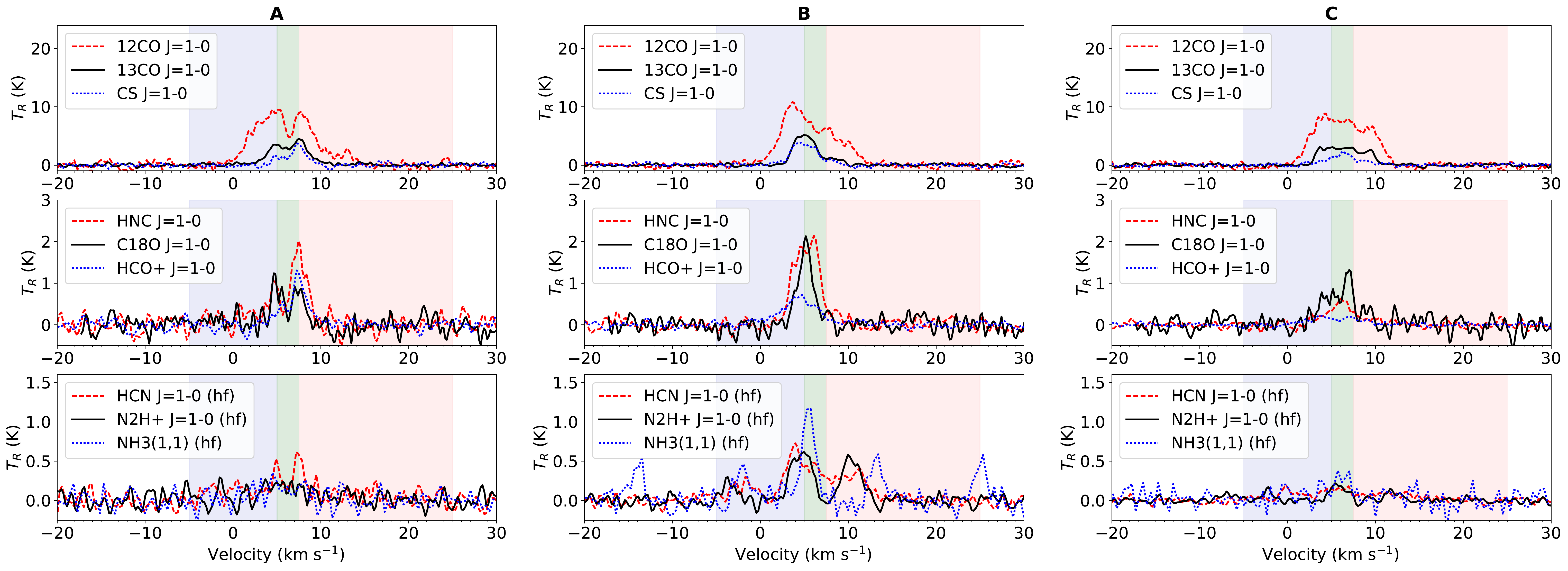}
\plotone{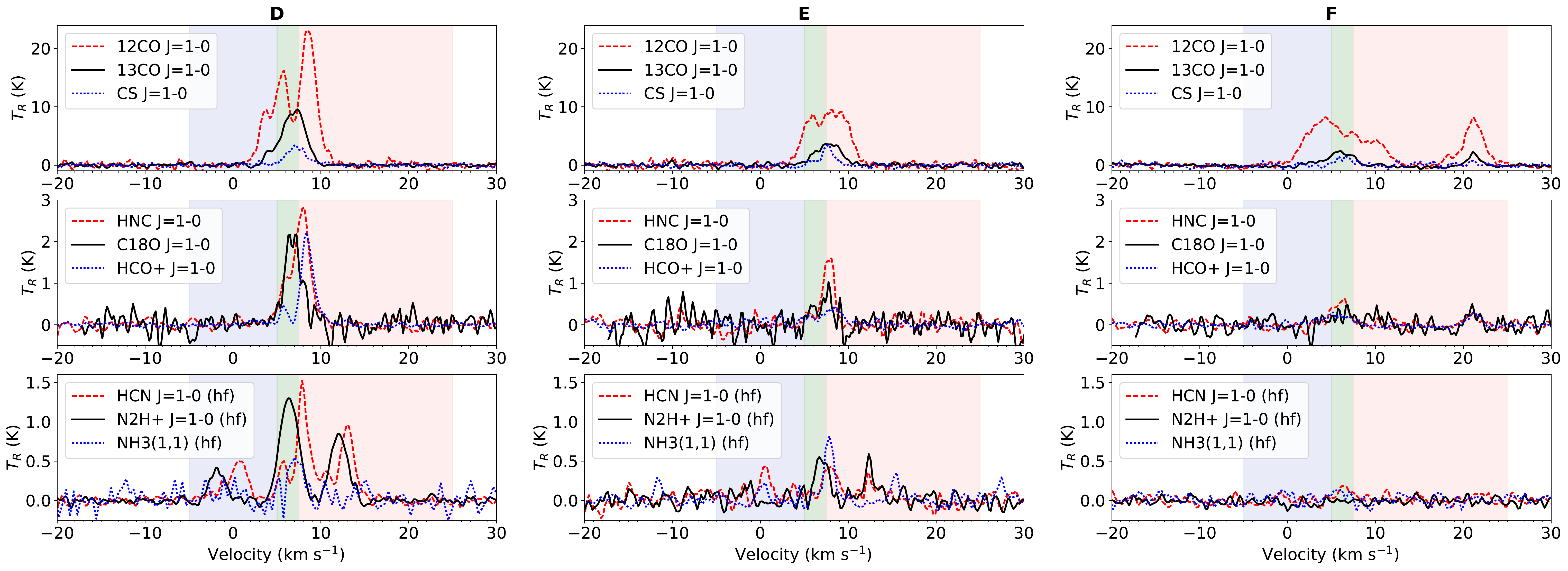}
\caption{Spectra extracted for nine molecular lines at the locations labeled in Figure \ref{fig:vel_struct}. (The colored bands indicate the velocity integration limits for the RGB images shown in Figure \ref{fig:vel_struct}). {\em Top panel:} $^{12}$CO, $^{13}$CO, and CS lines; {\em Middle panel:} C$^{18}$O, HNC, HCO$^{+}$; {\em Bottom panel:} HCN, N$_2$H$^+$~and NH$_3$(1,1) including their additional hyperfine structure (hf).  Note that location D is a sightline coincident with the cluster powering the \ion{H}{2}~region RCW\,36, while position F coincides with the location of background stellar cluster G266.0349+01.1450.\label{fig:spectra}
}
\end{figure*}

  Figure \ref{fig:vel_struct} shows RGB maps of both the C$^{13}$O \joz~line (top-left panel) and HNC \joz~line (top-right panel), the latter generally probing higher density molecular gas.  The cubes were Gaussian smoothed to 60\arcsec~FWHM resolution and each color represents an integration over a different velocity slice of the cube (blue, -5.0 to 5.0\,km\,s$^{-1}$; green, 5.0 to 7.5\,km\,s$^{-1}$; and red, 7.5 to 25\,km\,s$^{-1}$). The line-of-sight cloud velocity structure is shown in more detail in the lower panels, which are position-velocity diagrams sampled along the dotted path indicated on the RGB images. In Figure \ref{fig:spectra} we show the line profiles of all nine molecular lines at the positions labeled in Figure \ref{fig:vel_struct}.  

Overall, Figure \ref{fig:vel_struct} shows a trend of increasing line of sight velocity from East to West across Vela\,C, which is particularly prominent along the Centre-Ridge to the right of RCW 36 (position D).  However, Vela\,C also has complex line-of-sight velocity structure with (in many cases) multiple velocity peaks along the same sightline (e.g., at position A). These multi-peaked lines are seen in both optically thick ($^{12}$CO and $^{13}$CO) and thin (C$^{18}$O) tracers, and thus are likely the result of multiple velocity components in the molecular gas, rather than self absorption of the molecular line emission. 

Most of the line emission is observed to occur within the velocity range  $0\,<\,v_{\mathrm{LSR}}\,<\,12$\,km\,s$^{-1}$, however the $^{12}$CO $J$\,=\,1\,$\rightarrow$\,0~line in particular shows additional (lower brightness) emission at both $v_{\mathrm{LSR}}\,<\,$0\,km\,s$^{-1}$~and $v_{\mathrm{LSR}}>\,$12\,km\,s$^{-1}$.  This emission is likely associated with molecular gas at different distances along the line of sight.  The most obvious example is at the position labeled F in Figures \ref{fig:vel_struct} and \ref{fig:spectra}, where there is an additional line centered at $v_{\mathrm{LSR}}\,\simeq$\,21\,km\,s$^{-1}$, clearly seen not only in $^{12}$CO but also $^{13}$CO, C$^{18}$O, HNC, HCO$^+$, and CS. The spatial location of this second molecular line emission coincides with the location of a stellar cluster G266.0349+01.1450 identified in \cite{baba_2006}, who argue that because of the faintness of the sources the cluster is likely located in a distant molecular cloud beyond Vela\,C.  

\cite{hill_2011} previously showed that at $A_{\mathrm V}\simeq\,$7 Vela\,C  breaks-up into five sub-regions. Four of these regions are covered in our Mopra/BLASTPol survey (labeled in Figure \ref{fig:vel_struct}): two ``ridges'' (the South-Ridge and Centre-Ridge), which are each dominated by a high column density filament ($A_V$\,$>$\,100\,mag); and two ``nests'' (the South-Nest and Centre-Nest), which have many lower column density filaments with a variety of orientations. We note that molecular line emission appears over a larger range of $v_{\mathrm{LSR}}$~towards the South-Nest and Centre-Nest regions; most of the sightlines for which lines other than $^{12}$CO and $^{13}$CO show multiple velocity peaks occur toward these regions (for an example see the spectral line plots in Figure \ref{fig:spectra} at positions A and C).

\subsection{Moment Maps}\label{sect:mom_maps}
To further explore the emission and line of sight velocity structure of Vela\,C we calculate the first three moment maps for the cloud.  The zeroth-moment map is the integrated line-intensity:
\begin{equation}\label{eqn:mom0}
I\ =\ \int_{v_0}^{v_1}\,T_{\mathrm{R}}\,dv,
\end{equation}
where $T_{\mathrm{R}}$\ is the radiation temperature in velocity channel $v$. $T_{\mathrm{R}}$~can be calculated from the measured antenna temperature $T_{\mathrm{A}}$ corrected by the main beam efficiency for extended structure $\eta_{\mathrm{xb}}$~values determined from previous Mopra observations and listed in Table \ref{tab:mom_params} ($T_{\mathrm{R}}$\,=\,$T_{\mathrm{A}}/\eta_{\mathrm{xb}}$).  Here $v_0$\ and $v_1$\ are the minimum and maximum velocities over which the line data are integrated. These velocity integration limits are listed for each line in Table \ref{tab:mom_params}, and are generally within the 0\,km\,s$^{-1}<v_{\mathrm{LSR}}<$\,12\,km\,s$^{-1}$~range where the molecular line emission is likely associated with Vela\,C. For the HCN and N$_2$H$^+$ $J$\,=\,1\,$\rightarrow$\,0 lines we integrate over a larger velocity range to include additional hyperfine spectral components and increase the signal-to-noise.

We can use higher order moments to study the velocity structure of each data cube.  The first moment map gives the intensity weighted average line-of-sight velocity \mone:
\begin{equation}\label{eqn:mom1}
\mmone =\ \frac{\int_{v_0}^{v_1}\,T_{\mathrm{R}}\,v\,dv}{\int_{v_0}^{v_1}\,T_{\mathrm{R}}\,dv}.
\end{equation}
Similarly where the signal-to-noise of the line data is high enough we can calculate the second moment, which gives the line of sight velocity dispersion \mtwo:\footnote{Note that the second moment is written as $\sigma_{v}$~in some publications, but in this paper we use \mtwo~to avoid confusion with the measurement uncertainties which are labeled with $\sigma$.}
\begin{equation}\label{eqn:mom2}
\mmtwo =\ \left(\frac{\int_{v_0}^{v_1}\,T_{\mathrm{R}}\,\left(v\,-\,\mmone\right)^2\,dv}{\int_{v_0}^{v_1}\,T_{\mathrm{R}}\,dv}\right)^{1/2}.
\end{equation}
Note that in the case of a Gaussian line profile Equation~\ref{eqn:mom2} would return the Gaussian width ($\sigma$).

\begin{figure*}
\epsscale{0.95}
\plotone{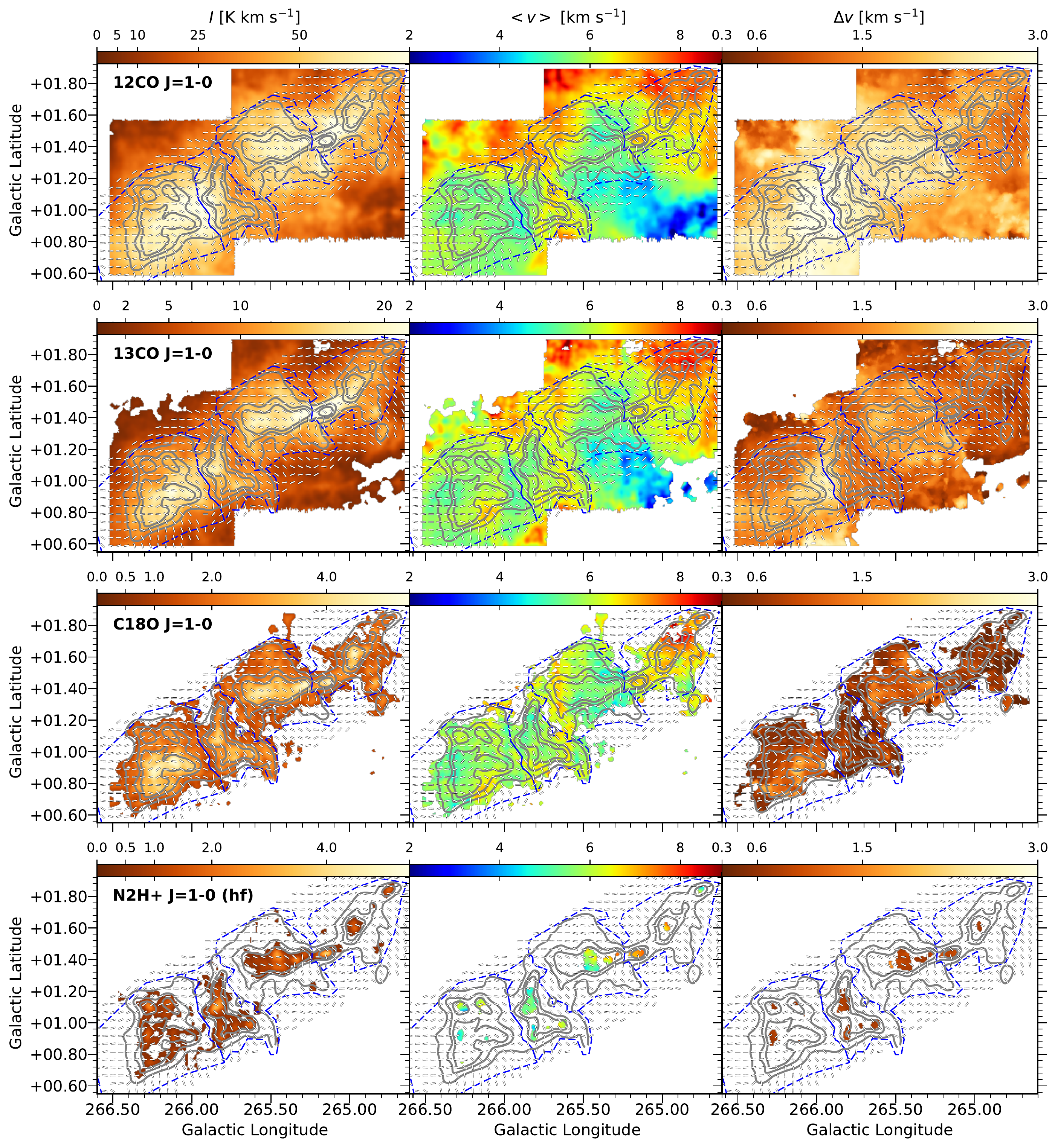}
\caption{Moment maps \mzero~(left panels), \mone~(center panels), and \mtwo~(right panels) for the $^{12}$CO, $^{13}$CO, C$^{18}$O, and  N$_2$H$^+$~lines calculated as described in Section \ref{sect:mom_maps}. The label hf indicates lines with significant hyperfine structure. Contours show \nh~levels of 1.2, 2.4, and 3.6 \,$\times$\,10$^{22}$\,cm$^{-2}$ derived from {\em Herschel}~dust emission maps (Section \ref{sect:col_dens_nh}), while dashed blue lines indicate the cloud subregions defined in \cite{hill_2011} and labeled in Figure \ref{fig:vel_struct}. Line segments show the orientation of the magnetic field projected on the plane of the sky inferred from BLASTPol 500\,\um\ data. \label{fig:mom_maps1}
}
\end{figure*}

Before calculating the moment maps we first smooth each channel map with a 2-D Gaussian kernel so that the resulting cube has 120\arcsec~FWHM resolution.\footnote{The exception is the NH$_3$~cube, which has an intrinsic FWHM resolution of 132\arcsec.  For this cube we smooth instead to 150\arcsec.} 
This smoothing is needed both to increase the signal to noise ratio of the extended structure and to minimize any narrow spurious map features due to differences in $T_{\mathrm{sys}}$~levels within the map; we show in Appendix \ref{sect:resolution} that the choice of smoothed resolution does not significantly change our final results.
Table \ref{tab:mom_params} lists the smoothed FWHM resolution ($\theta_{\mathrm{sm}}$) for each cube.  The pixel size for the smoothed cubes is the same as the pixel size in the original data cubes (see the last column in Table \ref{tab:mom_params}).

To estimate the uncertainty in $T_{\mathrm{R}}$ we select velocity channels in the spectra that have no apparent signal, and find both the standard deviation of all the voxels and the standard deviation for each pixel over all the velocity channels that have no signal.  We take as the per velocity channel uncertainty $\sigma_{T_{\mathrm{R}}}$~the maximum of these two standard deviations for each pixel.  Uncertainties in the moment maps $\sigma_{\mmzero}$, $\sigma_{\mmone}$, and $\sigma_{\mmtwo}$, are then estimated through a Monte Carlo method by taking the data cube and adding to each voxel in the cube a random number selected from a normal distribution centered at 0\,K, with a width of $\sigma_{T_{\mathrm{R}}}$.
We recalculate the moment maps using this method 1000 times, and take the per-pixel standard deviation in the resulting moment maps to be our uncertainty.
 
For the analysis of the \mzero~(zeroth-moment) maps we only use data points where  $\mmzero/\sigma_{\mmzero}\,>$\,8, except for the N$_2$H$^+$~and NH$_3$\,maps, which have relatively low signal-to-noise, where we relax the signal-to-noise requirement to 6 and 5 respectively.  For the \mone~(first moment) maps we additionally require that the uncertainty $\sigma_{\mmone}$~be less than 0.4\,km\,s$^{-1}$. More strict criteria are applied for the calculation of the \mtwo~(second moment) maps, which are very sensitive to noise spikes.  Here we only use spectral channels where $T_{\mathrm{R}}\,\geq\,3\sigma_{T_{\mathrm{R}}}$~and require the integrated line strength to be above a threshold \mzero~signal-to-noise level listed for each line in Table \ref{tab:mom_params}.

\begin{figure*}
\epsscale{0.95}
\plotone{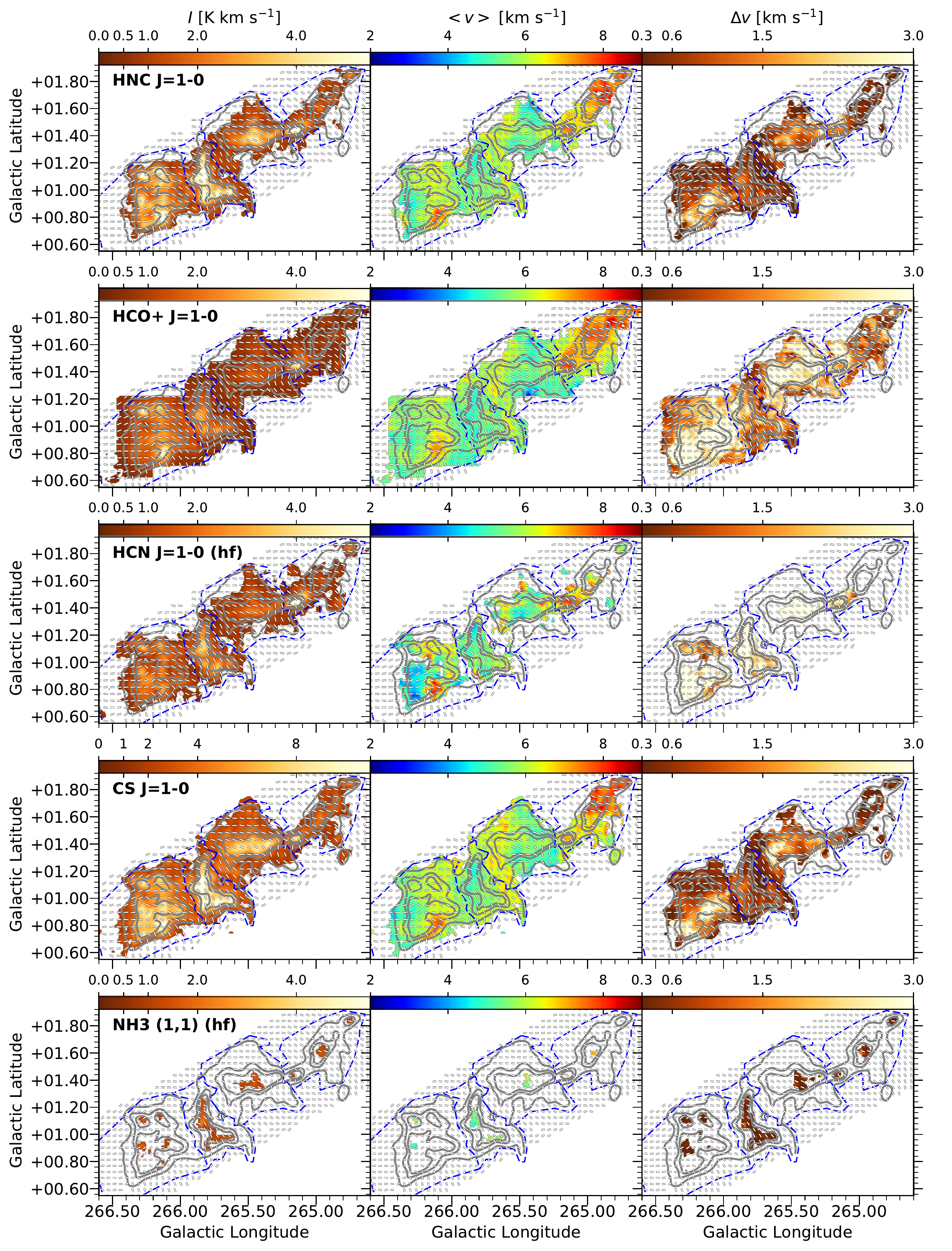}
\caption{Same as Figure \ref{fig:mom_maps1}~but for the HNC, HCO$^{+}$, HCN, CS, and NH$_3$~lines.\label{fig:mom_maps2}
}
\end{figure*}

Figures \ref{fig:mom_maps1} and \ref{fig:mom_maps2} show the calculated moment maps of nine different molecular lines, with contours of $N_{\mathrm{H}}$ (Section \ref{sect:col_dens_nh}). In general the molecular lines appear to trace different density, chemical, and excitation conditions within the cloud.  The $^{12}$CO $J$\,=\,1--0 \mzero\ map shows little correspondence to the column density structure of Vela\,C, which is consistent with the expectation that the emission is optically thick, such that only the lower density outer layers are probed by the line.  

We expect $^{13}$CO to have a lower optical depth than $^{12}$CO.  The \mzero\ map of $^{13}$CO~shows similar structure to \nh, but does not show the dense filamentary structure seen in the \herschel~observations. The even rarer isotopologue C$^{18}$O shows a very similar structure to $^{13}$CO, although with lower signal-to-noise ratio and more contrast towards the highest column density regions where $^{13}$CO~might be optically thick.

The HNC, HCO$^+$, HCN, and CS $J$\,=\,1\,$\rightarrow$\,0 lines show significant \mzero~detections only toward higher column density structures.  We note that these intermediate number density tracers show weaker emission in the Centre-Ridge sub-region to the right of RCW\,36 compared to the \herschel-derived \nh~map (contours in Figures \ref{fig:mom_maps1} and \ref{fig:mom_maps2}).  This could imply that molecular abundance or excitation conditions are different in the Centre-Ridge compared to the rest of Vela\,C. We also include two tracers that are often used to probe higher density gas, NH$_3$~(1,1) and N$_2$H$^+$~$J$\,=\,1\,$\rightarrow$\,0.  These lines tend to have low signal to noise ratios (Figure \ref{fig:spectra}), but show emission near the highest column density cloud regions.

Throughout the paper we refer to $^{12}$CO and $^{13}$CO as ``low density'' tracers because these molecules are optically thick towards high \nh~sightlines and have high enough abundance levels to be detected in the low density envelope of Vela\,C. We refer to N$_2$H$^+$, HNC, HCO$^+$, HCN, CS $J$\,=\,1\,$\rightarrow$\,0, and NH$_3$~(1,1) as intermediate or high density tracers because these molecules trace mostly higher column density regions, are not generally detected in the cloud envelope, and tend to have higher estimated characteristic densities (see discussion in Section \ref{sect:density_deriv}). C$^{18}$O $J$\,=\,1\,$\rightarrow$\,0 is also only detected toward higher column density structures, however radiative transfer modeling in Section \ref{sect:n_radex} suggest the line typically traces lower densities than our intermediate or high density tracers.

The first moment or \mone~maps 
 within the \nh~contours show that the molecular gas of Vela\,C has on average a line of sight velocity 1--2 km\,s$^{-1}$ higher in the Centre-Ridge compared to the rest of the cloud.  As discussed in Section \ref{sect:col_dens_nh}, many cloud sightlines, particularly towards the South-Nest and Centre-Nest sub-regions, have multiple spectral peaks centered at different line of sight velocities.  Some of the structure in the \mone~maps is therefore likely the result of variations in the relative intensity of the different spectral components that contribute to the total cloud sightline emission.  In addition, the HCN, N$_2$H$^+$, and NH$_3$~lines have hyperfine structure, and so \mone~maps calculated for these lines could be influenced by the optical depth of the different hyperfine components.

The second moment \mtwo~maps show large apparent velocity dispersions where there are two nearly equal strength spectral peaks at different line of sight velocities (for example location A in Figures \ref{fig:vel_struct} and  \ref{fig:spectra}).
   For the C$^{18}$O and the intermediate to high density tracers HNC, HCO$^+$, and CS, which do not have hyperfine line structure, we see that the two ``nest-like'' regions identified in \cite{hill_2011}~show much larger average values of \mtwo~than the two ``ridge-like'' regions.  This suggests that in addition to having filamentary structure with a variety of orientations, the South-Nest and Centre-Nest also have more complicated line of sight velocity structure than the South-Ridge and Centre-Ridge regions.

\section{Methods and Results}\label{sect:methods}

In this paper we quantify the relative orientation between the Mopra zeroth-moment maps shown in Figures \ref{fig:mom_maps1} and \ref{fig:mom_maps2} and the magnetic field orientation \bpos~inferred from BLASTPol data.   
We first calculate the relative orientation angles in Section \ref{sect:rel_orientation} and characterize their distribution using the methods first presented in \cite{soler_2013}. In Section \ref{sect:prs}, we evaluate a statistical measure of the relative orientation, the projected Rayleigh statistic, for different molecular tracers.
We estimate the characteristic densities traced by each molecular line in Section \ref{sect:density_deriv}.

\subsection{Calculating the Relative Orientation Angle} \label{sect:rel_orientation}
Similar to the methods described in \cite{soler_2013, soler_2017}, the orientation of structure in the Mopra moment maps is calculated by computing the gradient vector field of each map. The moment map is convolved with a Gaussian gradient kernel of FWHM width $\theta_{\mathrm{gr}}$, where $\theta_{\mathrm{gr}}$ was chosen to be larger than three map pixels to avoid spurious measurements of the gradient orientation due to map pixelization (see Table \ref{tab:mom_params}).  

The relative orientation angle $\phi$~between the plane of the sky magnetic field \bpos~and a line tangent to the local iso-\mzero~map contour is equivalent to the angle between the polarization direction $\mathbf{\hat{E}}$~and $\nabla\,\mmzero$:
\begin{equation}\label{eqn:phi}
\phi\,=\,\arctan\left(|\nabla\,\mmzero\,\times\,\mathbf{\hat{E}}|, \nabla\,\mmzero\,\cdot\,\mathbf{\hat{E}}\right)
\end{equation}
\citep{soler_2017}.
 With this convention, $\phi\,=\,$0\deg~indicates that the magnetic field and local \mzero~structure orientations are parallel, while $\phi\,=\,$90\deg~indicates that \bpos~is perpendicular to the local \mzero~structure.  Because dust polarization can be used to measure only the orientation of the magnetic field, not the direction, the relative orientation angle $\phi$~is unique only within the range $\left[0^{\circ},\,90^{\circ}\right]$. That is, $\phi\,=\,$20\deg~is equivalent to both $\phi\,=\,-$20\deg~and $\phi\,=\,$160\deg.  

\begin{figure*}
\epsscale{0.95}
\plotone{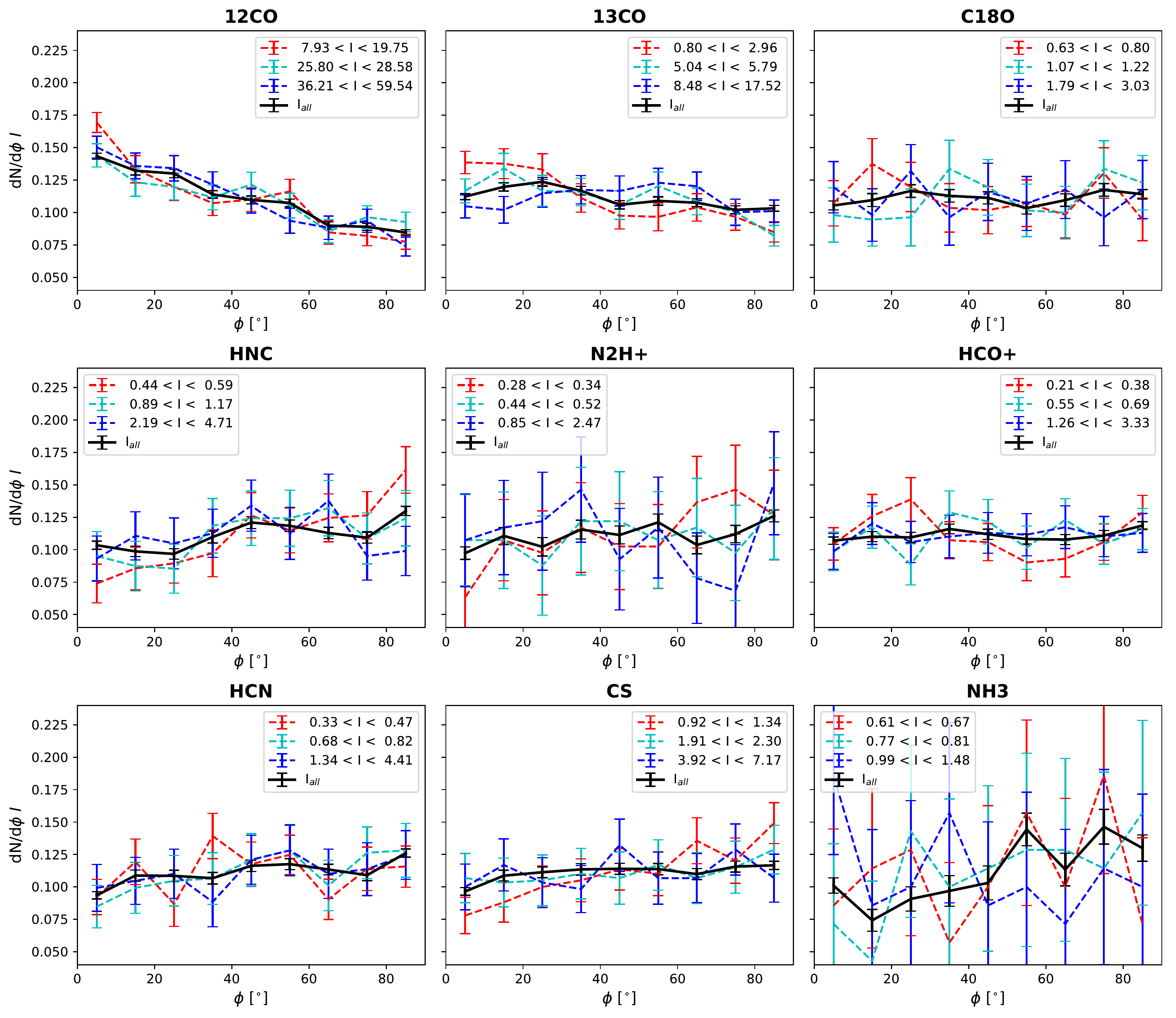}
\caption{Histograms of relative orientation (HROs), showing the fraction of map sightlines with a given angle $\phi$~between the inferred magnetic field orientation (\bpos) and the local iso-\mzero~contour calculated from Equation \ref{eqn:phi}.  Here $\phi\,=\,$0\deg (90\deg)~implies that the local structure in the \mzero~map~is parallel (perpendicular) to \bpos.  The black line shows the HRO for all sightlines. The dashed colored lines show the HROs for sightlines within different bins in \mzero.  \label{fig:hros_mom0}
}
\end{figure*}
\capstartfalse
\begin{deluxetable*}{cccccccc}
\tabletypesize{\footnotesize}
\tablecaption{Projected Rayleigh Statistics for Each Molecular Line.
\label{tab:prs}}
\tablewidth{0pt}
\tablehead{
\colhead{Molecular Line\footnotemark} & \colhead{$Z_{x}^{\prime}$\footnotemark[1]} & \colhead{$\langle Z_{x\,WN}^{\prime}\rangle$\footnotemark[2]} &
\colhead{$\sigma_{Z_{x\,WN}^{\prime}}$\footnotemark[2]} & \colhead{\nind\footnotemark[3]}& \colhead{$Z_{x}$\footnotemark[4]} & \colhead{$Z_{x\,noise}$\footnotemark[5]} & \colhead{med($N_{\mathrm{H}}$)\,[cm$^{-2}$]\footnotemark[6]}
}
\startdata
$^{12}$CO\,$J$\,=\,1\,$\rightarrow$\,0 & 61.706$\pm$0.990 & $-$0.412 & 6.473  & 3038 & 9.532 & $-$0.637 & 1.29E+22 \\
$^{13}$CO\,$J$\,=\,1\,$\rightarrow$\,0 & 18.528$\pm$0.994 & $-$0.393 & 6.481  & 3003 & 2.859 & $-$1.670 & 1.29E+22 \\
C$^{18}$O\,$J$\,=\,1\,$\rightarrow$\,0 & $-$4.420$\pm$1.000 & $-$0.205 & 6.212  & 1893 & $-$0.712 & $-$1.473 & 2.02E+22 \\
N$_2$H$^+$\,$J$\,=\,1\,$\rightarrow$\,0 & $-$6.330$\pm$0.992 & 0.115 & 6.007  & 631 & $-$1.054 & $-$0.806 & 3.68E+22 \\
HNC\,$J$\,=\,1\,$\rightarrow$\,0 & $-$19.177$\pm$0.995 & 0.096 & 6.350 & 1429 & $-$3.020 & $-$1.020 & 2.42E+22 \\
HCO$^+$\,$J$\,=\,1\,$\rightarrow$\,0 & $-$5.468$\pm$1.003 & 0.213 & 6.314 & 1967 & $-$0.866 & 1.780 & 1.93E+22 \\
HCN\,$J$\,=\,1\,$\rightarrow$\,0 & $-$14.285$\pm$0.993 & 0.219 & 6.301  & 1557 & $-$2.267 & 0.921 & 2.27E+22 \\
CS\,$J$\,=\,1\,$\rightarrow$\,0 &  $-$6.940$\pm$0.994 & 0.084 & 3.602  & 1404 & $-$1.927 & $-$0.928 & 2.06E+22 \\
NH$_3$\,(1,1) & $-$3.765$\pm$0.990 & $-$0.050 & 2.568 & 73 & $-$1.467 & 0.652 & 4.83E+22 \enddata
\footnotetext[1]{Projected Rayleigh statistic $Z_{x}^{\prime}$~using data sampled every pixel without correcting for oversampling.}
\footnotetext[2]{The mean and standard deviation of $Z_{x}^{\prime}$~calculated for 1000 white noise maps smoothed to the same resolution as the Mopra 120\arcsec\,FWHM \mzero~maps.}
\footnotetext[3]{Number of independent pixels $\mnind\,=\,n_{\mathrm{pix}}/\left(\sigma_{Z_{x\,WN}^{\prime}}\right)^2$.}
\footnotetext[4]{Oversampling corrected PRS  $Z_{x}$\,=\,$ Z_{x}^{\prime}/\sigma_{Z_{x\,WN}^{\prime}}$.}
\footnotetext[5]{Oversampling corrected PRS calculated for \mzero\,map made from spectral  cube channels that do not show line emission.}
\footnotetext[6]{Median value of hydrogen column density N$_{\mathrm{H}}$\,derived from Herschel maps (Section \ref{sect:col_dens_nh}) toward the sightlines where the \mzero~map has significant detections (as defined in Section \ref{sect:mom_maps}) and that were included in the calculation of \prs.}
\end{deluxetable*}
\capstarttrue

We calculate the relative orientation angle $\phi$~for each Mopra molecular line \mzero~map, sampling our data at the location of every Mopra map pixel (see Table \ref{tab:mom_params} for pixel size information). Figure \ref{fig:hros_mom0} shows the histograms of relative orientation (HROs).  The black solid line shows the normalized histogram for all values of $\phi$ that have passed the \mzero\,map cuts described in Section \ref{sect:mom_maps} and have 
\begin{eqnarray}\label{eqn:sigma_phi}
\sigma_{\phi} & = & \sqrt{\sigma_{\nabla I}^2\,+\,\sigma_{\mathbf{\hat{E}}}^2}~<~10\,^{\circ},
\end{eqnarray}
where $\sigma_{\nabla I}$~is the measurement uncertainty of the gradient angle and $\sigma_{\mathbf{\hat{E}}}$~is the measurement uncertainty of the polarization angle.

The HROs for the nine observed molecular lines show different trends.  For the $^{12}$CO HRO there are significantly more sightlines where the \mzero~structure is parallel to the magnetic field than perpendicular.  The other molecular lines show either slightly more sightlines parallel than perpendicular ($^{13}$CO), a flat HRO indicating no preferred orientation with respect to the magnetic field (C$^{18}$O and HCO$^+$), or more sightlines perpendicular to the magnetic field than parallel (HCN, HNC, CS, N$_2$H$^+$, and NH$_3$).

We test for changes in the shape of the HRO with \mzero~by dividing our sightlines into seven bins based on their \mzero~values, with the bins chosen such that each has the same total number of sightlines.  Figure \ref{fig:hros_mom0} shows no consistent trends in the shape of the HRO for different \mzero~bins, in contrast with the \cite{soler_2017} application of HRO analysis to \herschel-derived column density maps, where there was a clear transition to a more perpendicular alignment with increasing column density.  This could imply that our \mzero~maps are not a direct proxy column density, or the difference could be due to the low resolution of our Mopra \mzero~maps compared to the 36\arcsec~FWHM resolution \nh~maps used in \cite{soler_2017}. We discuss the change in relative orientation versus column density in Section \ref{sect:nh_or_dens}.

\subsection{The Projected Rayleigh Statistic}\label{sect:prs}
As discussed in \cite{jow_2018} given a set $\{\theta_i\}$~of n independent angles distributed within the range [0, 2$\pi$], the Rayleigh statistic $Z$~can be used to test whether the angles are uniformly distributed
\begin{equation} \label{eqn:rs}
Z\,=\,\frac{\left(\sum_i^{\mnind} \cos{\theta_i}\right)^2\,+\,\left(\sum_i^{\mnind} \sin{\theta_i}\right)^2}{\mnind},
\end{equation}
where \nind~is the number of independent data samples. This equation is equivalent to a random walk, with $Z$~characterizing the displacement from the origin if one were to take steps of unit length in the direction of each $\theta_i$.  If the distribution of angles is uniformly random then the expectation value for $Z$~is zero. 

To test for preferential parallel or perpendicular alignment we take $\theta\,=\,2\phi$, where $\phi$~is the relative orientation angle calculated as described in Section \ref{sect:rel_orientation}.  Here $\theta\,=\,0$~corresponds to parallel alignment, while $\theta\,=\,\pi$~corresponds to perpendicular alignment.  \cite{jow_2018}~showed that the projected Rayleigh statistic (PRS) $Z_{x}$~can be used to test for a preference for perpendicular or parallel alignment: 
\begin{equation} \label{eqn:prs}
Z_{x}\,=\,\frac{\sum_i^{\mnind} \cos{\theta_i}}{\sqrt{\mnind/2}}.
\end{equation} 

\prs~in Equation \ref{eqn:prs} represents the random walk component projected on the x-axis in a Cartesian coordinate system.  If a measurement of \bpos~is parallel to the local iso-\mzero\,map contour then $\cos{\theta_i}$\,=\,1. If the two orientations are perpendicular then $\cos{\theta_i}$\,=\,$-$1.  \cite{jow_2018} used Monte-Carlo simulations to show that for uniformly distributed samples of \{$\phi_i$\}~the expectation value of \prs~converges to 0 with $\sigma_{Z_x}$\,=\,1.  We also note that \prs~in Equation \ref{eqn:prs} will increase proportionally to $\mnind^{1/2}$.  The PRS can therefore be thought of as quantifying the significance of a detection of relative orientation.  Measurements of \prs\,$\gg$\,1 indicate a significant detection of parallel relative alignment, while measurements of \prs\,$\ll$\,$-$1 indicate a strong detection of perpendicular relative alignment.

Under the assumption that the uncertainty is dominated by the sample size, rather than by the measurement errors associated with the BLASTPol polarization angles or \mzero~gradient angles, the variance of the \prs~is
\begin{equation} \label{eqn:varprs}
\sigma_{Z_{x}}^2\,=\,\frac{2\,\sum_i^{\mnind} \left(\cos{\theta_i}\right)^2\,-\,\left(Z_{x}\right)^2}{\mnind}
\end{equation}
\citep{jow_2018}.
For the null hypothesis of a uniform distribution of angles (no alignment), $\sigma_{Z_{x\,\mathrm{uni}}} = 1$, which is the  standard against which $Z_{x}$ is tested. The behavior and convergence of the Rayleigh statistic and projected Rayleigh statistic are examined in detail in \cite{jow_2018}.

In practice finding \prs~for the set of relative orientation angles between the \blastpol~data and Mopra \mzero~maps (as calculated in Equation \ref{eqn:phi}) is complicated by the fact we measure $\theta_i$\, for every map pixel, therefore our data is highly oversampled.  
In Table \ref{tab:prs} we list the oversampled PRS $Z_{x}^{\prime}$~calculated for our measurements of \{$\theta_i$\} as 
\begin{equation} \label{eqn:prs_oversamp}
Z_{x}^{\prime}\,=\,\frac{\sum_i^{n_{pix}} \cos{\theta_i}}{\sqrt{n_{pix}/2}},
\end{equation}
where $n_{pix}$~is the number of map pixels of size indicated in Table \ref{tab:mom_params}.  

To correct for oversampling we calculate $Z_{x\,WN}^{\prime}$ for a series of relative orientation angles $\{\phi_{\mathrm{WN}\,i}\}$~where we replace $\nabla \mmzero$~in Equation \ref{eqn:phi} with $\nabla I_{\mathrm{WN}}$. $I_{\mathrm{WN}}$\,is a white noise map smoothed to the same resolution as the Mopra \mzero~maps.  The gradient angles of $I_{\mathrm{WN}}$~should be random but will also have the same degree of oversampling as the Mopra~\mzero~maps.  We calculate $Z_{x\,\mathrm{WN}}^{\prime}$~for 1000 $I_{\mathrm{WN}}$~realizations and list the mean ($\langle Z_{x\,WN}^{\prime}\rangle$) and standard deviation ($\sigma_{Z_{x\,WN}^{\prime}}$) in Table \ref{tab:prs}.  If every $\phi_{\mathrm{WN}}$~measurement was independent then $\sigma_{Z_{x\,WN}^{\prime}}$~should approach 1.  The value of $\sigma_{Z_{x\,WN}^{\prime}}$~therefore gives an estimate for the factor by which the data is oversampled.  We can therefore estimate the PRS corrected for oversampling by
\begin{equation} \label{eqn:prs_sampcorr}
Z_{x}\,=\,\frac{Z_{x}^{\prime}}{\sigma_{Z_{x\,WN}^{\prime}}},
\end{equation}
while the number of independent data samples in the map is
\begin{equation} \label{eqn:prs_samps}
\mnind\,=\,\frac{n_{\mathrm{pix}}}{\left(\sigma_{Z_{x\,WN}^{\prime}}\right)^2}.
\end{equation}
Both quantities are listed in Table \ref{tab:prs}.

\begin{figure}
\epsscale{1.2}
\plotone{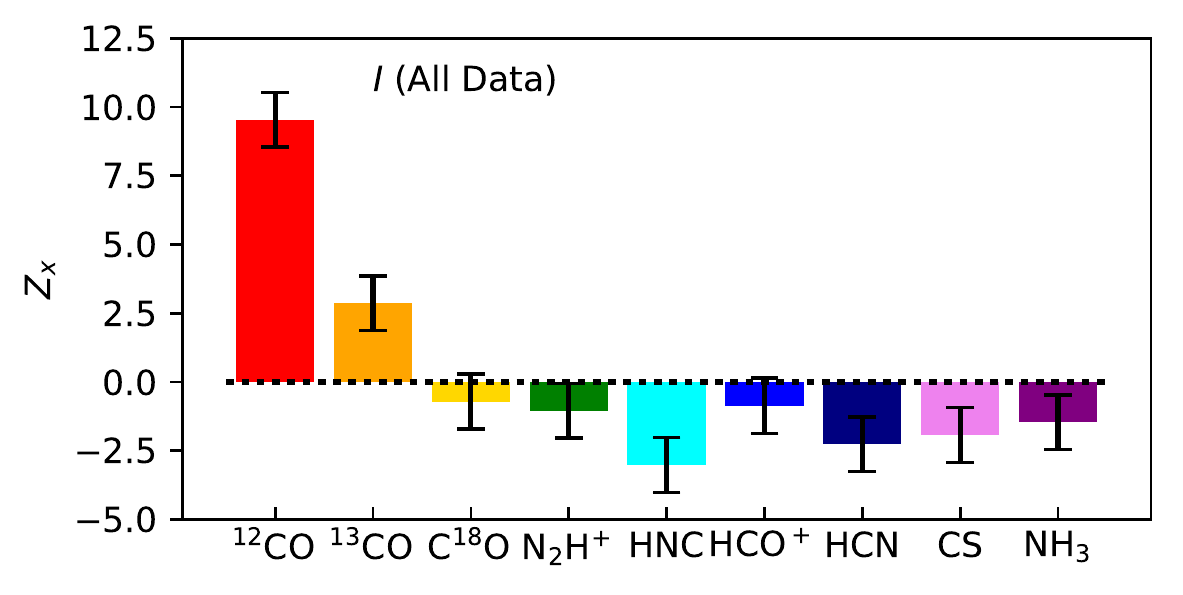}
\caption{Projected Rayleigh Statistic \prs~corrected for oversampling as discussed in Section \ref{sect:prs} for zeroth-moment \mzero~maps. \prs$\,>\,0$~indicates that \mzero~structures preferentially align parallel to \bpos, and \prs$\,<\,0$~indicates that \mzero~structures preferentially align perpendicular to \bpos.
\label{fig:prs_sys_test}}
\end{figure}

The statistical error bars for \prs~listed in Table \ref{tab:prs} are always $\simeq$\,1. However, these statistical error bars do not take into account potential systematic effects such as mapping artifacts associated with the Mopra telescope scanning strategy discussed in Section \ref{sect:mopra_obs}. 

To quantify this we replaced \mzero~in Equation \ref{eqn:phi} with \mzero$_{\mathrm{noise}}$, a ``zeroth-moment'' map made from velocity channels in the spectral data cube with no apparent molecular emission and recalculated \prs. The map gradient angles should be random, and so we would expect these calculated $Z_{x\,\mathrm{noise}}$ values to have a mean of 0 and a standard deviation of 1. The calculated values of \prs~listed in Table \ref{tab:prs}~have a mean of $-$0.35 and a standard deviation of 1.18, which is consistent with our expectations.

In Appendix \ref{sect:refs_and_res} we show our results are not sensitive to the resolution of the Mopra zeroth moment maps, the map sampling interval (provided the maps are sampled at least twice per smoothed Mopra beam FWHM $\theta_{\mathrm{sm}}$), or to the method used to remove the contribution of the diffuse ISM to the Vela\,C polarization maps. 

\subsubsection{Results from the PRS for individual molecular maps}
\label{sect:resultsprs}
Figure \ref{fig:prs_sys_test} shows the values of the oversampling corrected \prs~for each molecular line.  The $^{12}$CO emission tends to orient parallel to \bpos~(\prs$\,\gg\,$1).  We also see a weak preference  for $^{13}$CO to align parallel to \bpos (\prs$\,=\,$2.8). In contrast the \mzero~maps for the intermediate to higher density tracers tend to have no preferred orientation ($|Z_{x}|\,\leq\,$1), or show a weak preference to align perpendicular to the magnetic field (\prs\,=\,$-$2.3 for HCN and $-$3.0 for HNC).

\subsubsection{Results from the PRS in combination}
\label{sect:resultsprscom}

Even though the individual $|$\prs$|$~values are 3 or less for the intermediate to high density tracers N$_2$H$^+$, HNC, HCO$^+$, HCN, CS, and NH$_3$, we note that \prs~for each line is consistent with \prs\,$<$\,0, implying a preference for structures in these \mzero~maps to align perpendicular to \bpos.  We can statistically test whether intermediate and high density gas structures preferentially align perpendicular  as a whole. 

The PRS statistic in Equation \ref{eqn:prs} makes use of the set of angles measured for a given molecular line.  To construct a more sensitive PRS statistic for a combination of lines, in the numerator of Equation \ref{eqn:prs} each set of $n_\mathrm{ind,j}$ measurements $\theta_{i,j}$ for molecular line $j$ can be used for $n_\mathrm{lin}$ molecular lines (totalling $n_\mathrm{tot} = \sum_{j}^{n_\mathrm{lin}}\, n_\mathrm{ind,j}$ measurements), leading to
\begin{equation}
\label{eqn:avg_prs}
Z_{x\,\mathrm{com}} \,=\, \sum_{j}^{n_\mathrm{lin}} \sqrt{ n_\mathrm{ind, j}/n_\mathrm{tot} }\, Z_{x, j} \, .
\end{equation}

The variance for the null hypothesis of a uniform distribution of angles, but now anticipating that the sets of angles measured using different molecular lines might be correlated, is
\begin{equation} \label{eqn:varprscom}
\sigma_{Z_{x\,\mathrm{com}\,\mathrm{uni}}}^2 \,= 1 + 2\sum_{j k: j < k}^{n_\mathrm{lin}} \sqrt{ n_\mathrm{ind, j} \, n_\mathrm{ind, k}}/n_\mathrm{tot} \, \left< Z_{x, j} Z_{x, k}\right> \, ,
\end{equation}
where the angle brackets indicate the expectation value.  For this hypothesis
\begin{equation} \label{eqn:covar}
\left< Z_{x, j} Z_{x, k}\right> \, = \, 2\, \mathrm{cov}(\cos\theta_{i,j},\cos\theta_{i,k})\, ,
\end{equation}
which is unity when $j = k$ (the covariance is 0.5), so that $| \left< Z_{x, j} Z_{x, k}\right> | \le 1$ by the Cauchy-Schwartz inequality.\footnote{Because we are investigating whether the sets of gradient orientations $\psi_{i,j}$ and $\psi_{i,k}$ in two molecular line maps are independent, this measure of correlation can also be estimated from Equation~\ref{eqn:covar} with $\theta$ replaced by $2 \psi$  or from $Z_{x, j, k}/\sqrt{(Z_{x, j, j}Z_{x, k, k})}$, where this PRS is evaluated for angles $2\, (\psi_{i,j} - \psi_{i,k}$). The three approaches yield similar values.}

We consider two limiting cases.  
In the absence of correlation between the sets of angles, $\theta_{i,j}$ and $\theta_{i,k}$, $\left< Z_{x, j} Z_{x, k}\right> \,= 0$ and $\sigma_{Z_{x\,\mathrm{com}\,\mathrm{uni}}}= 1$; therefore, $Z_{x\,\mathrm{com}}$ will be a more sensitive statistic by virtue of the increase in $\sqrt{n_\mathrm{tot}}$.  
On the other hand, complete correlation would be like incorporating replicas of the same set of angles in the combination. For all lines, $n_\mathrm{ind, j} = n_\mathrm{tot}/n_\mathrm{lin}$. Compared to $Z_{x, j}$ for a single line, $Z_{x\,\mathrm{com}}$ would be larger by a factor $\sqrt{n_\mathrm{lin}}$.  But now all off-diagonal elements  $\left< Z_{x, j} Z_{x, k}\right> = 1$, so that $\sigma_{Z_{x\,\mathrm{com}\,\mathrm{uni}}} = \sqrt{n_\mathrm{lin}}$.  Thus the relevant figure of merit, $Z_{x\,\mathrm{com}}/ \sigma_{Z_{x\,\mathrm{com}}\,\mathrm{uni}} $ is unchanged, as expected because no new information has been added.  

Using the values of $n_{\mathrm{ind}}$ and $Z_x$ for the intermediate to high density tracers N$_2$H$^+$, HNC, HCO$^+$, HCN, CS, and NH$_3$\ in Table~\ref{tab:prs}, we find $Z_{x\,\mathrm{com}}\,=\,-$4.2 from Equation~\ref{eqn:avg_prs}.  For pairs of these tracers, we have calculated $ \left< Z_{x, j} Z_{x, k}\right> $ from the data using Equation~\ref{eqn:covar}, finding values with a mean of 0.29 and dispersion of 0.07.  Therefore, from Equation~\ref{eqn:varprscom}, $\sigma_{Z_{x\,\mathrm{com}\,\mathrm{uni}}} = 1.8$.
The relevant figure of merit, $Z_{x\,\mathrm{com}}/ \sigma_{Z_{x\,\mathrm{com}}\,\mathrm{uni}}  = -2.8$ implies that intermediate to high density gas structures are aligned preferentially perpendicular to the magnetic field, at the 
 2.8$\sigma$ confidence level, certainly much different than the parallel alignment revealed by the lowest density tracers.

\subsection{Characteristic Densities Traced by the Mopra Observations}\label{sect:density_deriv}

Here we quantify the characteristic density traced by each of our observed molecular lines, in order to understand how the Vela\,C cloud structure is aligned with respect to the magnetic field over different number density regimes.  
We do this in three ways, by using a simple non-LTE radiative transfer model to calculate the $n_{\mathrm{H}_{2}}$~needed to reproduce our \mzero~observations (Section \ref{sect:n_radex}), by calculating the critical density corrected for radiative trapping for the highly optically thick $^{12}$CO observations (Section \ref{sect:ncrit_co}), and by using the cloud width as a proxy for the depth in order to estimate characteristic number density from column density maps (Section \ref{sect:n_xsec}).

\subsubsection{Characteristic Densities Estimated from Radiative Transfer Models}\label{sect:n_radex}

We first estimate the characteristic $n_{\mathrm{H}_2}$~using an adaptation of the effective excitation density analysis presented in \cite{shirley_2015}, where the author found the typical density required to produce a 1\,K\,km\,s$^{-1}$~line for a number of molecular lines.

With only one observed line per molecular species we cannot calculate the excitation temperature (\tex), or kinetic temperature (\tkin) of the gas.  Instead we assume that \tex\,$=$\,\tkin~and that these temperatures are within the range of 10 to 20 K.\footnote{The {\tt RADEX}~radiative transfer models we use to estimate the characteristic density do not require \tex~as an input parameter, but do require an input molecular column density, which depends on \tex\,(see Equation \ref{eqn:cd_thin}).} We justify this by noting that the maximum $T_{\mathrm{R}}$~observed within Vela\,C in $^{12}$CO is typically 10\,K, as shown in Figure \ref{fig:spectra}, increasing to 20\,K near RCW\,36 (Spectrum D in Figure \ref{fig:spectra}).  The $^{12}$CO $J$\,=\,1\,$\rightarrow$\,0 emission should be optically thick over most of the cloud, and so we expect that $T_{\mathrm{R}}\,\approx\,$\tex~for $^{12}$CO.  Additionally, we note that the dust temperature in Vela\,C is generally in the range of 10 to 16\,K, except near the compact \ion{H}{2}~region RCW\,36 \citep{fissel_2016}.  At the moderately high densities traced by N$_2$H$^{+}$, HCO$^{+}$, HCN, HNC, CS, and NH$_3$\,(1,1) the gas should be collisionally coupled to the dust and therefore the dust temperature should be approximately equal to the gas kinetic temperature. 
\begin{figure*}
\epsscale{0.9}
\plotone{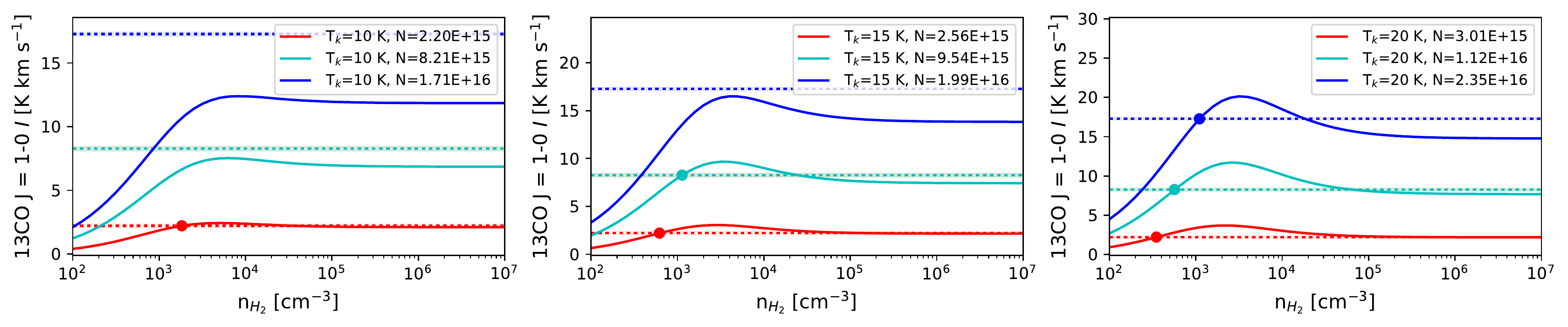}
\plotone{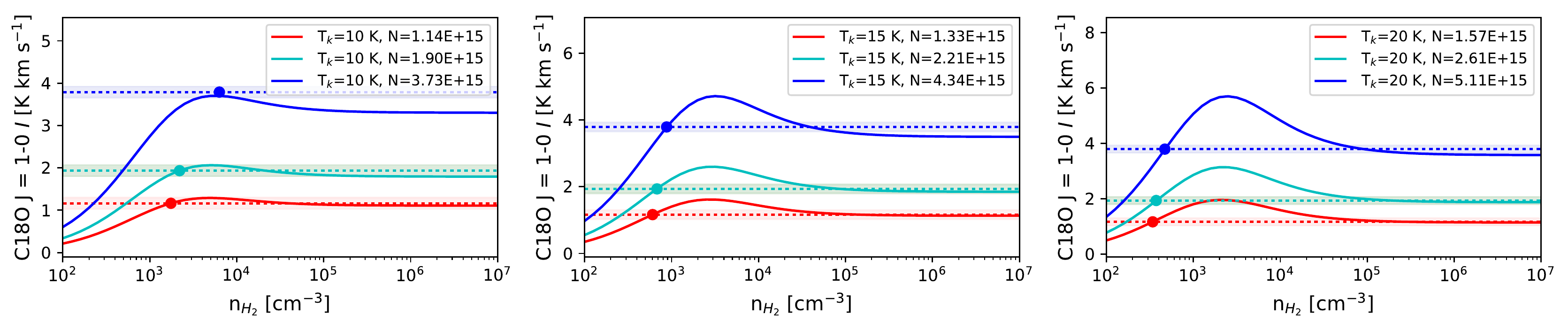}
\plotone{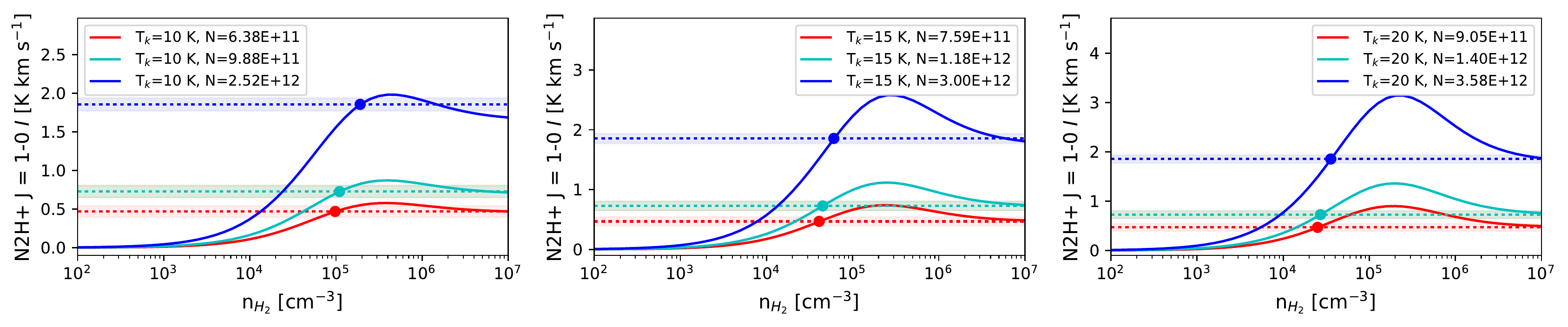}
\plotone{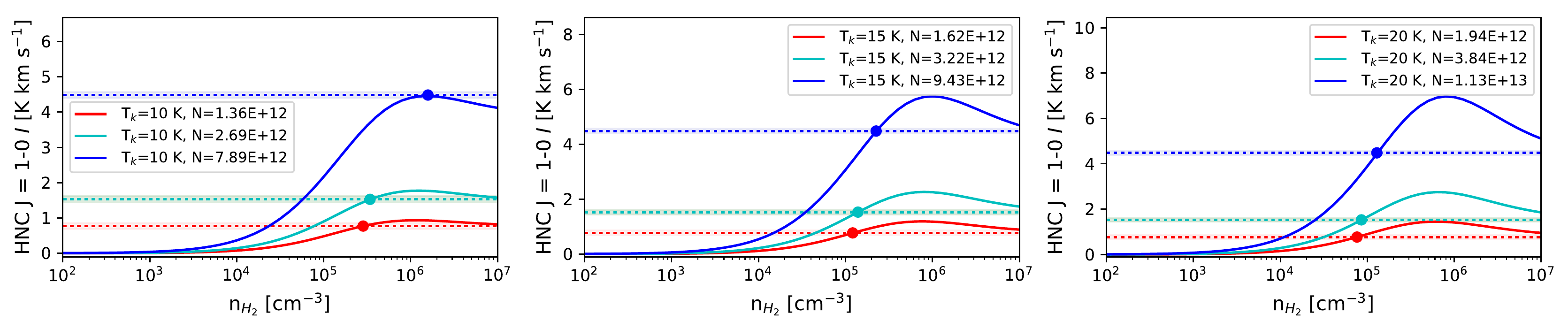}
\plotone{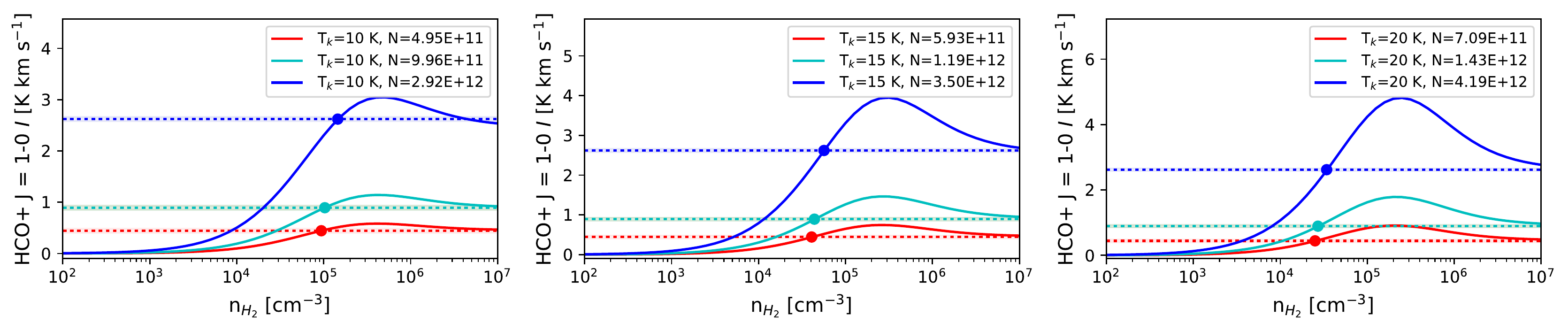}
\plotone{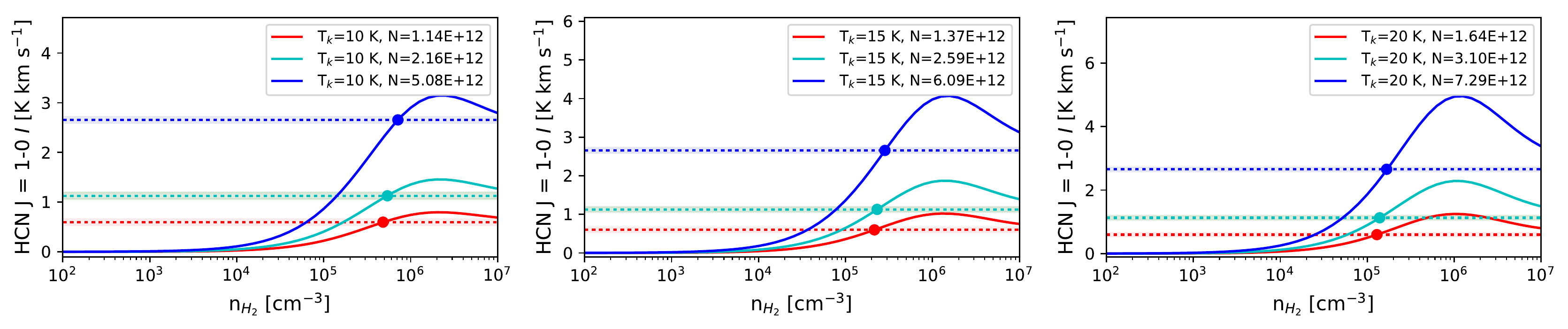}
\plotone{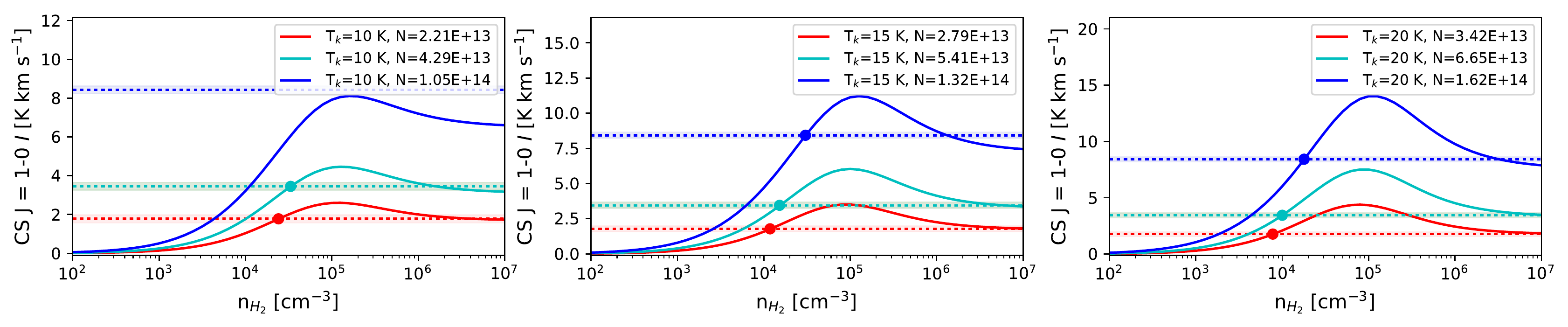}
\plotone{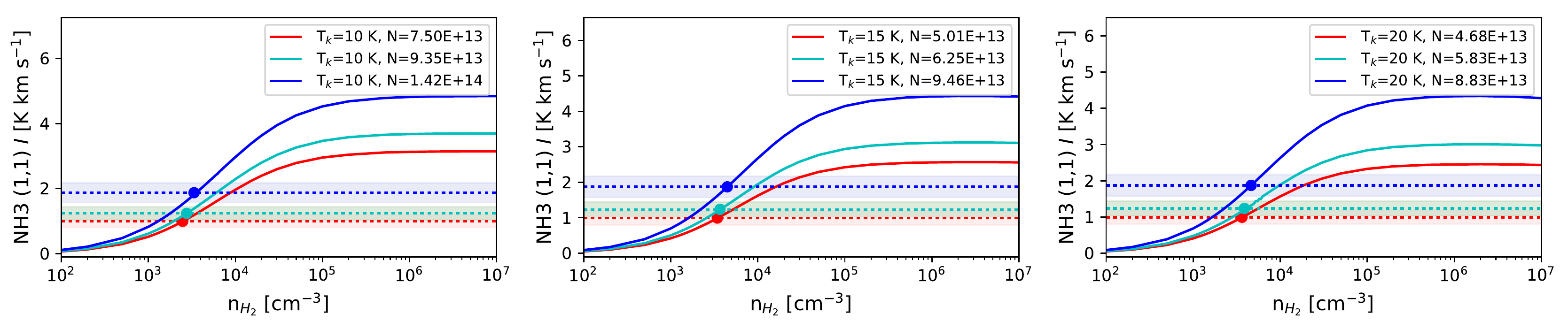}
\caption{Integrated line intensity predicted from {\tt RADEX}~models (solid lines)~compared to the measured \mzero~values at the corresponding percentile (dotted horizontal lines) for the 5th, 50th, and 95th column density percentiles (red, cyan, and blue, respectively), with shaded bands indicating the 1-$\sigma$~uncertainty range for \mzero.  The characteristic density is taken to be the lowest value of $n_{\mathrm{H}_2}$~for which the RADEX model intersects the observed \mzero~value (filled circles). 
\label{fig:radex_models}
}
\end{figure*}

We first calculate the column density $N_{\mathrm{tot}}^{\mathrm{thin}}$~of each molecule assuming \tex\,is in the set \{10, 15, and 20\,K\}, and using the methods outlined in \cite{mangum_2015}. We assume that the observed molecular lines are optically thin and in local thermodynamic equilibrium (LTE).  The details of these calculations are discussed in Appendix \ref{sec:app_cd}.  

Next we calculate the zeroth-moment for {\tt RADEX}~non-LTE radiative transfer models  \citep{vandertak_2007} over the $n_{\mathrm{H}_2}$ range [1.0\,$\times$\,10$^{2}$\,cm$^{-3}$, 1.0\,$\times$\,10$^{7}$\,cm$^{-3}$] for each line, as shown in Figure \ref{fig:radex_models}.  {\tt RADEX}~models require an input molecular column density, kinetic temperature, and a FWHM velocity width. We base the FWHM velocity width from the results of Gaussian fits to the single peaked line spectra at locations B and E shown in Figure \ref{fig:spectra}. We calculate {\tt RADEX} models for the 5th, 50th, and 95th percentiles of $N_{\mathrm{tot}}^{\mathrm{thin}}$~from Appendix \ref{sec:app_cd}, and kinetic temperatures $T_{\mathrm{k}}$\,=\,$T_{\mathrm{ex}}$ in the set $\{$10, 15, 20\,K$\}$, for a total of nine models calculated per molecular line.  

We take the lowest $n_{\mathrm{H}_2}$~value from {\tt RADEX}~that can reproduce the observed \mzero~value (dashed lines in Figure \ref{fig:radex_models}) as the characteristic number density \nradex~traced by the line.  For a few cases the {\tt RADEX}-model-predicted zeroth-moment does not reach the observed value.  In this case if \mzero$_{\mathrm{RADEX\,max}}$~is within the measurement uncertainty for \mzero, we take \nradex\, to be the $n_{\mathrm{H}_2}$~for which the {\tt RADEX}~model produces the largest zeroth-moment value; otherwise we cannot estimate \nradex~for those parameters.

The {\tt RADEX}-derived density values are listed in Table \ref{tab:dens_radex}.  In general, the models predict that the HCN, HNC, N$_2$H$^+$, CS, and HCO$^+$\,lines trace higher densities (\nradex$\,>$\,10$^4$\,cm$^{-3}$), while $^{13}$CO, C$^{18}$O, and NH$_3$~will be sensitive to gas densities \nradex$\,<$\,10$^4$\,cm$^{-3}$.  The spread in \nradex~values calculated for different assumptions of \tkin~and $N_{\mathrm{tot}}^{\mathrm{thin}}$~percentiles can be used as a rough estimate of the uncertainty of \nradex, which is typically an order of magnitude.  We have also tested the sensitivity of our derived densities to cases where \tex\,$<$\,\tkin~and found that the \nradex~values derived from these models do not differ significantly from the range of \nradex~values listed in Table \ref{tab:dens_radex}.

Note that the {\tt RADEX}~models do not account for variations in molecular abundance with density. In Section \ref{sect:prs_vs_dens} we discuss the possible effects of CO freeze-out and other abundance variations on the characteristic number density traced by each molecular line.  Our estimates of column density $N_{\mathrm{tot}}^{\mathrm{thin}}$ may also be underestimated if the lines have significant optical depth. This would result in an overestimate of the derived characteristic density, which scales roughly proportional to $N_{\mathrm{tot}}^{\mathrm{thin}}/N_{\mathrm{tot}}$, where $N_{\mathrm{tot}}$~is the true column density \citep{shirley_2015}.  However as shown in Section \ref{sect:ncrit_co}, we do not expect molecules other than $^{12}$CO to have $\tau\,\gg\,1$, so this should at worst result in an factor of a few error in our density estimates, which is much smaller than the range of densities traced by our target molecular lines.

\subsubsection{Estimates of the 12CO J = 1--0 Critical Density}\label{sect:ncrit_co}
The $^{12}$CO emission is likely to be so optically thick across Vela\,C that {\tt RADEX}~models are not applicable. In contrast we expect $\tau_{\mathrm{C}^{18}\mathrm{O}}\ll\,$1, such that:
\begin{eqnarray} \label{eqn:tau_c18o}
T_{\mathrm{R\,C}^{18}\mathrm{O}} & = & \tau_{\mathrm{C}^{18}\mathrm{O}} T_{\mathrm{ex}}.
\end{eqnarray} 
If we assume $T_{\mathrm{ex}}$\,=\,10 K, then $\tau_{\mathrm{C}^{18}\mathrm{O}}$~typically ranges from 0.015 to 0.18, with a median value of 0.026.  Assuming a [$^{13}$CO/C$^{18}$O] ratio of 10 and a [$^{12}$CO/C$^{18}$O]~ratio of 400, this implies a typical $\tau_{^{12}\mathrm{CO}}$\,=\,[$^{12}$CO/C$^{18}$O]$\tau_{\mathrm{C}^{18}\mathrm{O}}$~in the range of 6 to 72, and $\tau_{^{13}\mathrm{CO}}$~in the range of 0.15 to 1.8. The $^{12}$CO\,$J$\,=\,1\,$\rightarrow$\,0 emission is therefore extremely optically thick, while the next most abundant tracer $^{13}$CO has emission that is either optically thin or at most only moderately optically thick.  Since $^{13}$CO is much more abundant than all the other molecules probed in this study (except for $^{12}$CO), we expect that the other molecular lines will also not have $\tau\,\gg\,1$.

A useful estimate for the lower limit of the characteristic number density of $^{12}$CO is the critical density for $^{12}$CO $J$\,=\,1\,$\rightarrow$\,0 corrected for radiative trapping:
\begin{eqnarray}
n_{\mathrm{crit}}^{\mathrm{thick}} & = & n_{\mathrm{crit}}^{\mathrm{thin}} \bar{\beta},
\end{eqnarray}
where $n_{\mathrm{crit}}^{\mathrm{thin}}$~is the critical density calculated from the Einstein coefficients and collisional rates for $^{12}$CO without accounting for absorption or stimulated emission ($n_{\mathrm{crit}}^{\mathrm{thin}}$\,=\,900\,cm$^{-3}$\,for $^{12}$CO gas with $T_{\mathrm{ex}}$\,=\,10K), and $\bar{\beta}$~is the photon escape fraction.  For a static uniform sphere $\bar{\beta}$~can be approximated by:
\begin{eqnarray}
\beta & = & \frac{3}{4\tau}\,-\,\frac{3}{8\tau^{3}}\,+\,\exp{\left(-2\tau\right)\,\left(\frac{3}{4\tau^2}\,+\,\frac{3}{8\tau^{3}}\right)},
\end{eqnarray}
\citep{osterbrock_1989}.  Evaluating this correction factor for $^{12}$CO~gives $n_{\mathrm{crit}}^{\mathrm{thick}}$\,$\geq$\,9--111\,cm$^{-3}$.

\capstartfalse
\begin{deluxetable*}{cccc|ccc|ccc}[H!]
\tabletypesize{\footnotesize}
\tablecaption{Calculated Characteristic $n_{\mathrm{H}_2}$~Densities from {\tt RADEX} Models.
\label{tab:dens_radex}}
\tablehead{
\colhead{Molecular Line\footnotemark[1]} & \multicolumn{3}{c}{\nradex($N_{0.05}$)\footnotemark[2]\,[cm$^{-3}$]}& 
 \multicolumn{3}{c}{\nradex($N_{0.50}$)\footnotemark[2]\,[cm$^{-3}$]}& 
 \multicolumn{3}{c}{\nradex($N_{0.95}$)\footnotemark[2]\,[cm$^{-3}$]}\\ 
\colhead{} & \colhead{$T$\,=\,10\,K} & \colhead{$T$\,=\,15\,K} & \colhead{$T$\,=\,20\,K} & 
\colhead{$T$\,=\,10\,K} & \colhead{$T$\,=\,15\,K} & \colhead{$T$\,=\,20\,K} &
\colhead{$T$\,=\,10\,K} & \colhead{$T$\,=\,15\,K} & \colhead{$T$\,=\,20\,K} 
}
\startdata
$^{13}$CO\,$J$\,=\,1\,$\rightarrow$\,0 & 1.83E+03 & 6.18E+02 & 3.49E+02 & -- & 1.13E+03 & 5.66E+02 & -- & -- & 1.10E+03 \\
C$^{18}$O\,$J$\,=\,1\,$\rightarrow$\,0 & 1.75E+03 & 6.05E+02 & 3.41E+02 & 2.20E+03 & 6.80E+02 & 3.74E+02 & 6.31E+03 & 8.77E+02 & 4.70E+02 \\
N$_2$H$^+$\,$J$\,=\,1\,$\rightarrow$\,0 & 9.79E+04 & 4.08E+04 & 2.51E+04 & 1.10E+05 & 4.50E+04 & 2.71E+04 & 1.91E+05 & 6.05E+04 & 3.57E+04 \\
HNC\,$J$\,=\,1\,$\rightarrow$\,0 & 2.84E+05 & 1.21E+05 & 7.62E+04 & 3.41E+05 & 1.39E+05 & 8.56E+04 & 1.58E+06 & 2.25E+05 & 1.29E+05 \\
HCO$^+$\,$J$\,=\,1\,$\rightarrow$\,0 & 9.38E+04 & 4.08E+04 & 2.51E+04 & 1.03E+05 & 4.39E+04 & 2.71E+04 & 1.45E+05 & 5.66E+04 & 3.41E+04 \\
HCN\,$J$\,=\,1\,$\rightarrow$\,0 & 4.81E+05 & 2.15E+05 & 1.29E+05 & 5.40E+05 & 2.31E+05 & 1.39E+05 & 7.13E+05 & 2.84E+05 & 1.67E+05 \\
CS\,$J$\,=\,1\,$\rightarrow$\,0 & 2.41E+04 & 1.18E+04 & 7.78E+03 & 3.33E+04 & 1.52E+04 & 1.00E+04 & -- & 3.03E+04 & 1.79E+04 \\
NH$_3$\,(1,1)           &  2.48E+03 & 3.43E+03  & 3.64E+03 & 2.75E+03 & 3.68E+03 & 3.88E+03 & 2.35E+03 & 4.46E+03 & 4.63E+03  \\
\enddata
\tablecomments{Characteristic densities for each line are derived from the {\tt RADEX}~radiative transfer models shown in Figure \ref{fig:radex_models} and described in Section \ref{sect:n_radex}. 
\footnotetext[1]{{\tt RADEX} FWHM velocity width assumed: 3.0\,km\,s$^{-1}$~for $^{13}$CO and HCO$^+$\,$J$\,=\,1\,$\rightarrow$\,0; 2.0\,km\,s$^{-1}$~C$^{18}$O, HNC, HCN, CS\,$J$\,=\,1\,$\rightarrow$\,0; and 1.0\,km\,s$^{-1}$~for N$_2$H$^+$ \,$J$\,=\,1\,$\rightarrow$\,0, NH$_3$\,(1,1).}}
\footnotetext[2]{$N_{0.05}$, $N_{0.50}$, and $N_{0.95}$, refer to the 5th, 50th and 95th percentiles of the molecular column density $N_{\mathrm{tot}}^{\mathrm{thin}}$ (see Table \ref{tab:column_density} in Appendix \ref{sec:app_cd}).}
\end{deluxetable*}
\capstarttrue

\capstartfalse
\begin{deluxetable*}{cccc|ccc}
\tabletypesize{\footnotesize}
\tablecaption{Characteristic $n_{H_2}$~Densities Estimated from Molecular Column Density Cross-sections.
\label{tab:dens_depth}}
\tablewidth{0pt}
\tablehead{
\colhead{} & \colhead{} & \colhead{Centre-Ridge\footnotemark[1]} & \colhead{} &  \colhead{} & \colhead{South-Nest\footnotemark[1]} & \colhead{} \\
\colhead{Molecular Line} & \colhead{Width} & \colhead{$\bar{N}_{\mathrm{H_2 mol}}$} & \colhead{\ndepth}
& \colhead{Width} & \colhead{$\bar{N}_{\mathrm{H_2 mol}}$} & \colhead{\ndepth} \\
\colhead{} & \colhead{[pc]} & \colhead{[cm$^{-2}$]} & \colhead{[cm$^{-3}$]} & 
\colhead{[pc]} & \colhead{[cm$^{-2}$]} & \colhead{[cm$^{-3}$]}
}
\startdata
$^{12}$CO $J$\,=\,1\,$\rightarrow$\,0 & 11.4 & 4.2E+20 & 1.2E+01 & 10.3 & 5.1E+20 & 1.6E+01 \\
$^{13}$CO $J$\,=\,1\,$\rightarrow$\,0 & 11.4 & 5.2E+21 & 1.5E+02 & 9.8 & 6.9E+21 & 2.3E+02 \\
C$^{18}$O $J$\,=\,1\,$\rightarrow$\,0 & 2.6 & 1.2E+22 & 1.5E+03 & 5.8 & 1.5E+22 & 8.3E+02 \\
N$_2$H$^+$~$J$\,=\,1\,$\rightarrow$\,0 & 0.9 & 4.5E+22 & 1.5E+04 & 3.1 & 1.8E+22 & 1.9E+03 \\
HNC $J$\,=\,1\,$\rightarrow$\,0 & 2.9 & 1.7E+22 & 1.9E+03 & 6.1 & 2.1E+22 & 1.1E+03 \\
HCO$^+$ $J$\,=\,1\,$\rightarrow$\,0 & 4.9 & 1.3E+22 & 8.8E+02 & 6.4 & 2.2E+22 & 1.1E+03 \\
HCN $J$\,=\,1\,$\rightarrow$\,0 & 3.3 & 2.1E+22 & 2.0E+03 & 5.9 & 1.7E+22 & 9.1E+02 \\
CS $J$\,=\,1\,$\rightarrow$\,0 & 2.0 & 1.2E+22 & 2.0E+03 & 6.4 & 1.7E+22 & 8.3E+02 \\
NH$_3$~(1,1) & -- & -- & -- & 0.4 & 2.0E+22 & 1.5E+04 \\
\enddata
\tablecomments{
To convert from the molecular column densities $N_{\mathrm{tot}}^{\mathrm{thin}}$~given in Table \ref{tab:column_density} to $\bar{N}_{\mathrm{H_2 mol}}$~we use the derived median abundance ratios (also listed in Table \ref{tab:column_density}), except for $^{12}$CO (which is extremely optically thick) for which we assume [$N_{\mathrm{H_2}}$/$N_{\mathrm{12CO}}$]\,=\,1.1$\,\times\,10^{-4}$ \citep{millar_1997}.
\footnotetext[1]{Average molecular hydrogen column densities ($\bar{N}_{\mathrm{H_2 mol}}$), cloud widths, and inferred molecular hydrogen number densities (\ndepth) were calculated for two cloud cross-sections (shown in Figure \ref{fig:cdmaps}), one across the South-Nest, and one that crosses the highest column density peak in the Centre-Ridge.}
}
\end{deluxetable*}
\capstarttrue

\subsubsection{Characteristic Densities Estimated from Mopra Column Density Maps}\label{sect:n_xsec}
\begin{figure*}
\epsscale{0.40}
\plotone{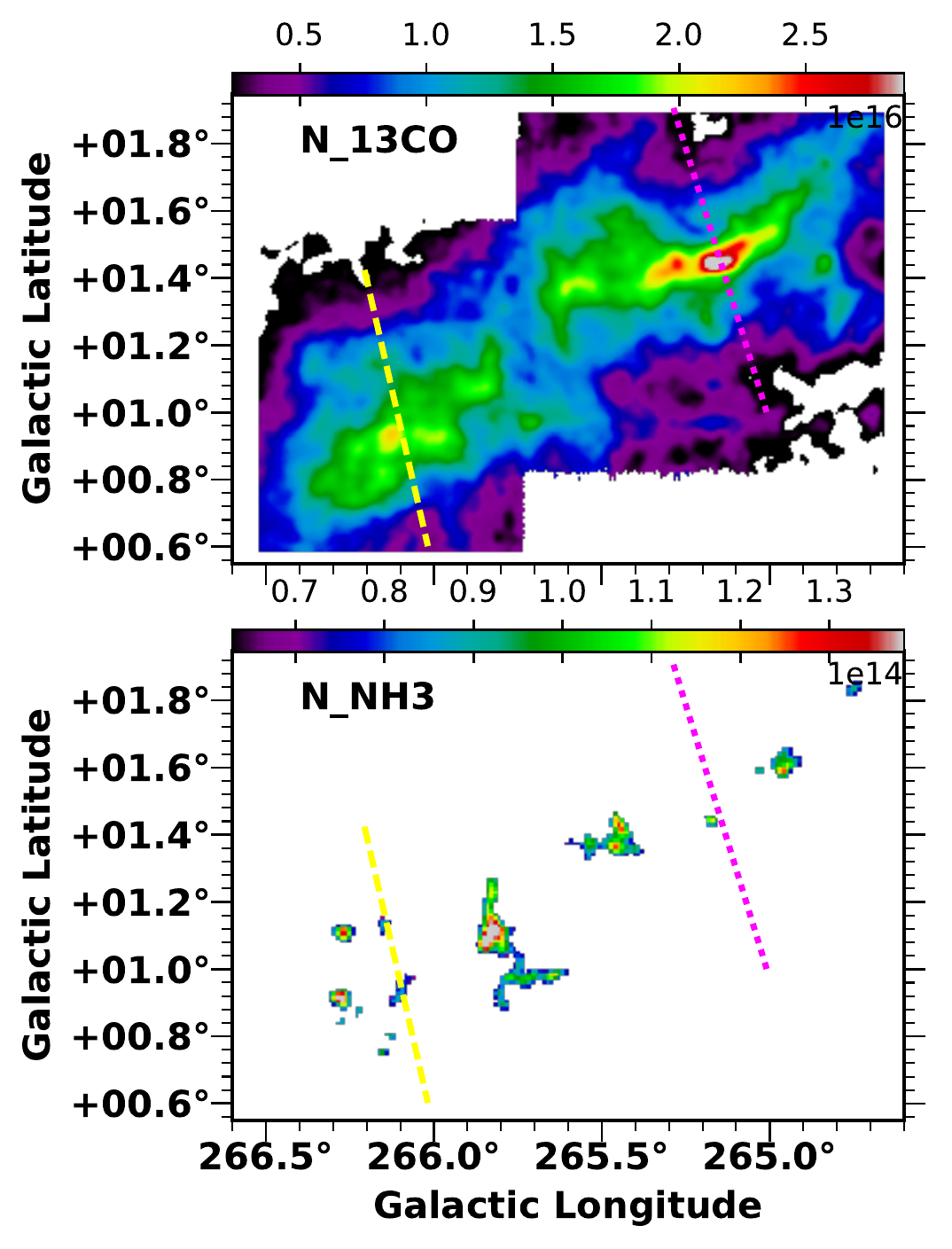}
\epsscale{0.37}
\plotone{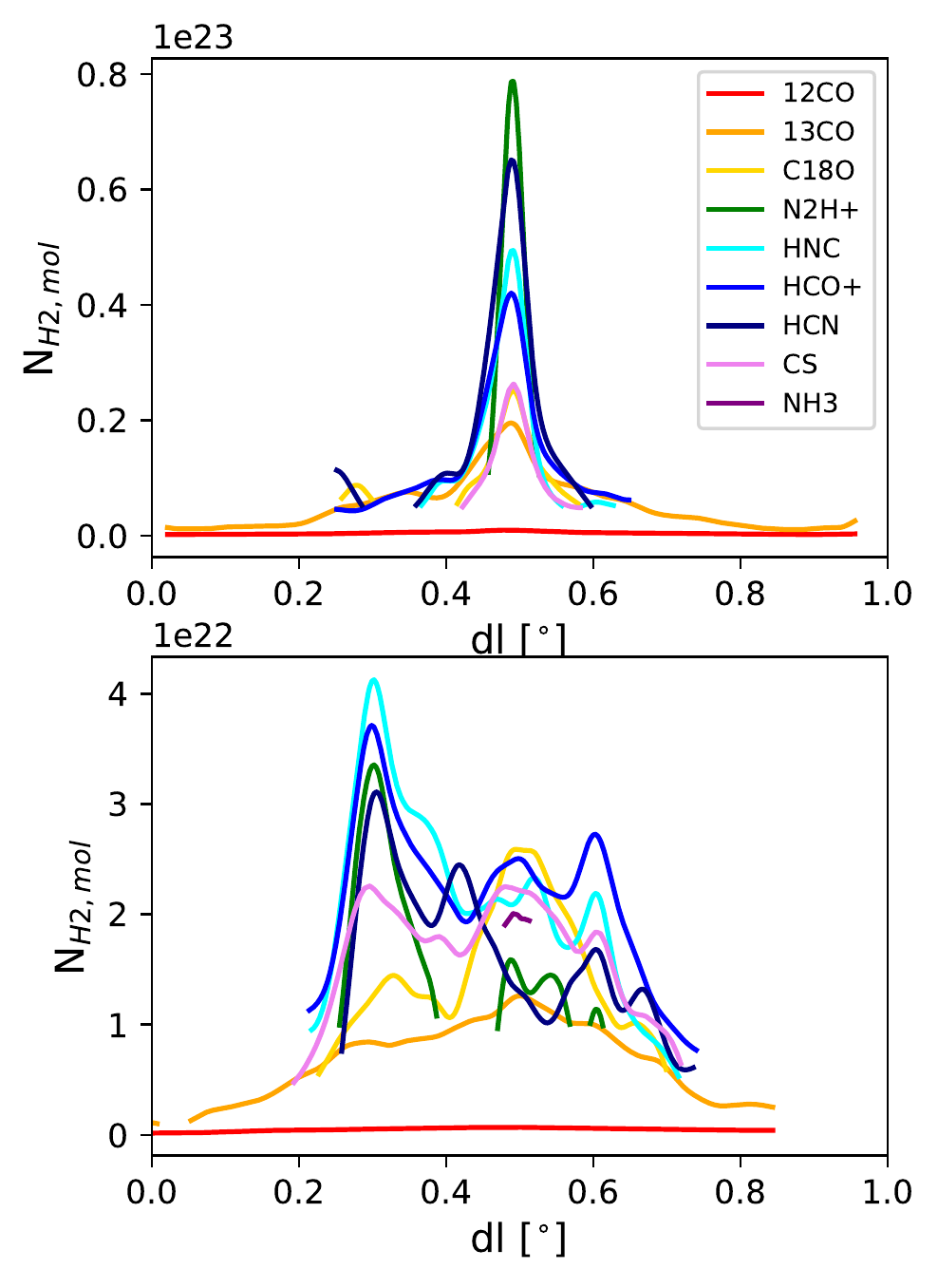}
\caption{{\em Left panels:} Maps of molecular hydrogen column density $N_{\mathrm{tot}}^{\mathrm{thin}}$~(calculated as described in Appendix \ref{sec:app_cd}) for $^{13}$CO\,$J$\,=\,1\,$\rightarrow$\,0~{\em (top panel)}~and NH$_3$(1,1)~{\em (bottom panel)} assuming an excitation temperature of 10\,K.  The dashed lines show the South-Nest (yellow) and Centre-Ridge (magenta) cross sections used to estimate the characteristic molecular density, as derived in Section \ref{sect:n_xsec}. {\em Right panels:} Estimated molecular hydrogen column density $N_{\mathrm{H_2,mol}}$~traced by each molecular line (see Equation \ref{eqn:nh2_mol}) for a cross-section of the Centre-Ridge ({\em top panel}) and South-Nest ({\em bottom panel}) assuming an excitation temperature of 10\,K.\label{fig:cdmaps}
}
\end{figure*}

We can also estimate the characteristic number density of the gas traced by each molecular line if the molecular abundance ratio [$N_{\mathrm{H_2}}$/$N_{\mathrm{tot}}^{\mathrm{thin}}$]~and cloud depth $\Delta z$~are known:
\begin{eqnarray} \label{eqn:n_xsect}
\mndepth & = & \frac{\bar{N}_{\mathrm{H_2,mol}}}{\Delta z} \\
\end{eqnarray}
where $\bar{N}_{\mathrm{H_2,mol}}$~is the molecular hydrogen column density traced by a line averaged over a cross-section through the cloud calculated by
\begin{eqnarray} \label{eqn:nh2_mol}
N_{\mathrm{H_2,mol}} & = & N_{\mathrm{tot}}^{\mathrm{thin}}\,\times\,\left[\frac{N_{\mathrm{H_2}}}{N_{\mathrm{tot}}^{\mathrm{thin}}}\right].
\end{eqnarray}
Here the molecular abundance ratios are calculated from the median ratio of the molecular hydrogen column density $N_{\mathrm{H_2}}$, assumed to be $N_{\mathrm{H}}/2$, where $N_{\mathrm{H}}$ is the hydrogen column density calculated from the \herschel~dust SED fits as described in Section \ref{sect:col_dens_nh}, to the molecular line column density $N_{\mathrm{tot}}^{\mathrm{thin}}$, which is derived for each molecule from the integrated zeroth-moment maps for different assumptions of excitation temperate, as described in Appendix \ref{sec:app_cd}.  The only exception is for the optically thick $^{12}$CO line for which we assume a conversion factor of [$N_{\mathrm{H_2}}$/$N_{\mathrm{tot\,12CO}}^{\mathrm{thin}}$]\,=\,1\,$\times$\,10$^4$~from the literature (e.g., \citealt{millar_1997}). 
The cloud line of sight depth $\Delta z$ cannot be measured, but as a first approximation we can assume that it is similar to the cloud width.  

We estimate the average density across two cross-sections of Vela\,C as shown in Figure \ref{fig:cdmaps}: one that crosses the highest column density location in Vela\,C on the Centre-Ridge; and one that crosses the more diffuse South Nest.  For each molecular column density map we use Equation \ref{eqn:n_xsect} to calculate \ndepth, using the mean molecular column density along the cross section as $\bar{N}_{\mathrm{tot}}^{\mathrm{thin}}$, and assuming that $\Delta z$~is approximately equal to the total length along the cloud cross-section for which we have significant detections of \mzero. The abundance ratio is assumed to be constant across the cloud.

The range of cloud depths and estimated densities \ndepth~from the cross-sectional estimates are given in Table \ref{tab:dens_depth}, assuming  T$_{\mathrm{ex}}$\,=\,10\,K.\footnote{Note that unlike the estimates of \nradex~from Section \ref{sect:n_radex} there is no significant difference between \ndepth~estimates for different assumptions of excitation temperature. This is because the abundance ratio is calculated from the average ratio of the molecular hydrogen column density (derived from \herschel~observations and discussed in Section \ref{sect:col_dens_nh}) to the molecular column density (see Table \ref{tab:column_density} in Appendix \ref{sec:app_cd}). The excitation temperature dependence of the abundance in Equation \ref{eqn:n_xsect} therefore cancels when multiplied by $\bar{N}_{\mathrm{tot}}^{\mathrm{thin}}$. Only $^{12}$CO (where an abundance ratio was assumed) shows a dependence of the estimated \ndepth~on the excitation temperature.} 
 Note that this method of estimating the number density requires more assumptions than the density estimates in Sections \ref{sect:n_radex} and \ref{sect:ncrit_co}, and so the estimates of \ndepth~in Table \ref{tab:dens_depth} are most useful as a consistency check rather than an equally valid determination of characteristic number density.
  The {\tt RADEX}~derived and cross-section density estimates are broadly consistent for $^{12}$CO, $^{13}$CO, and C$^{18}$O~$J$\,=\,1\,$\rightarrow$\,0, but the cross-section estimates are systematically lower for intermediate and higher density tracers HCN, HCO$^{+}$, HNC, N$_2$H$^+$, and CS.  We discuss the discrepancies between the different methods for calculating the characteristic number density in more detail in Section \ref{sect:prs_vs_dens}.

No estimate of \ndepth~for NH$_3$\,(1,1) was made for the Centre-Ridge cross-section as there was no detection of \mzero~that passed the signal-to-noise selection criteria  described in Section \ref{sect:mom_maps}. The NH$_3$~\ndepth~calculated for the South-Nest is higher than the \ndepth~estimates for any other molecule, because the  width over which the NH$_3$~emission was detected is smaller than the cross-sectional width of detected emission for the other molecular lines. This indicates that even though NH$_3$\,(1,1) is expected to trace intermediate gas (see \nradex~in Table \ref{tab:dens_radex}), in our observations we only have the sensitivity to detect NH$_3$\,(1,1) toward the highest column density regions of Vela\,C.

\section{Discussion} \label{sect:discussion}

The most striking feature of the above projected Rayleigh statistic (PRS) analysis is that the average orientation of structures in zeroth-moment (\mzero)~maps relative to the  magnetic field orientation inferred from BLASTPol polarization data \bpos~is substantially different for the different molecular line tracers.  In this section we discuss the cause of these differences and the extent to which our PRS results can tell us about the role magnetic fields play in the formation of structure within molecular clouds.


\subsection{Changes in Relative Orientation with Column Density?}\label{sect:nh_or_dens}

Unlike the \herschel~derived column density maps used in the analysis of \cite{soler_2017} and \cite{jow_2018}, the \mzero~maps in this work do not necessarily reflect the structure of the total gas column density.  Instead the zeroth-moment maps shown in the left panels of Figures \ref{fig:mom_maps1} and \ref{fig:mom_maps2} are sensitive to the column density of the emitting molecules, the number density and average speed of particles colliding with the molecules (usually assumed to be H$_2$), the line optical depth, and the excitation temperature that characterizes the populations of the various rotational energy levels.

The \mzero~maps shown in Figures \ref{fig:mom_maps1} and \ref{fig:mom_maps2} exhibit noticeable differences in total sky area passing our signal-to-noise threshold requirements (described in Section \ref{sect:mom_maps}).  Emission from the lower density tracers $^{12}$CO and $^{13}$CO (which show on average a tendency to align parallel to the magnetic field) covers almost the entire map, while C$^{18}$O and the intermediate or high density tracers, HCN, HNC, HCO$^{+}$, and CS mostly show emission within the column density contour of \nh\,$=$\,1.2\,$\times$\,10$^{22}$\,cm$^{-2}$~(this corresponds to the lowest \nh~contour shown in Figures \ref{fig:mom_maps1} and \ref{fig:mom_maps2}), and the weaker NH$_3$~and N$_2$H$^+$~lines only show emission towards the highest \nh~peaks.

\begin{figure}
\epsscale{1.2}
\plotone{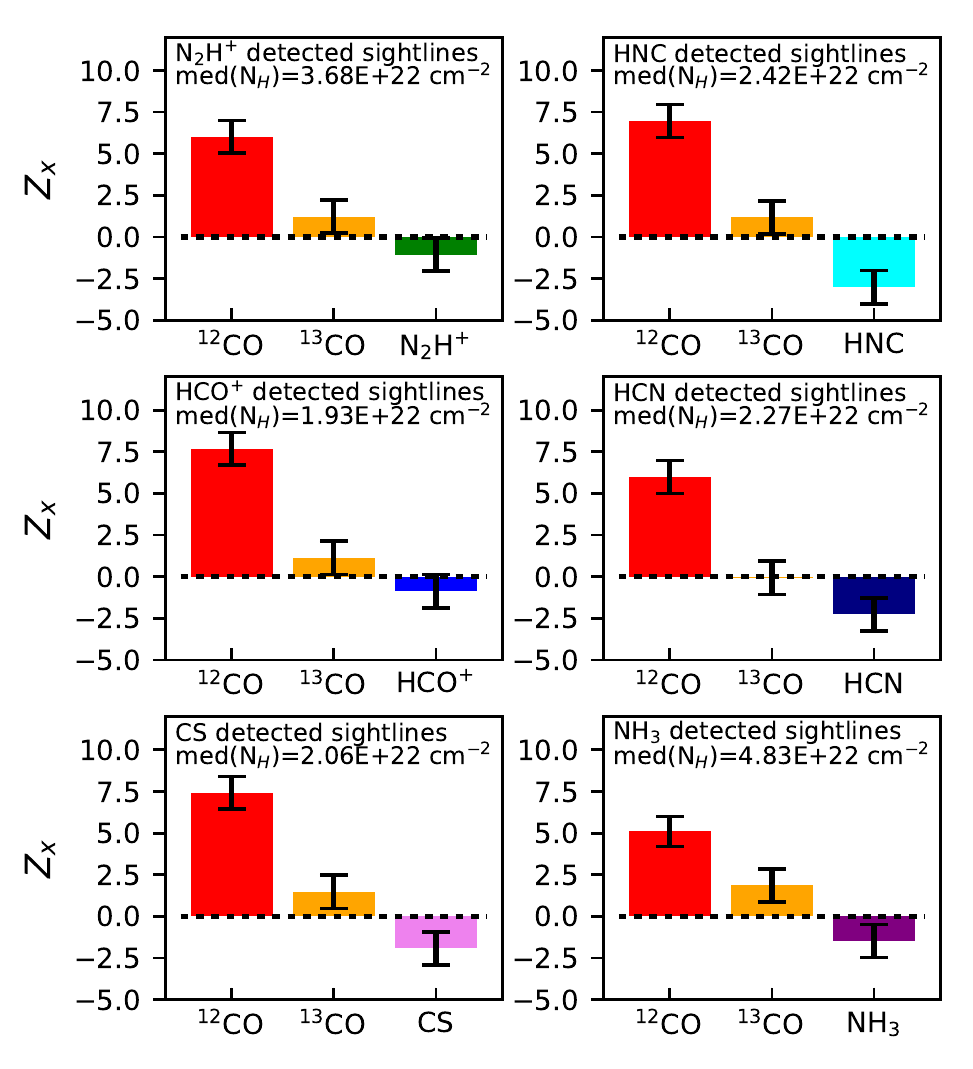}
\caption{Comparison of the projected Rayleigh statistic \prs~calculated for $^{12}$CO and $^{13}$CO~when restricted to sightlines where our intermediate to high density molecular lines (N$_2$H$^+$, HNC, HCO$^+$, HCN, CS, and NH$_3$) are detected.  We also list the median Herschel-derived $N_{\mathrm{H}}$~values for those sightlines in each panel.\label{fig:prs_vs_coverage}
}
\end{figure}

Given the difference in map extent for each of our molecular line \mzero~maps it is possible that the change in relative orientation between our molecular tracers is simply showing the same trend of \prs~with \nh~observed by \cite{soler_2017} and \cite{jow_2018} in Vela C.  \cite{soler_2017} found that below \nh\,$\simeq\,1.2\,\times\,10^{22}$\,cm$^{-2}$ \bpos~is on average parallel to the \nh~iso-contours.  Since only  $^{12}$CO and $^{13}$CO have significant emission at \nh\,$<\,1.2\,\times\,10^{22}$\,cm$^{-2}$ the differences in relative orientation between our observed lines could just be due to the difference in average \nh~sampled by each line.  

To test this hypothesis, in Figure \ref{fig:prs_vs_coverage} we recalculate \prs~for $^{12}$CO and $^{13}$CO only for the sightlines where our intermediate and high density tracers were detected.  We see that even when restricting $^{12}$CO and $^{13}$CO  to the sightlines where higher density tracers are detected the behavior of \prs~shows the same trends: structures in the $^{12}$CO \mzero~map align preferentially parallel to \bpos; $^{13}$CO structures show a weak tendency to align parallel to \bpos, and intermediate to high density tracers show a weak preference to align perpendicular to \bpos.  This suggests that the $^{12}$CO and $^{13}$CO preferentially trace lower density gas in outer cloud regions compared to the higher density molecular tracers.  The only systematic difference in the \prs~values for $^{12}$CO and $^{13}$CO shown in Figures \ref{fig:prs_sys_test} and \ref{fig:prs_vs_coverage} is that \prs~is lower in Figure \ref{fig:prs_vs_coverage}, which is expected as the intermediate and high density tracers have lower values of \nind\,(see Table \ref{tab:prs}) and Equation \ref{eqn:prs} shows that \prs~is proportional to $\sqrt{\mnind}$.  

\begin{figure}
\epsscale{1.25}
\plotone{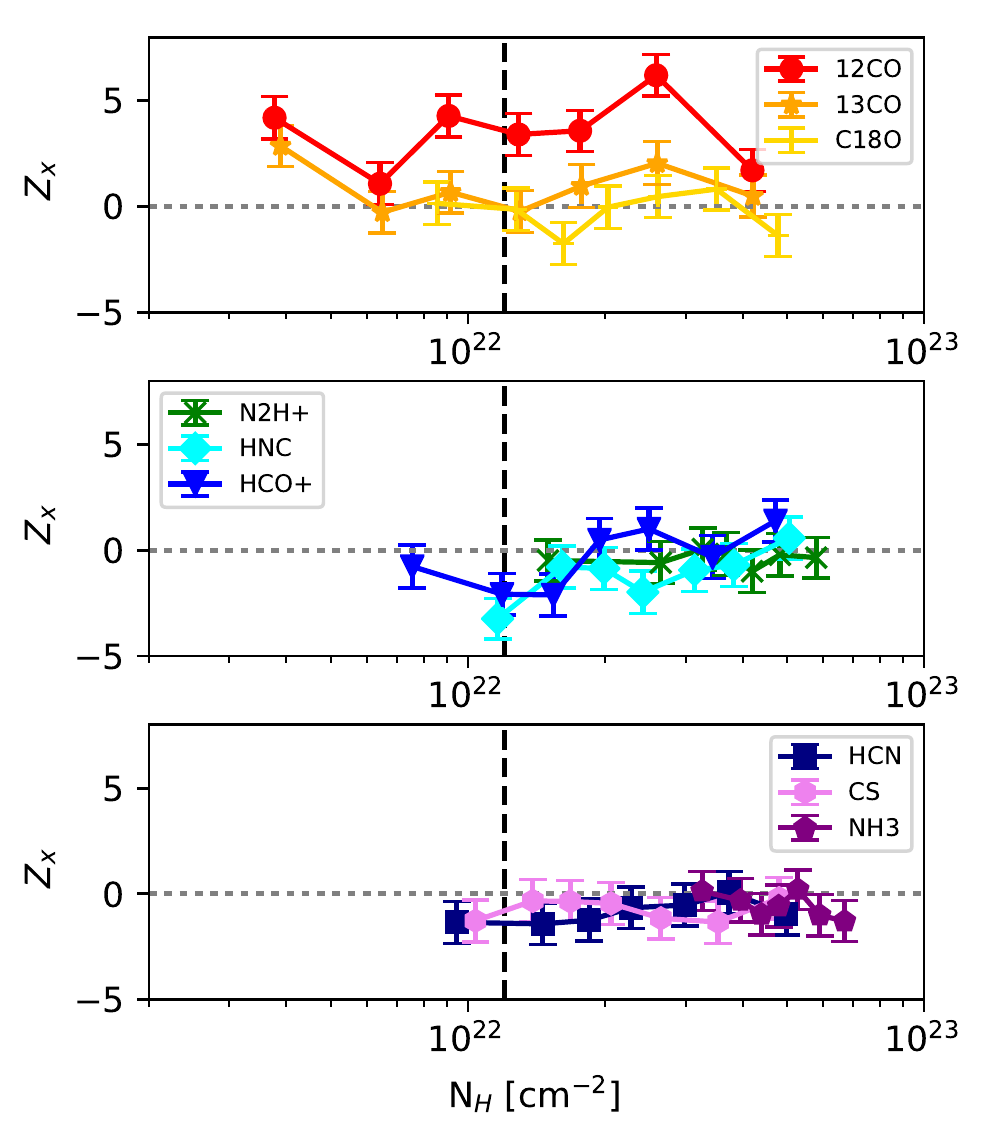}
\caption{Projected Rayleigh statistic \prs~vs.~$N_{\mathrm{H}}$~(as calculated from {\em Herschel}~dust spectral fits) for our sample of nine molecular lines.  The dashed vertical line indicates the $N_{\mathrm{H}}$~intercept in the $\xi_{N_{\mathrm{H}}}$~vs.~$\log_{10}\left(N_{\mathrm{H}}\right)$~fit from \cite{soler_2017}.\label{fig:prs_vs_nh}
}
\end{figure}

We can also directly test for changes in \prs~with column density by dividing our relative orientation angle $\phi$~data into seven groups binned by \nh.  The bins are chosen such that for a given molecular line each group has the same number of sightlines.  We then calculate \prs~for the sightlines in each group.  Figure \ref{fig:prs_vs_nh} shows the change in relative orientation \prs~with increasing \nh.  Overall this figure gives the same impression as Figure \ref{fig:hros_mom0}, in that there is no consistent trend of relative orientation~vs \nh. The average \prs~decreases with \nh~for some tracers (e.g.,~$^{13}$CO and NH$_3$), but increases for other tracers (e.g., HCO$^{+}$\,and HNC).  

In summary, our results are not consistent with a trend in relative orientation versus hydrogen column density, but suggestive of some relationship to volume density and/or excitation conditions.   The magnetic field orientation probed by BLASTPol is always a sum along the line of sight weighted by the dust density, emissivity, and grain alignment efficiency within the volume probed by the telescope beam. For example, if the grain alignment efficiency and temperature were higher in low density cloud regions, the magnetic field orientation measured by BLASTPol could be more sensitive to the field direction in the low density rather than high density cloud regions within the sightline. This averaged \bpos~orientation measurement is what is compared to the orientation of the molecular structures, whether from a low density tracer or a high density tracer, wherever they happen to be along the line of sight. Thus it is important to keep in mind that the preference for intermediate and high density structures to appear aligned perpendicular to the magnetic field measured by BLASTPol does not imply that the magnetic field orientation is that of a field entirely within the volume highlighted by the molecules.

\subsection{Changes in Relative Orientation as a Function of Characteristic Density}\label{sect:prs_vs_dens}
\begin{figure}
\epsscale{1.25}
\plotone{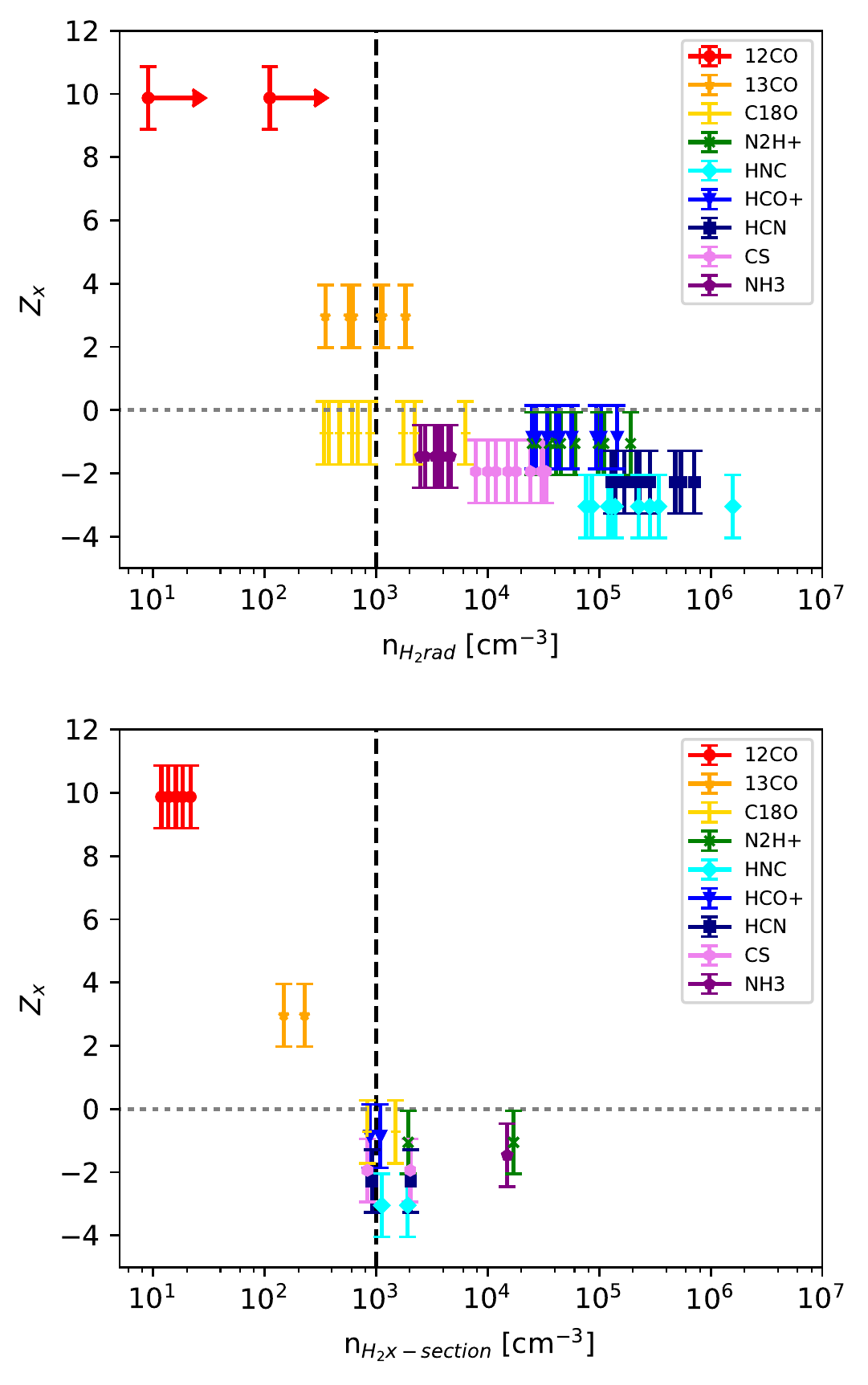}
\caption{Projected Rayleigh statistic \prs,~characterizing the relative orientation of the magnetic field compared to the orientation of elongated structures in the zeroth-moment maps of nine different molecular lines versus molecular hydrogen number density.  {\em Top panel:} Characteristic number density estimated from: the critical density corrected for radiative trapping ($^{12}$CO, lower limits) and {\tt RADEX}~radiative transfer models (all other molecules) as described in Sections \ref{sect:n_radex} and \ref{sect:ncrit_co}. {\tt RADEX}~models were calculated for \tkin\,=\,10, 15, and 20\,K, and for the 5th, 50th, and 95th percentile column densities for a maximum of nine estimates of \nradex~per molecule.   The spread in the calculated values should be taken as a rough estimate of the uncertainty in determining \nradex. {\em Bottom panel:} Characteristic number density estimated from column density cross-sections shown in Figure \ref{fig:cdmaps} and described in Section \ref{sect:n_xsec}. The transition from preferentially parallel (\prs\,$>$\,0) to perpendicular (\prs\,$<$\,0) occurs at approximately $n_{\mathrm{H}_2}\,\sim\,$10$^{3}$\,cm$^{-3}$ (vertical dashed line) for both methods of estimating characteristic density.\label{fig:xsi0_vs_dens}}
\end{figure}

Using the density estimates presented in Sections \ref{sect:n_radex}~to \ref{sect:n_xsec} we can probe the characteristic number density at which the relative orientation of the cloud structure changes with respect to the magnetic field, as traced by the \mzero~maps.  Figure \ref{fig:xsi0_vs_dens}~shows \prs~versus $n_{\mathrm{H}_2}$~for our two number density estimation techniques. The top panel shows \prs~vs.~\nradex, which was derived from the {\tt RADEX}~models (or in the case of $^{12}$CO\,$J$\,=\,1\,$\rightarrow$\,0 the critical density corrected for radiative trapping). The bottom panel shows \ndepth, where we use the molecular column density cross-sections shown in Figure \ref{fig:cdmaps} to estimate the cloud depth and give a rough estimate of the average molecular hydrogen density along the cross-section.  In both panels the values for \prs~are the same as those discussed in Section \ref{sect:prs}, listed in Table \ref{tab:prs}, and shown in Figure \ref{fig:prs_sys_test}.

Figure \ref{fig:xsi0_vs_dens}~shows a transition from a clear detection of preferentially parallel alignment to \bpos (\prs\,$\gg$\,0) for $^{12}$CO to no preferred orientation or a weakly perpendicular alignment (\prs\,$<$\,0) for intermediate and high density tracers. As discussed in Section \ref{sect:prs}~while the intermediate and high density tracers with characteristic densities $n_{\mathrm{H}_2 }\geq\,$10$^3$\,cm$^{-3}$\,tend to individually have low significance values of \prs, this is partially explained by the lower number of independent relative orientation angle measurements \nind\,compared to $^{12}$CO~and $^{13}$CO.  When we calculate the averaged projected Rayleigh statistic accounting for the correlations in \mzero~map structure between~ N$_2$H$^{+}$, HCO$^{+}$, HCN, HNC, CS, and NH$_3$\,we obtain an average \prs\,$=\,-4.03$, showing that on average intermediate and high density gas structures do preferentially align perpendicular to \bpos.

Our results show that the change in \prs~from cloud structures aligned parallel to structures aligned perpendicular to the magnetic field takes place at  molecular gas densities between between those traced by $^{13}$CO and C$^{18}$O. For both number density estimation methods, this transition number density $n_{\mathrm{H}_2 \mathrm{tr}}\sim\,$10$^3$\,cm$^{-3}$, though with the spread in density estimates the uncertainty in the value of $n_{\mathrm{H}_2 \mathrm{tr}}$~could be up to a factor 10.  

Above $n_{\mathrm{H}_2}\sim\,$10$^3$\,cm$^{-3}$, there are significant inconsistencies between the characteristic density estimated for the same molecules in the two panels of Figure \ref{fig:xsi0_vs_dens}.  For the molecules N$_2$H$^+$, HCO$^+$, HNC, HCN, and CS \ndepth~is at least an order of magnitude lower than \nradex.  For HNC and HCN \ndepth~is more than a factor of 100 lower than \nradex.  This discrepancy may in part be due to the estimated density being averaged over the width of the cross-section, and also partly because we assume a molecular gas volume filling factor of unity.  If the molecular gas filling factor is less than unity then \ndepth~will be less than the true characteristic number density probed by the molecular line. In the astrochemical models of a molecular cloud simulation presented in \cite{gaches_2015}, the volume filling factor for these molecules ranges from 0.005 (N$_2$H$^+$~$J$\,=\,1\,$\rightarrow$\,0) to 0.40 (HCN~$J$\,=\,1\,$\rightarrow$\,0).

Molecular abundance variations with density are not accounted for in either technique for estimating the characteristic density.  For example, CO, the primary reservoir of carbon with molecular clouds, is expected to ``freeze-out'' onto dust grains at intermediate densities. In pre-stellar cores \cite{bacmann_2002} estimate that freeze-out becomes important above $n_{\mathrm{H}}\,\sim\,$10$^{4}$\,cm$^{-3}$~(corresponding to $n_{\mathrm{H_2}}\sim$\,5\,$\times$\,10$^{3}$\,cm$^{-3}$).  Lower levels of carbon in the molecular phase can then reduce the abundance of other carbon-bearing molecules such as CS, HCN, HNC, and HCO$^{+}$ \citep{bergin_2007}. In contrast nitrogen-bearing molecules such as  N$_2$H$^+$~and NH$_3$~are not expected to freeze-out onto dust grains, and because these molecules tend to be destroyed in interactions with CO and HCO$^+$, their abundance can increase towards high densities where CO is depleted \citep{aikawa_2001,tafalla_2002,jorgensen_2004}.  These abundance variations most likely are not important at our estimated transition density $n_{\mathrm{H}_2 \mathrm{tr}}\,\sim\,10^3$\,cm$^{-3}$. However, studies of whether the observed trend of decreasing \prs~continues with increasing $n_{\mathrm{H_2}}$~continues beyond $n_{\mathrm{H}_2 \mathrm{tr}}$~will need to consider the possibility of molecular abundance variations with density.

\subsection{Magnetization of Vela\,C Implied by Relative Orientation Analysis}\label{sect:implications}
In the PRS analysis presented in this work we have shown for the first time a clear change in the average orientation of gas structures of different characteristic number density with respect to the magnetic field.  Previous comparisons with synthetic observations of magnetized cloud formation show that this change of relative orientation has implications for the magnetization of Vela\,C. This was first shown by \cite{soler_2013}, who analyzed three {\tt RAMSES}-MHD adaptive mesh refinement simulations with self gravity for low, intermediate, and high magnetization cases (specifically, initial thermal to magnetic pressure ratio $\beta$\,=\,($c_{\mathrm s}/v_{\mathrm A}$)$^2$ = 100.0, 1.0, and 0.1).  After beginning the simulation and allowing turbulence to decay they found that only the highest magnetization simulation (initially sub-Alfv\'enic) showed a change in relative orientation from parallel to perpendicular with increasing density/column density.  The intermediate magnetization simulation, where the turbulence was initially close to equipartition with the magnetic field, showed the alignment changing from preferentially parallel at low values of $n$~or $N$, to showing no preferred orientation at high densities.

Our PRS results thus imply that the cloud-scale magnetic field in Vela\,C is at least trans-Alfv\'enic in strength, and therefore strong enough to have played an important role in the formation of global cloud structure.  This same conclusion was also reached in the studies of \cite{soler_2017} and \cite{jow_2018}, which revealed a change in relative orientation of column density iso-contours and magnetic field orientation with increasing column density (see Section \ref{sect:intro}). 

Does the observation of a change in the project Rayleigh statistic \prs~for gas tracers of different densities give us any additional information about the cloud magnetic field structure compared to the studies of \prs~vs.~\nh~presented in \cite{soler_2017}?  One advantage of studying the change in relative orientation with density rather than column density is that the observed column density distribution will change for different cloud viewing angles.  
This is shown in Figure 10 of \cite{soler_2013}, where different viewing angles resulted in different transition column densities $N_{\mathrm tr}$, even though in both cases the magnetic field is parallel to the plane of the sky.  Studies of \prs~versus~$n_{\mathrm H_2}$~remove this projection effect;
however, this method is still sensitive to yet another projection effect, because the polarization data is only sensitive to \bpos, the orientation of the magnetic field projected on the plane of the sky.  If the mean direction of the cloud magnetic field is exactly parallel to the line of sight then \bpos~will only measure the disordered components of $\mathbf{B}$~and no average correlation of the \bpos~direction with cloud structure is expected.  

Comparisons of the probability distribution functions of the fractional polarization $p$, and the dispersion of polarization angle $S$~on 0.7-pc scales with those from synthetic observations of cloud-forming simulations suggest either that the magnetic field in Vela\,C is highly turbulent and disordered, or that the mean-field direction is highly inclined with respect to the plane of the sky \citep{king_2018}.  The first explanation of a disordered (i.e., relatively weak) magnetic field is in conflict with the PRS observations presented in this work and \cite{soler_2017}. The latter explanation of a highly inclined magnetic field is therefore more likely and might explain why the \prs~versus~$N_{\mathrm{H}}$~trend in Vela\,C appears to be shallower than the same curves for many of the clouds discussed in \cite{planck2016-XXXV}. However, we note that the simulations considered in \cite{king_2018} are highly idealized and did not cover a wide range of cloud physical parameters. A more comprehensive parameter study is being conducted and will be published in a separate paper.

\subsubsection{Origin of the Transition}\label{sect:origins_nh2tr}
The threshold number density $n_{\mathrm H_2 tr}$~at which \prs~changes from positive (parallel) to negative (perpendicular) has been shown to depend on the magnetization level of the cloud, with simulations with a lower Alfv\'en Mach number $\mathcal{M}_{A}$ having a correspondingly lower value of $n_{\mathrm tr}$~\citep{soler_2013, chen_2016}.  \cite{chen_2016} studied the significance of $n_{\mathrm tr}$~in their {\tt Athena}~1\,pc$^3$~simulations of dense cores and filaments formed in the post-shock layer resulting from the collision of two lower density super-Alfv\'enic gas flows.  In their simulations the post-shock layer is initially sub-Alfv\'enic, restricting the gas to mostly flow parallel to the magnetic field direction.  The change in relative orientation from parallel to perpendicular happens where the magnetic field comes into equipartition with the kinetic energy of the gas, i.e., where the gas transitions from sub-Alfv\'enic (magnetic field dominated) to super-Alfv\'enic (dominated by motions generated by self-gravity).  If this change in dominant energy is responsible for the observed change in orientation within Vela\,C with density, then the value of critical density (at $n_{\mathrm H_2 tr}\,\sim\,10^3\,$\,cm$^{-3}$) could be used to estimate the magnetic field strength near the transition region (i.e., $E_{B}\,\approx\,E_{k}$).
We note however that the simulations of \cite{chen_2016} might not be comparable to our observations of Vela\,C as their simulations are for a 1 pc$^3$~volume and are designed to test models of magnetized core formation, while the FWHM resolution of the BLASTPol polarization observations is 0.7\,pc.  Furthermore all of their simulations are sub-Alfv\'enic, while (as shown above) Vela\,C could also be consistent with trans-Alfv\'enic gas motions.

A similar explanation for the origin of the change in relative orientation has been proposed in \cite{yuen_2017}~and \cite{lazarian_2018}. In their simulations of sub-Alfv\'enic non-self-gravitating gas, turbulent eddies form parallel to the local magnetic field, leading to elongated density features parallel to the magnetic field.  At higher densities near self-gravitating regions the gas acceleration will be largest parallel to the magnetic field (as the accelerations perpendicular to the magnetic field are counteracted by magnetic forces). If the magnetic field is dynamically important, the resulting plasma flows can lead to the formation of dense structures orthogonal to the local magnetic field.

However, self-gravity is not the only explanation for the change in relative orientation. \cite{yuen_2017}~note that similar changes in orientation can also occur within shocks.  More generally \cite{soler_2017b} have shown that both the parallel and perpendicular orientations of the density gradient with respect to the magnetic field represent equilibrium states in the ideal MHD turbulent transport equations, and as such tend to be over-represented compared to a random distribution of relative orientations.  In their analysis the change in relative orientation from parallel to perpendicular is associated with divergence in the velocity field in the presence of a strong magnetic field, which could be due to gravitational collapse, but could also be caused by shocks, or other convergent gas flows.

\subsubsection{Relationship to Zeeman-splitting Observation of the B-n Scaling }\label{sect:zeeman}
  We have noted that our derived threshold density for the change in relative orientation $n_{\mathrm H_2 tr}$\,is approximately\ 10$^{3}$\,cm$^{-3}$. The transition density where the powerlaw scaling of the magnetic field changes from $B\,\propto\,n^{0}$~to $B\,\propto\,n^{2/3}$ is $n_{\mathrm{H}}\sim\,300$\,cm$^{-3}$, as derived from Zeeman-splitting observations of HI, OH, and CN \citep{crutcher_2010}. This is a factor of seven lower than our estimate of $n_{\mathrm H_2 tr}$, assuming $n_{\mathrm{H}_2}\,=\,n_{\mathrm{H}}/2$, though as noted in Section \ref{sect:nh_or_dens},  $n_{\mathrm H_2 tr}$~is probably only constrained to within a factor of order 10.  
The change in power-law and increase in magnetic field strength with density coincides with a transition in the average mass-to-flux ratio ($\mu$) from $<$\,1 (sub-critical, implying that the magnetic pressure is sufficiently strong to support the cloud against gravity)~to $\mu\,>\,1$ (super-critical, where the magnetic field alone is not strong enough to support the cloud against collapse).  

A significant difference between the transition density for the $B$--$n$~scaling, and $n_{\mathrm{H_2 tr}}$, our measured threshold density for the change in relative orientation, could imply that different physical processes are responsible for each transition.  This comparison would benefit from a more precise determination of the characteristic values of n$_{\mathrm{H_2}}$~probed by our different molecular line tracers.  This should be possible in future studies if additional rotational lines can be observed for each molecule, as this will allow a better characterization the optical depth, excitation temperatures, and kinetic temperatures of the gas traced by the different molecules.

\subsection{Regional Variations in Relative Orientation}\label{sect:reg_variations}
\begin{figure*}
\epsscale{1.0}
\plotone{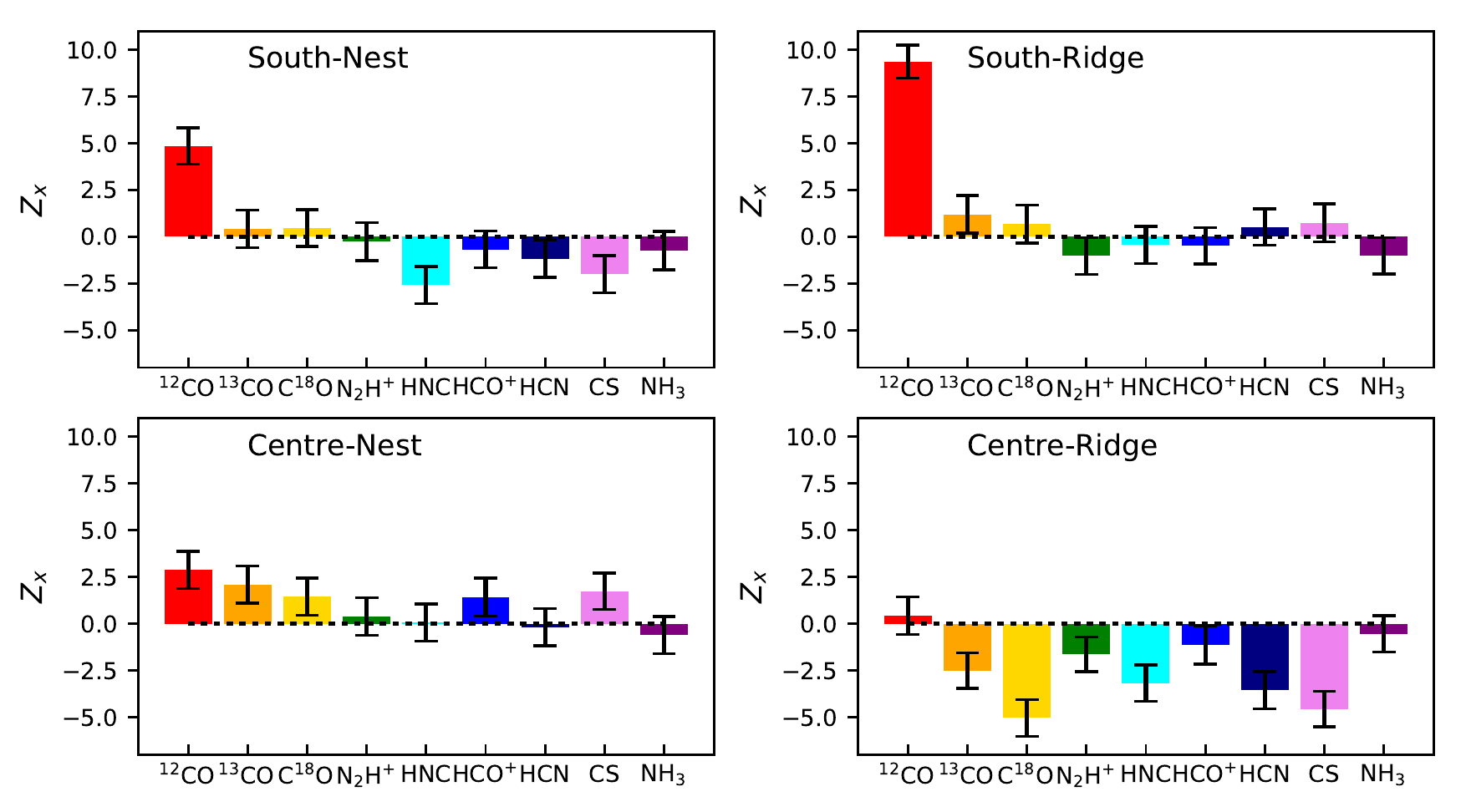}
\caption{Projected Rayleigh statistic \prs~versus molecular line for the Vela\,C sub-regions identified in \cite{hill_2011}~and labeled in Figure \ref{fig:vel_struct}.  
\label{fig:prs_by_hill_reg}}
\end{figure*}

Finally, we look for differences in relative orientation between the magnetic field and cloud structure for each of the four sub-regions identified in \cite{hill_2011},  which are labeled in Figure \ref{fig:vel_struct} and were previously discussed in Section \ref{sect:molecular}. 
\cite{hill_2011} showed that the column density probability distribution functions for the Centre-Ridge and South-Ridge sub-regions extend to higher values and show a shallower power-law slope at high column densities. 
As noted in Section \ref{sect:mom_maps}, the nest-like regions also have on average higher values of \mtwo~for intermediate density tracers without hyperfine line structure, caused by a complicated line-of-sight velocity structure with more than one spectral peak along many sightlines, while the ridge-like regions generally only show one velocity peak.

\cite{soler_2017} and later \cite{jow_2018} both found significant differences in the trends of the relative orientation~as a function of $N_{\mathrm{H}}$ for the different sub-regions within Vela\,C.  The South-Ridge and Centre-Ridge show a much steeper change from positive to negative \prs, compared to the South-Nest or Centre-Nest.  In addition, the change from no preferred orientation to perpendicular occurs at a much lower $N_{\mathrm{H}}$ for the Centre-Ridge, which is the most evolved star forming region in Vela\,C, harboring a young roughly 1-Myr-old OB cluster associated with the compact bipolar \ion{H}{2}~region RCW\,36 \citep{ellerbroek_2013} as well as most of the high mass ($M\,>\,8\,M_{\sun}$) cores in Vela\,C\,\citep{giannini_2012}.

\capstartfalse
\begin{deluxetable}{cccccc}
\tabletypesize{\footnotesize}
\tablecaption{\prs~Comparison for Different Vela\,C Sub-regions.
\label{tab:prs_reg}}
\tablewidth{0pt}
\tablehead{
\colhead{Hill Reg.\footnotemark[1]}  & \colhead{\prs$_{\mathrm{12CO}}$} & \colhead{\prs$_{\mathrm{13CO}}$} &  \colhead{\prs$_{\mathrm{C18O}}$} & \colhead{\prs$_{\mathrm{avg}}$\footnotemark[2]} & \colhead{\nind\footnotemark[3]}
}
\startdata
SN  & ~4.85 & ~0.41 & ~0.46 & $-$2.86 & 5192 \\
SR  & ~9.36 & ~1.22 & ~0.67 & $-$0.46 & 2413 \\
CN  & ~2.87 & ~2.05 & ~1.45 & ~1.30 & 4130 \\
CR  & ~0.44 & $-$2.43 & $-$5.04 & $-$5.59 & 3257
\enddata
\footnotetext[1]{Vela C subregions as defined by \cite{hill_2011}~(see Section \ref{sect:molecular}): SN, South-Nest; SR, South-Ridge; CN, Centre-Nest; CR, Centre-Ridge.}
\footnotetext[2]{Average \prs~calculated for the intermediate to high density tracers N$_2$H$^{+}$, HCO$^{+}$, HCN, HNC, CS, and NH$_3$~as described in Equation \ref{eqn:avg_prs}.}
\footnotetext[3]{Number of independent detections of relative orientation (Equation \ref{eqn:prs_samps}).}
\end{deluxetable}
\capstarttrue

We plot \prs~calculated for our molecular line \mzero~maps for the individual Hill sub-regions in Figure \ref{fig:prs_by_hill_reg}, and list the \prs~values for $^{12}$CO, $^{13}$CO, C$^{18}$O, and the average \prs$_{\mathrm{avg}}$~calculated for the intermediate to high density tracers in Table \ref{tab:prs_reg}. 
The \prs~values for the Centre-Nest, South-Ridge, and South-Nest show similar trends to those seen when the analysis is applied to the entire cloud (Figure \ref{fig:prs_sys_test}). In these sub-regions the $^{12}$CO is on average parallel (\prs\,$>\,$0) while the structure in the intermediate and high density tracers \mzero~maps, has either no strong preferred alignment (e.g.,~the Centre-Nest and South-Ridge) or has a weak preference to align perpendicular to the magnetic field (South-Nest).  

In contrast to the other sub-regions, for the Centre-Ridge we see a preference towards perpendicular alignment between the \mzero~map structure and magnetic field for most lines. The exceptions are $^{12}$CO and HCO$^{+}$\,$J$\,=\,1\,$\rightarrow$\,0, which both show no preferred orientation between \bpos~and \mzero.  According to our RADEX models HCO$^{+}$\,$J$\,=\,1\,$\rightarrow$\,0~is an intermediate density tracer, but it is also commonly used as a tracer of shocked gas, and so the zeroth-moment map for HCO$^{+}$\,$J$\,=\,1\,$\rightarrow$\,0~could be strongly affected by the active star formation in the Centre-Ridge.\footnote{We note that \prs~for HCO$^+$~also appears to be systematically higher when compared to other intermediate and high density tracers for both the South-Nest, and Centre-Nest sub-regions, as well as when the \prs~is calculated for all Vela\,C data (Figure \ref{fig:prs_sys_test}), even though HCO$^+$ has more independent samples than any other intermediate or high density tracer (Table \ref{tab:prs}).} Since both $^{13}$CO or C$^{18}$O have \prs$\,\ll\,$0, it appears that in the Centre-Ridge the transition from mostly parallel to perpendicular happens at lower densities ($n_{\mathrm H_2}\,\lesssim\,$10$^2$\,cm$^{-3}$) compared to the Centre-Nest, South-Ridge and South-Nest, where \prs~typically approaches zero at densities traced by $^{13}$CO or C$^{18}$O ($n_{\mathrm H_2}\,\sim\,$10$^3$\,cm$^{-3}$, as discussed in Section \ref{sect:prs_vs_dens}).  
This implication that $n_{\mathrm{H}_2\,\mathrm{tr}}$ is lower for the Centre-Ridge is consistent with the finding by \cite{soler_2017}~that the transition from parallel to perpendicular occurs at a much lower \nh~for the Centre-Ridge compared to the other \cite{hill_2011} regions.  

Why does the relative orientation of the cloud structure compared to the magnetic field as a function of density show a different behavior towards the Centre-Ridge?  One possibility is that the field in the Centre-Ridge has been affected by the active star formation in the sub-region.  In particular, the field geometry near the OB cluster that powers RCW\,36, a roughly 1-pc bipolar \ion{H}{2}~region aligned perpendicular to the main filament, might be affected by the associated expanding shell of ionized gas \citep{minier_2013}. However, the Centre-Ridge filament extends approximately 5\,pc beyond RCW\,36, where \bpos~is also nearly orthogonal to the main filament, and so this explanation seems unlikely to explain the preference towards perpendicular orientations over the entire sub-region.

Numerical models show that the transition density $n_{\mathrm{H}_2\,\mathrm{tr}}$~is lower in more strongly magnetized clouds \citep{soler_2013,chen_2016,soler_2017}. A strong magnetic field could be expected to slow the progress of star formation by inhibiting collapse in the directions normal to \bpos, but the Centre-Ridge appears have more active star formation than the other Vela\,C sub-regions. 
Another possibility is that a stronger magnetic field in the Centre-Ridge region has allowed more material to gather along the field lines. 

\cite{hill_2011}~speculate that the high column density filaments ($A_V\,>\,$100\,mag) seen in the Centre-Ridge and South-Ridge indicate that these regions were formed via convergent flows.  \cite{soler_2017} note that in numerical simulations of magnetized cloud formation, regions of high density gas are more efficiently created when the matter-gathering flows are directed nearly parallel to the magnetic field, resulting in dense structures oriented perpendicular to the local magnetic field, which then become unstable to gravitational collapse and subsequently form stars \citep{inutsuka_2015,ntormousi_2017, soler_2017b}. They speculate that the Centre-Ridge could be the result of a flow mostly parallel to \bpos~that efficiently formed dense gas and has already collapsed, while the South-Ridge could be at an earlier stage of collapse and the Centre-Nest and South-Nest could be regions formed from convergent flows that were less well aligned with the \bpos, resulting in less high density material being created.  

Our observations of the change in \prs~with density are consistent with this interpretation for the Centre-Ridge. However, our results are less consistent with the interpretation proposed by \cite{soler_2017} for the other Vela\,C sub-regions, because we do not see a clear change to perpendicular alignment for intermediate and high density tracers toward the South-Ridge subregion (\prs$_{\mathrm{avg}}$\,=\,$-$0.46). Indeed Figure \ref{fig:prs_by_hill_reg} shows that, if anything, the intermediate and high density structures in the South-Nest (which has the most disordered magnetic field morphology of the four sub-regions observed with BLASTPol) are more likely to align perpendicular to the magnetic field (\prs$_{\mathrm{avg}}$\,=\,$-$2.88).  This discrepancy with the results of \cite{soler_2017} could be due to the range of spatial scales probed in our Mopra \mzero~maps.  The \herschel~column density maps used in \cite{soler_2017} have $\sim$0.1-pc FWHM resolution, which is the characteristic width of filaments in Vela\,C \citep{hill_2012}, and so the $\nabla$\nh~maps measure the orientation of narrow filamentary structures that cannot be resolved in the Mopra \mzero~maps.  It should be noted though that the \nh~and Mopra \mzero~maps are also not necessarily tracing the same structures: some features in the \nh~maps might be due to projection of multiple cloud density structures along the line-of-sight, while some structures in the Mopra \mzero~map might be due to changes in excitation conditions or molecular abundance variations rather than density gradients.

\section{Summary}\label{sect:summary}
We present a Mopra telescope survey of nine molecular rotational  lines toward the young giant molecular cloud Vela C, which we compare with BLASTPol 500-$\mu$m polarization data in order to study the density, velocity, and magnetic structure of the cloud. We use the projected Rayleigh Statistic (PRS) \prs~to quantify the orientation of gas structures in our molecular line maps (as traced by gradient fields of zeroth-moment, \mzero, maps) with respect to the cloud magnetic field orientation (inferred from BLASTPol data, \bpos).  Each of the mm-molecular lines observed with Mopra is sensitive to different density and excitation conditions, allowing us to test whether there is a systematic difference in relative orientation of cloud structures with respect to the local magnetic field for molecular gas of different densities.  

Our main findings are as follows.
\begin{enumerate}
\item We see a significant change in the average relative orientation between structures in the \mzero~maps and \bpos~for the nine different molecular lines (Section~\ref{sect:prs}). Structures observed with tracers of lower density molecular gas, such as $^{12}$CO and $^{13}$CO tend to align parallel to the magnetic field, while intermediate or higher density tracers (N$_2$H$^{+}$, HNC, HCO$^+$, HCN, CS, and NH$_3$) on average show a weak preference toward orienting perpendicular to the magnetic field. 
The transition from preferentially parallel to no preferred orientation (corresponding to \prs\,=\,0) appears to occur between the densities traced by $^{13}$CO and C$^{18}$O.
\item The change in average relative orientation of \bpos~compared to \mzero~map structures for different molecular lines cannot be solely explained by the tendency previously reported by \cite{soler_2017} for higher column density gas structures to align perpendicular to \bpos~(Section \ref{sect:nh_or_dens}).  When we restrict our calculation of \prs~to only the cloud sightlines that are detected in intermediate and high density tracers, we still find that structures in $^{12}$CO and $^{13}$CO \mzero~maps tend to align parallel to the magnetic field, and within maps of individual molecular lines we see no trend in \prs~as a function of $N_{\mathrm{H}}$.  The differences between the \prs~values appear more likely to be caused by changes in alignment of molecular gas structures of different characteristic densities with respect to the magnetic field. 
\item We estimate the characteristic densities for each of our molecular lines and find that the transition from parallel to weakly perpendicular coincides with a molecular hydrogen number density $n_{\mathrm H_2}\,\sim\,$10$^3$\,cm$^{-3}$\,(Section \ref{sect:prs_vs_dens}).  Given the  assumptions made in calculating the characteristic densities for our molecular observations, this transition density, $n_{\mathrm H_2 tr}$~is likely uncertain by a factor of 10.  Within these large uncertainties, our transition density for the change in orientation of the density structures with respect to the magnetic field is consistent with the $n_{\mathrm H }$~threshold above which Zeeman splitting observations show that $B\,\propto\,n^{2/3}$, which is thought to indicate the density transition where molecular clouds become self-gravitating (Section \ref{sect:zeeman}).
\item We observe regional differences in the line-of-sight velocity structure of the cloud (Section \ref{sect:molecular}).  The ``Centre-Nest'' and ``South-Nest'' sub-regions, which have lower column density filamentary structure with no preferred direction of filament orientation, also tend to have more complicated line-of-sight velocity structure, with line profiles often showing multiple spectral peaks, in contrast to the ``Centre-Ridge'' and ``South-Ridge'' sub-regions, which tend to be dominated by a single high column density filament and usually show a single-peaked spectral line profile. 
\item We measured the relative orientation for each of the four observed sub-regions of Vela\,C identified in \cite{hill_2011} (Section \ref{sect:reg_variations}). The Centre-Ridge, which is the most evolved of these sub-regions and harbors several late type OB stars, shows a strong preference for perpendicular relative orientation of structures in intermediate to high density tracers, C$^{18}$O, and even the relatively low gas density tracer $^{13}$CO.  The transition density $n_{\mathrm{H_2 tr}}$~appears to be lower for the Centre-Ridge, occurring at densities between $^{13}$CO and $^{12}$CO ($n_{\mathrm{H_2 tr}}\sim 10^2$cm$^{-3}$, compared to $n_{\mathrm{H_2 tr}}\sim 10^3$\,cm$^{-3}$~in the other three cloud subregions).  This might represent a dependence of $n_{\mathrm{H_2 tr}}$~on the cloud formation history, or alternatively the orientation of the magnetic field might be affected by feedback from the young stars that have formed and are currently forming within the Centre-Ridge. 
\item Comparing to the simulations of \cite{soler_2013} and \cite{chen_2016}~the observed change in relative orientation with molecular density indicates that the magnetic field in Vela\,C must be globally at least trans-Alfv\'enic (Section \ref{sect:implications}).  This is consistent with previous results from a study of the change in relative orientation of the magnetic field with structures in column density ($N_{\mathrm H}$) maps of Vela\,C by \cite{soler_2017}. 
\end{enumerate}

Our results imply that there is a connection between the structure of dense gas on small scales and the larger-scale cloud magnetic field.  We note that while the analysis in this work represents a significant advance in the study of the relationship between molecular cloud morphology and magnetic field structure, we have only utilized the maps of the simplest observable, namely the zeroth-moment map.  Molecular line data cubes contain a great deal of additional information on the dynamic structure of the cloud.  Future studies of the relative orientation of the magnetic field and gradients in higher order moment maps,  velocity centroids, or velocity channel maps, as well as higher resolution molecular line observations will allow us to better understand both the physical state of clouds like Vela\,C and the role that the magnetic field plays in forming such clouds.

The BLASTPol collaboration acknowledges support from NASA through grant numbers NNX13AE50G, 80NSSC18K0481, NAG5-12785, NAG5-13301, NNGO-6GI11G, NNX0-9AB98G, and the Illinois Space Grant Consortium, the Canadian Space Agency, the Leverhulme Trust through the Research Project Grant F/00 407/BN, the Natural Sciences and Engineering Research Council of Canada, the Canada Foundation for Innovation, the Ontario Innovation Trust, and the US National Science Foundation Office of Polar Programs.  The Mopra radio telescope is part of the Australia Telescope National Facility, which is funded by the Australian Government for operation as a National Facility managed by CSIRO. L.M.F.~is a Jansky Fellow of the National Radio Astronomy Observatory (NRAO). NRAO is a facility of the National Science Foundation (NSF operated under cooperative agreement by Associated Universities, Inc.). J.D.S~acknowledges the support from the European Research Council (ERC)
under the Horizon 2020 Framework Program via the Consolidator Grant
CSF-648505. F.P.~thanks the European Commission under the Marie Sklodowska-Curie Actions within the H2020 program, Grant Agreement number: 658499 PolAME H2020-MSCA-IF- 2014. We would like to thank Jeff Mangum, Brett McGuire, and Helen Kirk for their helpful advice on interpreting the density and chemical structure of Vela\,C.  We would also like to thank Alex Lazarian and Ka Ho Yuen for their advice on interpreting the relationship between intensity gradients and the magnetic field.  This research made use of {\tt APLpy}, an open-source plotting package for Python \citep{robitaille_2012}, and {\tt spectral-cube}, an open-source Python package for the reading, manipulation, and analysis of data cubes.
We thank the Columbia Scientific Balloon Facility staff for their outstanding work.
 \newcommand{\noop}[1]{}

\appendix
\section{A Gaia-Informed Distance to Vela\,C}\label{sect:gaia_dist}
We compute a Gaia-informed distance to Vela C based on the methodology presented in \cite{zucker_2018}. That work determines a distance to the Perseus Molecular Cloud in a two step process. They start by inferring the distance and reddening to individual stars based on their near-infrared (2MASS) and optical (Pan-STARRS1) photometry, using a technique presented in \cite{green_2018}. Parallax measurements from Gaia DR2 are incorporated into the stellar distance estimates when available. \cite{zucker_2018} then model the cumulative distribution of dust along the line-of-sight towards the stars as a linear combination of emission in CO velocity slices. By fitting these per-star distance-reddening measurements they determine distances to the velocity slices towards star-forming regions across Perseus.

\begin{figure*}
\epsscale{1.0}
\plotone{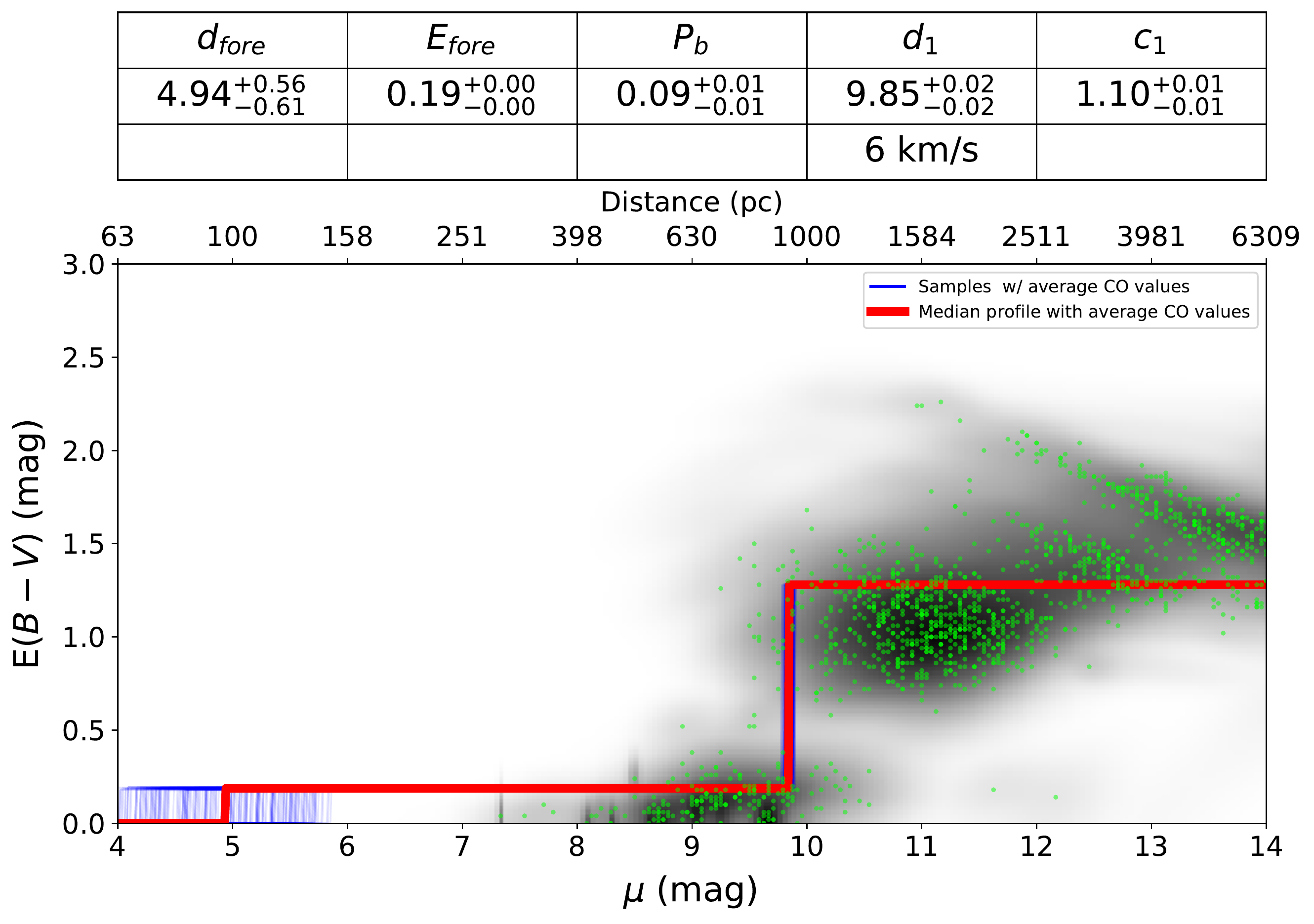}
\caption{Line-of-sight reddening ``profile'' (cumulative reddening as a function
of distance) towards the Vela C cloud. The background grayscale shows
our distance-reddening PDFs for individual stars towards the cloud
stacked on top of each other. Each green point marks the most probable
distance and reddening to each star. The red line is the typical
reddening profile we infer, using the median of our samples
(summarized in the table at top) and adopting the average CO value in
the velocity slice. The cloud distance $d_1$ is our primary free
parameter of interest, placing Vela C at a distance modulus of 9.85
mag, or 933\,pc. The blue line shows random samples from the same run,
and is meant to reflect the underlying statistical uncertainty of our
parameters.
\label{fig:vela_dist}}
\end{figure*}

The method we adopt here is almost identical to that of \cite{zucker_2018}, with the following exceptions. Instead of Pan-STARRS1 optical photometry (which is unavailable at the declination of Vela\,C) we use deep optical photometry from the DECam Galactic Plane Survey \citep{schlafly_2018} to infer the distance and reddening to individual stars. We fit a single velocity template centered at 6 km/s, containing all 12CO emission coincident with Vela,C along the line of sight. We have chosen a representative area towards the middle of the cloud (a circle of radius 0.2\deg, centered on $l\,=\,$265.4\deg, $b\,=\,$1.7\deg) in a region where CO does not saturate. Our primary free parameter of interest is the distance to the CO velocity slice ($d_1$), but we also determine values for various nuisance parameters, including the distance and reddening to an unassociated foreground cloud ($d_{\mathrm{fore}}$ and $E_{\mathrm{fore}}$), a term describing how CO emission is converted to reddening in our CO velocity slice ($c_1$), and a term quantifying the fraction of outlier stars ($P_b$). See Section 5 in \cite{zucker_2018} for a full description of these parameters. We sample for these free parameters using a Monte Carlo analysis. The results of our distance determination procedure is given in Figure \ref{fig:vela_dist}, which shows the ``reddening profile", or cumulative distribution of dust along the line-of-sight towards Vela C. 


We find a distance to Vela\,C of $\mu\,=\,$ 9.85$\,\pm\,$0.02\,mag, or 933$\,\pm\,$9\,pc. While the statistical uncertainty is very low, we estimate there to be additional systematic uncertainty. \cite{zucker_2018} estimated this to be 5$\%$, due to the reliability of their stellar models and their adoption of a fixed extinction curve, which are used to derive the individual distance-reddening estimates. Given the simplicity of our line-of-sight dust model (a single velocity slice, covering all the emission towards Vela C, for a cloud near the Galactic plane) we conservatively recommend the adoption of a 10$\%$~systematic uncertainty, to be added in quadrature with the statistical uncertainty. This produces a distance to Vela C of 933$\,\pm\,$94\,pc. 

\section{Reference regions and Resolution}\label{sect:refs_and_res}

In the analysis presented above we have Gaussian-smoothed the Mopra data to a resolution of 2\arcmin~FWHM before characterizing the orientation of map structures by calculating the gradient for every location in the Mopra zeroth-moment (\mzero) map, as described in Section \ref{sect:mom_maps}.  In this appendix we test whether the method for removing the contribution of the diffuse ISM polarized emission to our polarization maps affects our measurements of \prs~(Section \ref{sect:refregion}).  We also test whether changing the resolution of the smoothed Mopra \mzero~maps or choice of sampling interval significantly changes the value of \prs~calculated for each molecular line (Section \ref{sect:resolution}). 

 \subsection{Dependence of \prs on Diffuse Emission Subtraction Method}\label{sect:refregion}

Vela\,C is located near the Galactic Plane ($b_{\mathrm{center}}\,=\,$1.4\deg) and forms part of the larger Vela Molecular Ridge.  Therefore to study the magnetic field morphology of Vela\,C it is necessary to separate the polarized dust emission originating in Vela\,C from the emission due to dust grains in the diffuse ISM along the same sightlines. 

\begin{figure}
\epsscale{1.0}
\plotone{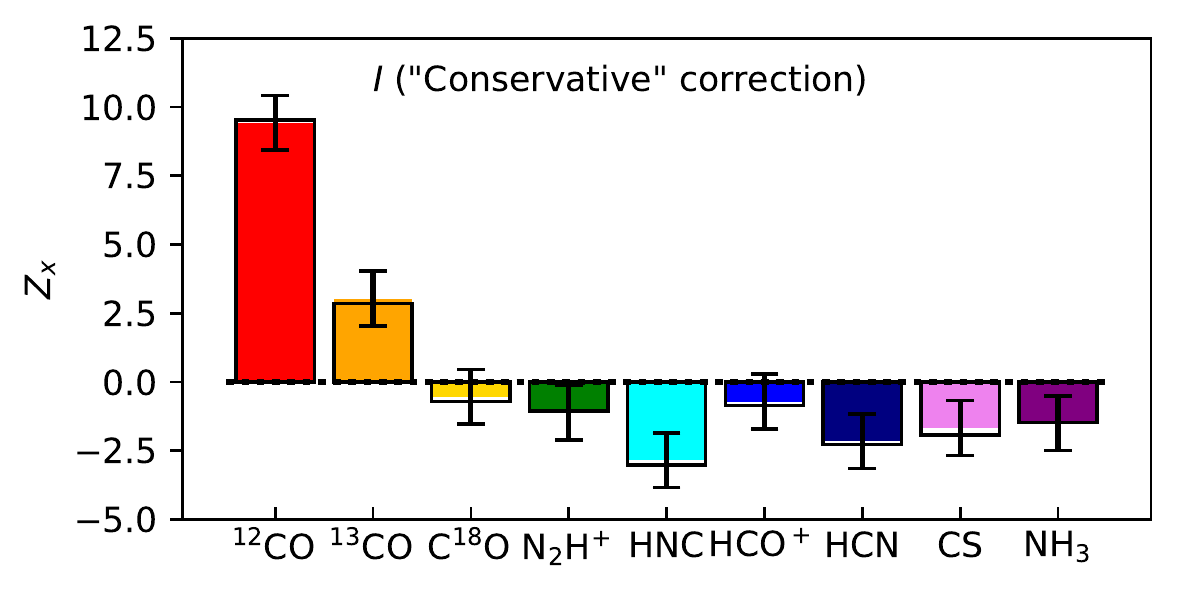}
\plotone{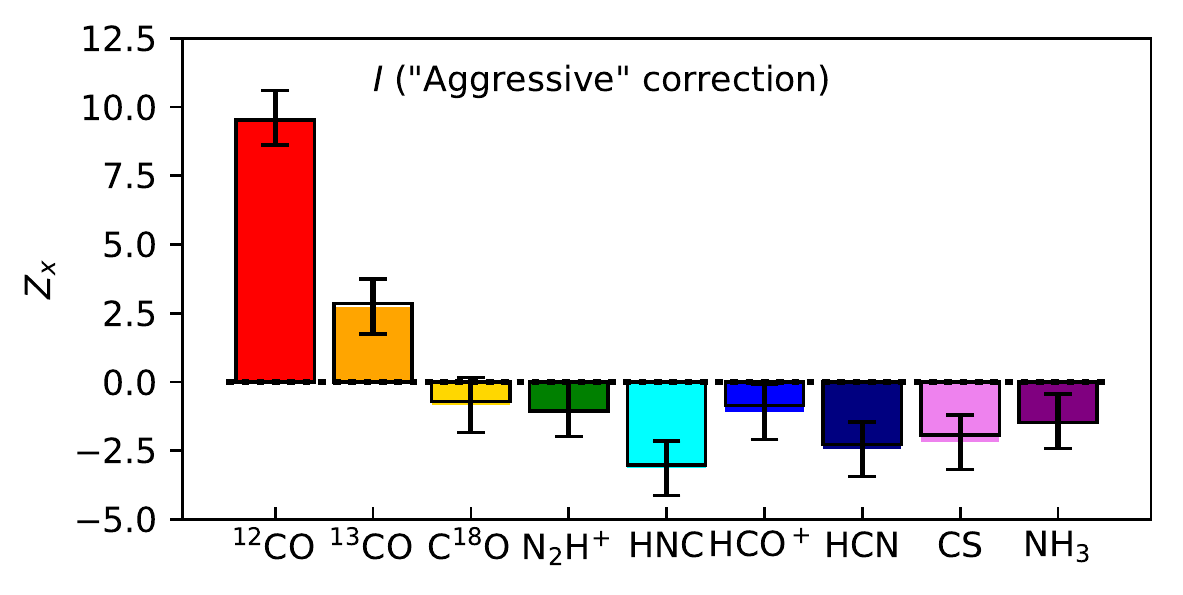}
\caption{Projected Rayleigh statistic \prs~calculated for each molecular line using BLASTPol data that has had two different methods of diffuse polarized emission subtraction applied: a conservative method ({\em top panel}) where a diffuse region to the north of Vela\,C is taken as a template for diffuse emission, and a more aggressive method ({\em bottom panel}), which uses a planar fit to two rectangular regions on either side of Vela\,C as an estimate of the diffuse ISM contribution to the polarized dust emission.  In both plots the \prs~results for the arithmetic mean of the two diffuse emission subtraction methods used in the main paper is indicated with black outlines. 
\label{fig:prs_ref}}
\end{figure}

\cite{fissel_2016}~presented two different methods for removing the diffuse dust emission from the BLASTPol maps.  In the first method the average $I$, $Q$, and $U$~values from a low intensity region to the north of Vela\,C were assumed to represent the contribution of the diffuse ISM to the cloud maps.  This method is ``conservative'' in the sense that it assumes that the diffuse ISM contribution to the Stokes $I$, $Q$, and $U$~maps is uniform and that all of the diffuse emission surrounding Vela\,C is associated with the cloud and not with foreground or background material.  The second, ``aggressive'', method defines two narrow regions close to the Vela\,C cloud, and fits a linear planar model of the diffuse ISM across Vela\,C. 

In \cite{fissel_2016} the analysis was performed on the arithmetic mean of $I$, $Q$, and $U$~maps from the ``conservative'' and ``aggressive'' methods (the  ``intermediate''~case), while the analysis was repeated with the other two diffuse emission subtraction methods to estimate the systematic errors due to diffuse emission correction.  In this paper we have also used the \bpos~maps calculated from BLASTPol data with the ``intermediate'' correction applied. 

Figure \ref{fig:prs_ref}~shows the results of using \bpos~maps made with the ``aggressive''~and~``conservative'' diffuse emission subtraction methods applied.  For each line the \prs~values calculated with the ``intermediate''~diffuse emission subtraction method (used in the main paper) are overlaid with black lines. The \prs~values calculated for the ``conservative'' and ``aggressive'' methods are consistent to within the statistical uncertainties.

\subsection{Dependence of \prs~on Resolution}\label{sect:resolution}
\begin{figure*}
\epsscale{1.0}
\plotone{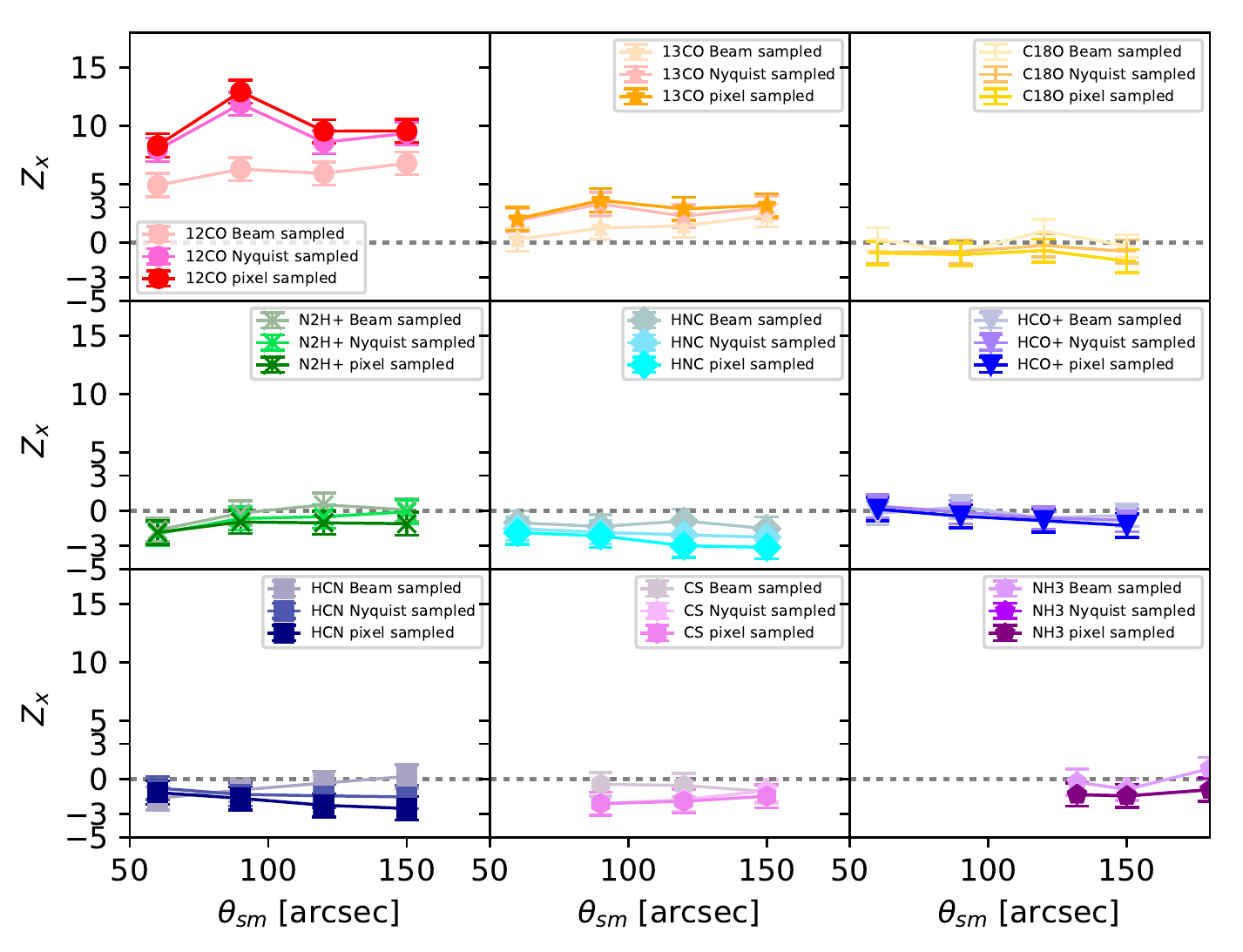}
\caption{Projected Rayleigh statistic \prs~calculated for each molecular line as a function of map FWHM resolution, for different sampling strategies (line color). 
\label{fig:prs_res}}
\end{figure*}

We repeat our projected Rayleigh statistic analysis on gradient maps made from Mopra cubes smoothed to $\theta_{\mathrm{sm}}$\,= 1, 1.5, and 2.5\arcmin~FWHM resolution, or 2\farcm5 and 3\arcmin~FWHM for the lower resolution NH$_3$(1,1) data. The size of the Gaussian gradient kernel, $\theta_{\mathrm{gr}}$, remains the same as the values listed in Table \ref{tab:mom_params}.  The results are shown with solid lines in Figure \ref{fig:prs_res}.  The color of the lines in each panel show the resulting \prs~values for different sampling strategies: in addition to sampling every pixel we also test sampling approximately twice per beam (and so close to Nyquist sampling), and once per smoothed beam FWHM $\theta_{\mathrm{sm}}$. By sampling once per $\theta_{\mathrm{sm}}$~we are clearly missing information, and the calculated \prs~has a lower amplitude compared to \prs~calculated when sampling approximately twice per beam, or sampling every map pixel.  There is also little improvement in the resulting \prs~amplitude from sampling twice per $\theta_{\mathrm{sm}}$~to sampling once every map pixel; the improvement seems to saturate at higher than Nyquist sampling frequencies.

Figure \ref{fig:prs_res} also shows that while the values of \prs~change with resolution the overall trends are not affected: the structure in the $^{12}$CO map always shows a preference towards parallel alignment with the magnetic field, while the intermediate density and high density tracers show a weak preference for perpendicular orientation rather than parallel orientation.  We note that the \prs~derived from moment maps with 1\arcmin~FWHM resolution often have lower absolute values, even though the number of independent data points \nind~should be larger. This could be because \mzero~maps calculated from cubes smoothed to 1\arcmin~FWHM resolution have higher $\sigma_{\mmzero}$~levels, so that there is more randomness in the calculated gradient angles.  It is also possible that the cloud structure on smaller scales within Vela\,C is less well aligned with respect to the magnetic field traced by BLASTPol.

\section{Details of the column density calculations}\label{sec:app_cd}

\capstartfalse
\begin{deluxetable*}{cccccccccc}
\tabletypesize{\footnotesize}
\tablecaption{Column Density Calculation Parameters and Constants
\label{tab:cd_constants}}
\tablewidth{0pt}
\tablehead{
\colhead{Molecular Line} & \colhead{$E_{\mathrm{u}}$} &  \colhead{$\mu$}  &\colhead{S} &  \colhead{$g_{\mathrm{u}}$} &\colhead{$R_{i}$}\footnotemark[1] & \colhead{$B_0$} & \colhead{$C_0$} & \colhead{$Q_{\mathrm{rot}}$} & \colhead{$Q_{\mathrm{rot}}$}\\
\colhead{} & \colhead{[K]} & \colhead{[10$^{-18}$ esu\,cm]} & \colhead{} & \colhead{} &  \colhead{} & \colhead{[MHz]} &  \colhead{[MHz]} & \colhead{ (10\,K)} & \colhead{(20\,K)}
}
\startdata
$^{12}$CO $J$\,=\,1\,$\rightarrow$\,0  & 5.53 & 0.110 & $1/3$ & 3 & 1 & 57636.0 &  -- & 3.94 & 7.56 \\
$^{13}$CO $J$\,=\,1\,$\rightarrow$\,0  & 5.29 & 0.110  & $1/3$ & 3 & 1 & 55101.0 & --  & 4.11 & 7.89 \\
C$^{18}$O $J$\,=\,1\,$\rightarrow$\,0  & 5.27 & 0.111  & $1/3$ & 3 & 1 & 54891.4 & -- & 4.13 & 7.92 \\
N$_2$H$^+$~$J$\,=\,1\,$\rightarrow$\,0 &  4.47 & 3.40 & $1/3$ & 3 & 1 & 46586.9 & -- & 4.80 & 9.26  \\
HNC $J$\,=\,1\,$\rightarrow$\,0        &  4.32 & 3.05  & $1/3$ & 3 & 1 & 45332.0 & -- & 4.92 & 9.52 \\
HCO$^+$ $J$\,=\,1\,$\rightarrow$\,0    &  4.28 & 3.89  & $1/3$ & 3 & 1 & 44594.4 & -- & 5.00 & 9.67 \\
HCN $J$\,=\,1\,$\rightarrow$\,0        &  4.25 & 2.98  & $1/3$ & 3 & 1 & 44316.0 & -- & 5.03 & 9.73\\
CS $J$\,=\,1\,$\rightarrow$\,0         &  2.35 & 1.96  & $1/3$ & 3 & 1 & 24495.6d & -- & 8.83  & 17.32 \\
NH$_3$~(1,1)           &  24.35 & 1.47  & $1/2$ & $12/8$ & 0.5 & 298192.9 & 186695.9 & 0.58 & 1.42
\enddata
\footnotetext[1]{Fraction of the total line intensity for the hyperfine components included in the velocity integration range in Equation \ref{eqn:mom0}.  This is always equal to 1 for lines without hyperfine structure, and for the N$_2$H$^{+}$~and HCN lines, where we integrate over all hyperfine lines (see Table \ref{tab:mom_params}). For NH$_3$ we integrate only over the central hyperfine components, which account for half of the total line strength.}  

\end{deluxetable*}
\capstarttrue

We assume local thermal equilibrium (LTE) and that our lines are optically thin. Following the outline in \cite{mangum_2015} we can relate the properties of the resulting emission line to total column density of the molecule:
\begin{eqnarray} \label{eqn:cd_thin}
N_{\mathrm{tot}}^{\mathrm{thin}}\,&=&\,\left(\frac{3h}{8\pi^3 S\mu^2 R_{i}}\right)\left(\frac{Q_{\mathrm{rot}}}{g_Jg_Kg_I}\right)
\frac{\exp\left(\frac{E_u}{kT_{\mathrm{ex}}}\right)}{\exp\left(\frac{h\nu}{kT_{\mathrm{ex}}}\,-\,1\right)} \nonumber \\ 
 & & \times\,\frac{1}{J_{\nu}\left(T_{\mathrm{ex}}\right)\,-\,J_{\nu}\left(T_{bg}\right)}\int\,\frac{T_{\mathrm{R}}\,dv}{f}
\end{eqnarray}
(equation 80 from \citealt{mangum_2015}).
Here $\mu$~is the dipole moment of the molecule, $g_J$, $g_K$, and $g_I$ are the degeneracies of the upper energy level, $S$~is the line strength, $R_{i}$~is the fractional strength of the hyperfine components included in the integral, $Q_{\mathrm{rot}}$~is the rotational partition function of the line, $T_{\mathrm{bg}}$~is the background temperature (we assume $T_{\mathrm{bg}}\,=\,T_{\mathrm{CMB}}\,$=\,2.73\,K), $f$~is the filling fraction of the emitting molecular gas within the telescope beam (assumed to be 1), and 
\begin{eqnarray}
J_{\nu}\left(T\right) & \equiv & \frac{\frac{h\nu}{k}}{\exp{\left(\frac{h\nu}{kT}\right)}\,-\,1},
\end{eqnarray}
is the Rayleigh-Jeans equivalent temperature.  (The constants used in these calculations are listed in Table \ref{tab:cd_constants}).  The integral $\int \frac{T_{\mathrm{R}}\,dv}{f}$, is equivalent to the zeroth-moment (\mzero) map shown in Figures \ref{fig:mom_maps1} and \ref{fig:mom_maps2} (assuming that the filling fraction $f$\,=\,1).  We calculate $N_{\mathrm{tot}}^{\mathrm{thin}}$~for values of \mzero~at the 5th, 50th, and 95th percentiles of the maps.  These are referred to as the $N_{\mathrm{tot}}^{\mathrm{thin}}$~percentiles.

\capstartfalse
\begin{deluxetable*}{cccc|ccc|ccc|cc}
\tabletypesize{\footnotesize}
\tablecaption{Integrated Line-intensity and Estimated Molecular Column Density Values.
\label{tab:column_density}}
\tablewidth{0pt}
\tablehead{
\colhead{Molecular Line} & \multicolumn{3}{c}{\mzero \footnotemark[1]} & \multicolumn{3}{c}{$N_{\mathrm{thin}}^{\mathrm{tot}}$ (10\,K)\footnotemark[1]} &  \multicolumn{3}{c}{$N_{\mathrm{tot}}^{\mathrm{thin}}$ (20\,K)\footnotemark[1]}    
& \colhead{$<$[$N_{\mathrm{tot}}^{\mathrm{thin}}$/N$_{\mathrm{H2}}$]$>$}
& \colhead{$<$[$N_{\mathrm{tot}}^{\mathrm{thin}}$/N$_{\mathrm{H2}}$]$>$}
\\
\colhead{} & \multicolumn{3}{c}{K\,km\,s$^{-1}$} & \multicolumn{3}{c}{cm$^{-2}$} & \multicolumn{3}{c}{cm$^{-2}$}  &   \colhead{(10\,K)} & \colhead{ (20\,K)} 
}
\startdata
$^{12}$CO $J$\,=\,1\,$\rightarrow$\,0 & 14.46 & 43.42 & 70.92 &
 1.4E+16 & 4.1E+16 & 6.6E+16 & 1.8E+16 & 5.5E+16 & 9.0E+16 &  8.4$\pm$3.9E-06 & 1.1$\pm$0.5E-05 \\
$^{13}$CO $J$\,=\,1\,$\rightarrow$\,0 & ~2.21 & ~8.27 & 17.27 &
 2.2E+15 & 8.2E+15 & 1.7E+16 & 3.0E+15 & 1.1E+16 & 2.3E+16 & 1.6$\pm$0.5E-06 & 2.1$\pm$0.7E-06 \\
C$^{18}$O $J$\,=\,1\,$\rightarrow$\,0 & ~1.16 & ~1.93 & ~3.79 &
 1.1E+15 & 1.9E+15 & 3.7E+15 & 1.6E+15 & 2.6E+15 & 5.1E+15  & 1.9$\pm$0.4E-07 & 2.6$\pm$0.6E-07  \\
N$_2$H$^{+}$ $J$\,=\,1\,$\rightarrow$\,0 & ~0.47 & ~0.73 & ~1.86 &
 6.4E+11 & 9.9E+11 & 2.5E+12 & 9.1E+11 & 1.4E+12 & 3.6E+12  & 6.7$\pm$2.0E-11 & 9.4$\pm$2.8E-11 \\
HNC $J$\,=\,1\,$\rightarrow$\,0 & ~0.77 & ~1.53 & ~4.48 & 1.4E+12 & 2.7E+12 & 7.9E+12 & 1.9E+12 & 3.8E+12 & 1.1E+13  & 2.4$\pm$0.6E-10 & 3.5$\pm$0.8E-10 \\
HCO$^{+}$ $J$\,=\,1\,$\rightarrow$\,0 & ~0.44 & ~0.89 & ~2.62 & 
5.0E+11 & 1.0E+12 & 2.9E+12 & 7.1E+11 & 1.4E+12 & 4.2E+12  & 1.2$\pm$0.4E-10 & 1.7$\pm$0.6E-10\\
HCN $J$\,=\,1\,$\rightarrow$\,0 & ~0.60 & ~1.13 & ~2.66 &
1.1E+12 & 2.2E+12 & 5.1E+12 & 1.6E+12 & 3.1E+12 & 7.3E+12  & 2.0$\pm$0.6E-10 & 2.9$\pm$0.8E-10\\
CS $J$\,=\,1\,$\rightarrow$\,0 & ~1.78 & ~3.45 & ~8.43 & 
2.2E+13 & 4.3E+13 & 1.0E+14 & 3.4E+13 & 6.6E+13 & 1.6E+14  & 4.3$\pm$0.8E-09 & 6.7$\pm$1.2E-09 \\
NH$_3$ (1,1) & ~0.99 & ~1.24 & ~1.87 & 
7.5E+13 & 9.4E+13 & 1.4E+14 & 4.7E+13 & 5.8E+13 & 8.8E+13  & 4.2$\pm$0.9E-09 & 2.6$\pm$0.5E-09 
\enddata
\footnotetext[1]{$N_{\mathrm{tot}}^{\mathrm{thin}}$~column density and zeroth-moment \mzero~ranges are listed for the 5th, 50th, and 95th~percentile values of \mzero, respectively.}  
\end{deluxetable*}
\capstarttrue

Most of our observed molecules (with the exception of NH$_3$) are linear.  To estimate the rotational partition function we use the first two terms of the Taylor expansion in equation 52 of \cite{mangum_2015}
\begin{eqnarray}
Q_{rot} & \simeq & \frac{kT}{hB_0}\,+\,\frac{1}{3},
\end{eqnarray}
where $B_0$~is the rigid rotor rotation constant.  In addition, for linear molecules $g_J$\,=\,2$J_{u}$\,+\,1, $g_{I}$\,=\,1, and $g_{K}$\,=\,1, and $S$\,=\,$\frac{J_u}{2J_u\,+\,1}$.  The total degeneracy of the upper energy level $J_{u}$~$g_{u}$~is the product $g_{I}\,g_J\,g_{K}$.  The values of $\mu$~and $B_{0}$~were obtained from the online catalogs published by the JPL Molecular Spectroscopy Team\footnote{\url{https://spec.jpl.nasa.gov/}.  The value of $C_0$~used in Equation \ref{eqn:qrot}~is from the same catalog.} \citep{pickett_1998}.  

For NH$_3$, a prolate symmetric rotor molecule, we use the approximation for $Q_{\mathrm{rot}}$~from \cite{mcdowell_1990}:
\begin{eqnarray} \label{eqn:qrot}
Q_{\mathrm{rot}} & \simeq & \frac{\sqrt{m\pi}}{\sigma}\exp{\left(\frac{hB_0\left(4\,-\,m\right)}{12kT}\right)}
\left(\frac{kT}{hB_0}\right)^{3/2} \nonumber \\
 & & \times\,\left[1\,+\,\frac{1}{90}\left(\frac{hB_0\left(1\,-\,m\right)}{kT}\right)^2\right],
\end{eqnarray}
where $\sigma$~is the number of identical nuclei in the molecule ($\sigma$\,=\,3 for NH$_3$), and $m$~is $B_0/C_0$, the ratio of the rotational angular momentum constants.  The degeneracies for the para-NH$_3$~(1,1)~line are $g_J$\,=\,2$J_u$\,+\,1, $g_K$\,=\,2, $g_I$\,=$2/8$, and the line strength $S$\,=\,$\frac{K^2}{J_u\left(J_u\,+\,1\right)}$.

We calculate the column density assuming  $T_{\mathrm{ex}}$\,=\,$T_{\mathrm{rot}}$\,=\,$T_{\mathrm{kin}}$\,=\,$T$,\,for $T$\,=\,{10, 15, and 20\,K}. Table \ref{tab:column_density} lists the range of column densities calculated for $T$\,=\,10\,and 20\,K.  These are used in Section \ref{sect:n_radex}.

In addition we estimate the abundance of each molecule compared to H$_2$:
\begin{eqnarray}
\left[\frac{N}{N_{\mathrm{H}_2}}\right] & \equiv & 2\,\times\,\frac{N_{\mathrm{tot}}^{\mathrm{thin}}}{N_{\mathrm{H}}},
\end{eqnarray}
where values of $N_{\mathrm{H}}$~are taken from the 2\farcm5 FWHM resolution $N_{\mathrm{H}}$~map from \cite{fissel_2016} based on fits to \herschel~dust emission maps (see Section \ref{sect:col_dens_nh}).  The median values of $\left[\frac{N}{N_{\mathrm{H}_2}}\right]$~and associated median absolute deviations are listed in Table \ref{tab:column_density}.  We emphasize that these derived abundance ratio maps assume that the molecular emission is optically thin.  This assumption is probably reasonable for less abundant molecules like N$_2$H$^+$, particularly since the much more abundant molecular $^{13}$CO is only, at most, marginally optically thick (Section \ref{sect:ncrit_co}). However, it is very likely that the $^{12}$CO~$J$\,=\,1\,$\rightarrow$\,0 line is highly optically thick across the cloud, so that the actual $\left[\frac{N}{N_{\mathrm{H}_2}}\right]$~is significantly higher than the measured value.  For the analysis in Section \ref{sect:n_xsec} we assume a standard abundance ratio of $\left[\frac{N_{12\mathrm{CO}}}{N_{\mathrm{H}_2}}\right]\,=\,1.0\,\times\,10^{-4}$ instead of using the value listed in Table \ref{tab:column_density}.
For the $J$\,=\,1\,$\rightarrow$\,0 transition of the less abundant tracer $^{13}$CO, our analysis in Section \ref{sect:ncrit_co}, indicates an 95th percentile range in optical depth of 0.15 to 1.8.
\end{document}